\documentclass[11pt,a4paper]{article}
\usepackage{jheppub}
\usepackage{graphicx}
\usepackage{url}
\usepackage{epsfig}
\usepackage{amssymb}
\usepackage{amsmath}
\usepackage{dsfont}
\usepackage{multirow}
\usepackage{axodraw4j}
\usepackage{color}
\usepackage{subfigure,amstext,alltt,setspace}
\usepackage{amsbsy}

\newcommand{\beq}{\begin{equation}}
\newcommand{\eeq}{\end{equation}}
\newcommand{\bea}{\begin{eqnarray}}
\newcommand{\eea}{\end{eqnarray}}

\newcommand{\as}{\alpha_s}

\def\gsim{\mathrel{\rlap{\lower4pt\hbox{\hskip1pt$\sim$}}
    \raise1pt\hbox{$>$}}}         
\def\lsim{\mathrel{\rlap{\lower4pt\hbox{\hskip1pt$\sim$}}
    \raise1pt\hbox{$<$}}}         
\def\raw{\rightarrow}
\def\ie{{\it i.~e.}}
\def\nn{\nonumber}
\def\Aslash{\not{\hbox{\kern-4pt $A$}}}
\def\Eslash{\not{\hbox{\kern-4pt $E$}}}
\def\pslash{\not{\hbox{\kern-4pt $p$}}}
\def\kslash{\not{\hbox{\kern-4pt $k$}}}
\def\lslash{\not{\hbox{\kern-4pt $l$}}}
\def\pone{\not{\hbox{\kern-4pt $p_{1}$}}}
\def\ptwo{\not{\hbox{\kern-4pt $p_{2}$}}}
\def\pthree{\not{\hbox{\kern-4pt $p_{3}$}}}
\def\partialslash{\not{\hbox{\kern-4pt $\partial$}}}

\title{b-initiated processes at the LHC: a reappraisal}
\author[a]{Fabio Maltoni,}
\author[b]{Giovanni Ridolfi,}
\author[c]{Maria Ubiali}
\affiliation[a]{Centre for Cosmology, Particle Physics and Phenomenology CP3,
Universit\'{e} Catholique de Louvain, Chemin du Cyclotron, 1348 Louvain--la--Neuve, Belgium}
\affiliation[b]{Dipartimento di Fisica, Universit\`a di Genova \& INFN, Sezione di Genova \\
Via Dodecaneso 33, 16146 Genova, Italy}
\affiliation[c]{Institut f\"ur Theoretische Teilchenphysik und Kosmologie, RWTH Aachen University,\\ 
D-52056 Aachen, Germany}
\emailAdd{fabio.maltoni@uclouvain.be}
\emailAdd{giovanni.ridolfi@ge.infn.it}
\emailAdd{ubiali@physik.rwth-aachen.de}
\abstract{Several key processes at the LHC in the standard model and beyond that involve $b$ quarks, such as single-top, Higgs, and weak vector boson associated  production, can be described in QCD either in a 4-flavor  or 5-flavor scheme.   In the former, $b$ quarks appear only in the final state and are typically considered massive. In 5-flavor schemes,  calculations include $b$ quarks in the initial state, are simpler and allow the resummation of possibly large initial state logarithms of the type $\log \frac{{\cal Q}^2}{m_b^2}$ into the $b$ parton distribution function (PDF), ${\cal Q}$ being the typical scale of the hard process.  In  this work we critically reconsider the rationale for using 5-flavor improved schemes at the LHC. Our motivation stems from  the observation that  the effects of initial state logs are rarely very large in hadron collisions: 4-flavor computations are pertubatively well behaved and a substantial agreement between predictions in the two schemes is found. We identify two distinct reasons that explain this behaviour, i.e.,  the resummation of the initial state logarithms  into the $b$-PDF is relevant only at large Bjorken $x$ and the possibly large ratios ${\cal Q}^2/m_b^2$'s are always accompanied by universal phase space suppression factors. Our study paves the way to using both schemes for the same process so to exploit their complementary advantages for different observables, such as employing a 5-flavor scheme to accurately predict the total cross section at NNLO and the corresponding  4-flavor computation at NLO  for  fully exclusive studies.}

\keywords{heavy quarks, LHC phenomenology QCD}
\arxivnumber{xxxx.xxxx}
\subheader{\small CP3-12-15,TTK-12-11}
\begin{document}

\maketitle

\section{Introduction}

Processes involving third generation quarks play a key role in collider phenomenology.
The reason is twofold:  on the one hand,  bottom and top quarks have a very peculiar signatures with respect to light quarks that make 
possible an easy flavor identification.  Experimentally, bottom quarks can be efficiently identified, for instance, through displaced vertices in the detector
while top quarks can be detected indirectly via their decay to bottom-quarks and a $W$ boson. On the other hand, they bring about special theoretical attention. Being the heaviest among all quarks,  they do interact most strongly with the electroweak symmetry breaking sector in the standard model (SM) and in many extensions thereof.  A light Higgs boson, for example, predominantly decays into bottom-antibottom quark pairs, providing the main decay channel where the Higgs associated production mechanisms ($WH,ZH,ttH$) can  be searched for.  In addition, in models like supersymmetry, where
the coupling of the Higgs to bottom quarks can be comparable to that of  top quark in the SM, associated production with bottom quarks can provide the largest
production mode. Many of the searches for the Higgs and for new physics in general start from final state signatures which include bottom quarks.  The need for  accurate theoretical predictions for such processes, both signal and backgrounds, is therefore very well motivated. 
 
Calculations of high-energy processes involving the production of  bottom quarks (and heavy quarks in general) are
typically performed in two different ways. One option (which, for simplicity, will be
referred to as the "massive" or 4-flavor (4F)  scheme in the following) is the most
straightforward from the conceptual point of view:  bottom quarks are
significantly heavier than the proton and therefore can be only created in pairs (or singly in association with a top quark, in the case of weak interactions) 
in high-${\cal Q}^2$ interactions. In this approach, the heavy quark does not contribute to the
proton wave-function, and can only be generated as a massive final
state, where the heavy quark mass can act as an infrared cut-off for the more inclusive observables. 
More technically, for this approach to be reliable, one assumes  the heavy quark mass to be of 
the same order  as the other hard scales involved in the process. At the same time,  $b$-parton distribution functions are 
set to zero.   In practice, the massive scheme amounts to
employ an effective theory with $n_l$ light quarks, where the heavy
quarks are decoupled and do not enter in the computation of the
running coupling constant and in the evolution of the
PDFs.  There are many (differential) calculations available at NLO in QCD performed in this scheme relevant for hadron collider phenomenology, including among the most important ones,  
$pp \to bb+X$~\cite{Nason:1989zy,Mangano:1991jk}, 
$pp \to 4b+X$~\cite{Greiner:2011mp}, 
$pp \to ttbb+X$~\cite{Bevilacqua:2009zn,Bredenstein:2010rs},  
$pp \to tbj+X$~\cite{maltoni:stop2},
$pp \to H^\pm tb+X$~\cite{Dittmaier:2009np},
$pp \to \phi bb+X$ with $\phi=H,A$~\cite{Dawson:2003kb,Dittmaier:2003ej,Dawson:2004sh,Dawson:2004wq}, 
and $pp \to V bb+X$ with $V=W,Z$~\cite{Ellis:1998fv,Campbell:2000bg,FebresCordero:2008xi,reinaZWBBtev,Cordero:2009kv,Badger:2010mg,Frederix:2011qg}.
However, no prediction at NNLO in QCD for any process of relevance at the LHC is available in this scheme to date.

Alternatively, one may face a situation where the typical scale ${\cal Q}$ of the process, such as the $p_T$ of the bottom quark itself or the mass weak boson in the final state, 
is way higher than the mass of the heavy quark and logarithms of the type $\log\frac{ {\cal Q}^2}{m_b^2}$ (which can be of initial or final state nature) 
appear that might spoil the convergence of a fixed-order perturbative  expansion. In this case, it is natural to consider a scheme where heavy quark mass are considered as small
parameters, power corrections of the type $(m_b^2/{\cal Q}^2)^n$ are pushed to higher orders and towers
of $\log^m \frac{{\cal Q}^2}{m_b^2}$  appearing at all orders explicitly resummed via usual Altarelli-Parisi evolution equations. We dub this class of approaches and their systematic
improvements as ``massless" or 5-flavor (5F)  schemes.\footnote{We draw the reader's attention to the fact  that the naming 
``massive" and  ``massless" is conventional and {\it it does not} imply anything on  whether power-like bottom mass effects are included or not in actual calculations. There are many accurate predictions available performed in the massive scheme which do not include mass effects because they are small for  the observables of interest. On the other hand, in massless "improved" schemes mass effects are normally taken into account when higher-order terms are included. }  In so doing, initial-state large logarithms are resummed into a $b$ distribution function and final-state ones into perturbative fragmentation functions. In fact initial- and final-state logarithms have a very different status, the latter being not only better understood but also possibly avoided by introducing more inclusive observables than those related to $B$-mesons, such as $b$-jets~\cite{Frixione:1996nh} or even $bb$-jets~\cite{Banfi:2007gu} which are not affected in the first place by such large logs. Different is the case for initial state $b$'s, on which we focus in this work.  Many are the (differential) NLO calculations available in this scheme too that are relevant at LHC: 
$pp \to tj+X$~\cite{Harris:2002md,Kidonakis:2006bu, Campbell:2004ch,Campbell:2005bb,Cao:2004ap,Cao:2005pq}
$pp \to tW+X$~\cite{Campbell:2005bb, Frixione:2008yi}, 
$pp \to H^\pm t+X$~\cite{Plehn:2002vy,Weydert:2009vr}
$pp \to \phi (bb)+X$~\cite{Maltoni:2003pn},
$pp \to \phi b (b) + X$ with $\phi=H,A$~\cite{Campbell:2002zm},
$pp \to V b+X$~\cite{WBmaltoni,ZBmaltoni}, 
and $pp\to   V b j +X$~\cite{Campbell:2005zv,Campbell:2006cu}  with $V=W,Z$. 
In addition,  $pp \to \phi (bb)+X$ with $\phi=H,A$~\cite{Harlander:2003ai} and  $pp \to Z (bb)+X$~\cite{Maltoni:2005wd} are also available at NNLO.

Both schemes present advantages and disadvantages and are typically adopted
in complementary regimes depending on the relative size of the heavy
quark masses relative to the scale $Q$ of the process. 
In turns out that  in the massless schemes calculations are highly simplified: in the Born configuration
 the number of external legs as well as the scales in the processes are reduced. 
 Mass and ``spectator heavy quarks" effects appear at higher orders and can be systematically added. 
 In addition, as mentioned  above, potentially large logarithms of the ratio ${\cal Q}^2/m_b^2$ 
arising from  collinear splitting of the initial heavy quarks and gluons, are consistently
resummed in the heavy quark PDF .  In the massive scheme instead, the
computation is more complicated due to the presence of massive final
states and higher multiplicity; however, the full kinematics of the heavy quarks is 
 taken into account already at the leading order and can be
accurately studied via a  next--to--leading order computation. In the latter case,
the implementation in parton shower codes is also straightforward as no complications
due to the inherent arbitrariness in the description of  massive effects arise. The 
downside is that possibly large logarithms developing in the
initial (and also the final state)  are not resummed and could lead to a poorly-behaved
perturbative expansion. In this work we will argue that in fact, at hadron collider, this is 
hardly the case for a large class of processes/observables, and such resummation effects are
quite mild in general and often smaller
than other approximations inherent in fixed order calculations.

To all orders in perturbation theory the two schemes can be defined such as to be exactly identical.
On the other hand, the way of ordering the perturbative expansion is different and at
any finite order the results might not necessarily match. This is the case for the many 5F improved 
schemes available which typically differ order-by-order in perturbation theory due to the different 
possible choices in including subleading 
terms. For some processes the difference between leading order calculations performed in the two schemes may
be very significant and  in extreme cases have yielded to (only apparently) puzzling predictions
differing up to an order of magnitude as in the case of the $b$--initiated Higgs 
production~\cite{Campbell:2002zm,Dittmaier:2003ej,Maltoni:2003pn, Dawson:2006dm,Maltoni:2005wd,kramer}.

In this work we critically reconsider the motivations for using 5F or massless improved schemes in which 
potentially large logarithms are resummed.  Our aim {\it is not} to analyse in detail the technical differences between 
the various approaches to 5F improved schemes, but to assess the relevance of the large logarithms in first place for 
processes of interest to LHC phenomenology and the need or not to employ
such schemes.  Our argument is based on two main points. First on a simple yet quantitive study of the importance of the
resummation effects in $b$ distribution functions with respect to having only explicit leading and subleading logs at fixed 
order.
Second on the systematic assessment of the size of such logs in 4F calculations. To this aim, we perform a thorough analysis  of 
their origin and role in various processes initiated by a single bottom. We analyse in
detail some representative processes that can be treated analytically
and cover a broad spectrum of possibilities.  In the single heavy quark category fall both lepton-hadron (e.g. DIS)
and hadron-hadron collider processes. The first category is mostly interesting
for pedagogical reasons, and because it is the place where heavy flavor
schemes have been profusely discussed in the literature. It might also be
interesting in view of the LHeC electron proton collider~\cite{Klein:2010zz}.  In 
the hadron-hadron processes we consider both $2\to2$ processes like associate 
bottom and vector boson production and $2\to 3$ processes like single top production. 

As a result, we will argue that at the LHC, the logs that are resummed in $b$ distribution
functions, apart from extreme cases which we will clearly identify, are not very large 
because the possibly large ratios ${\cal Q}^2/m_b^2$'s are always accompanied by universal phase space suppression factors 
at all orders in perturbation theory.  Such logs, therefore, do not spoil the convergence of the perturbative series of the 4F flavor 
scheme whose results are in general well-behaved.  5F schemes calculations, on the other hand, do typically display smaller scale uncertainties 
leading to more accurate predictions for very inclusive observables, such as total rates. However, for differential distributions and more
exclusive observables their predictions might be less accurate, technically more involved and in general less suitable to 
theoretical-experimental comparisons with respect to  those performed via NLO computations in the 4F scheme interfaced to parton showers, employing  MC@NLO~\cite{Frixione:2002lr} or the POWHEG~\cite{powheg0} methods.

The paper is organized as follows. In Section \ref{sec:schemes} we provide a brief overview on the various possibilities available in the literature for
the implementing 5F (improved) schemes. The section is meant for a reader who is not familiar with the subject and would like to learn the basic
concepts/techniques as well as where to find more details in selected references. In Section \ref{sec:resummation} we show that, apart from 
extreme kinematical situations,  the resummation effects in the $b$-distribution functions 
are mild compared to the contributions from the explicit leading and subleading logs present at fixed order in  4F scheme calculations.  
In Section \ref{sec:dis} the DIS case
is discussed. This allows to set the stage and illustrate our analysis approach in a simple and very instructive case, where  $\log \frac{{\cal Q}^2}/{m_b^2}$  are clearly
identified and whose relevance can be easily studied as a function of the kinematical setup. In Section \ref{sec:pp} we finally attack processes at the LHC and show that initial collinear logarithms are typically not very large. The last section presents a summary together with our conclusions.

\section{Heavy quark schemes: a short review}
\label{sec:schemes}

Processes involving heavy quarks often provide a good example of
multi--scale processes.  First, let us define a heavy quark to be one
whose mass $m$ is large enough that the effective coupling at the
scale of a heavy quark mass $\as(m^2)$ is in the perturbative
region. With this definition, the bottom and top quarks are definitely
heavy quarks.  The charm quark mass, on the other hand, is rather
close to the boundary region and QCD perturbative corrections at the
scale of the charm mass are generally large, and can only be taken as
an indication.

Whenever a cross section is characterised by two different scales,
here the hard scale ${\cal Q}$ and the mass of a heavy quark
$m$, perturbative calculations typically display both logarithms and
powers of the ratio $m^2/{\cal Q}^2$. The former contributions may
spoil the accuracy of the calculation if ${\cal Q}\gg m$. It is easy
to trace back the dynamical origin of such logs to the process of
gluon splitting into a heavy quark-antiquark pair. Such splitting can
take place either in the initial or in the final state. As already mentioned,
final state logs can be resummed into fragmentation
functions or cancelled by defining more inclusive observables such as
heavy-quark jets~\cite{Frixione:1996nh}.  In the following, we focus
on the logarithms arising from initial state gluon splitting.  By
defining a heavy quark parton density, one would automatically resum
the logarithms to all orders in perturbative QCD via the
Dokshitzer-Gribov-Lipatov-Altarelli-Parisi (DGLAP) evolution
equations: when computing the heavy quark parton densities from the
1--loop DGLAP evolution equations, all possible multiple collinear
gluon emission from the initial state are in effect summed up to all
orders in $\as$ in the leading--log approximation. However, this
approach is only adequate at large ${\cal Q}$, where the mass of the
heavy quark is small with respect to the typical scale of the
process. At smaller $\cal Q$, powers of $\log({\cal Q}^2/m^2)$ do
not spoil perturbation theory, and power corrections $m^2/{\cal Q}^2$
need to be consistently included.

Theoretical predictions in QCD have been performed according to a
variety of schemes for dealing with heavy quark masses.  We
restrict our discussion to two (classes of) schemes for dealing with
the $b$ distribution, which we denote by 5F or massless and 4F or
massive schemes. Technically, our definition of the 4F scheme
corresponds to the so-called Fixed Flavor Number (FFN) scheme with
$n_f=4$. In this scheme the bottom quark appears only among the final
state particles, and is not associated to a PDF.  Calculations in the
4F scheme are meant to provide reliable results when the scale of the process
is not much larger than $m$. At any finite order in a perturbative
calculation, the FFN result is expected to break down as the scale
becomes large compared to $m$.

Natural extensions of the FFN scheme are the so-called Variable Flavor
Number (VFN) schemes, which consist of a sequence of $n_f$-flavor FFN,
each in its region of validity, consistently matched at the transition
thresholds. There are, however, several ways of implementing a VFN scheme.
The simplest implementation of the VFN scheme is the zero-mass (ZM)
approximation, where all quarks are treated as massless. Heavy quarks
are absent at scales $\mu^2<m^2$, and they are radiatively generated
at the transition points $\mu^2=m^2$ by the subprocess $g\raw Q\bar{Q}$, 
but apart fom this
they are treated as massless partons.
The ZM-VFN scheme is therefore a combination of massless $\overline{\rm MS}$
schemes characterized by different numbers of light flavors
$n_l$. Basically the only mass effects are due to the change of number
of flavors in the QCD $\beta$ function and in the anomalous dimensions
as one crosses the heavy quark thresholds. The coefficient functions
are calculated under the assumption that all active partons are
massless, with the associated singularities subtracted in the
$\overline{\rm MS}$ scheme. This scheme becomes inaccurate for scales
close to the thresholds. This does not depend uniquely on the fact
that $\mathcal{O}(m^2/{\cal Q}^2)$ terms are neglected in the
coefficient functions, but also on the approximate treatment of the
phase space and the arbitrariness in the definition of the scaling variable,
i.e. the fraction of energy in the collision taken by the heavy
quark. Due to the absence of mass effects, the ZM formalism 
is not a good approximation in the region 
near the physical thresholds.

In the following, with 5F we indicate a scheme where the collinear
logarithms appearing in each term of the FFN scheme computation are
resummed to all orders in the heavy quark PDF, and where powers of
$m^2/{\cal Q}^2$ are consistently included at higher orders.  These
schemes are called General--Mass VFN scheme and all partons, including
the bottom, are associated to a parton distribution function above
threshold. The GM-VFN scheme represents an improvement over the ZM-VFN
scheme. However, it also introduces new ambiguities with respect to
the shifting of higher-order terms into the lower-order
expressions. In the GM-VFN schemes the mass of the heavy quarks is
taken into account in the partonic cross sections, and the schemes are
designed to interpolate between the FFN scheme, which gives a correct
description of the threshold region, and ZM-VFNS which accounts for
large energy logarithms. The first proposed matching technique for the
inclusion of mass-suppressed contributions, built upon the CWZ
renormalization scheme~\cite{CWZ}, was developed long ago; it is
the so--called ACOT scheme~\cite{acot}. It provides a mechanism to
incorporate the heavy quark masses in theoretical calculations both
kinematically and dynamically. It yields the complete quark mass
dependence from the low to high energy regimes; for $m\gg {\cal Q}$ it
ensures manifest decoupling, and in the limit $m\ll {\cal Q}$ it
reduces precisely to the $\overline{{\rm MS}}$ scheme without any
finite renormalization term.  Several variants of this method were
subsequently proposed, such as S-ACOT~\cite{sacot} or 
ACOT-$\chi$~\cite{Tung:2001mv},
which include some improvements and/or simplifications with respect the
original ACOT scheme.  A different approach, the Thorne-Roberts~\cite{trHQ} 
VFN scheme or TR'~\cite{TRp} in its latest version, emphazises
the correct threshold behaviour and includes higher-order terms in
order to smoothen the function at the transition points. Namely, the
NLO computation includes $\mathcal{O}(\as^2)$ terms, which formally
belong to the next-to-next-to-leading order, which are introduced in order to
cure the discontinuities of the physical observables at the heavy
quark mass threshold.
An alternative approach considers
both massless and massive scheme calculations as power expansions
in the strong coupling constant, and replaces the coefficient of the
expansion in the former with their exact massive counterpart in the
latter. A first step in this direction was taken in 
Ref.~\cite{Buza:matchPDF}.
More recently, another method, called FONLL, was introduced 
in Ref.~\cite{fonll} in the context of
hadroproduction of heavy quarks, and was recently applied to deep--inelastic
structure functions in Ref.~\cite{ForteNason}. 
Most of the up-to-date parton
sets~\cite{NNPDF21,NNPDF21nnlo,MSTW08,CT10,HERA} implement one of the
above mentioned matched scheme as a default scheme.  On top of that,
each collaboration also provides FFNS parton sets with
$n_f=3,4,5$. Other PDF fitting collaborations~\cite{ABM11,JR09}
provide as default sets FFNS parton sets with $n_f=3,4,5$.  At a fixed
order $n$ in perturbation theory, the difference between these schemes
amounts to adding different ${\cal O}(\as^{n+1})$ higher-order terms,
therefore the difference is reduced as one increases the perturbative
order. A benchmark comparison is available in Ref.~\cite{Binoth:2010ra}.


\section{Impact of resummation}
\label{sec:resummation}

In the previous section we have distinguished the 4F schemes, where
$b$-PDFs are not present and collinear logarithms of the mass of the
bottom are not resummed, and the 5F schemes, where instead this class
of logarithms is resummed. The question we address in this section is
the following:
\begin{itemize}
\item What is the typical size of the effects of the resummation of
  initial state collinear logs of the type $\log {\cal Q}^2/m_b^2$
  with respect to an approximation where only logs at a finite order
  in pertubation theory are kept?
\end{itemize}
While a detailed answer clearly depends on the kinematic regime
considered, it is interesting to explore the possibility that general
features might be present at colliders in general and at the LHC in
particular. Quite surprisingly, while already addressed in the past in
several specific contexts, to our knowledge it has never been answered
for $b$-PDF's. 

In order to assess the relevance of the resummation of logarithms of
${\cal Q}^2/m^2$ performed in  the 5F scheme, one must compare the 
exact solution of DGLAP equations, and its expansion in powers
of $\as$ up to a given fixed order. This comparison can be 
easily performed analytically at leading order. In this case
the full $b$ parton density is a solution of the evolution equation
(in Mellin moment space)
 \begin{equation}
\frac{d}{d\log\mu^2}b(N,\mu^2) 
=\frac{\as(\mu^2)}{2\pi}
\left[\gamma^{(0)}_{qq}(N)b(N,\mu^2)+\gamma^{(0)}_{qg}(N)g(N,\mu^2)\right],
\end{equation}
where $g(N,\mu^2)$ is the Mellin-transformed gluon distribution,
$\gamma^{(0)}_{ij}$ are the leading order Altarelli-Parisi anomalous dimensions,
and the initial condition is $b(N,m_b^2)=0$. This equation can be 
easily solved if the gluon scale dependence is neglected, an approximation which
is valid at this order if $N$ (i.e., $x$) is not too small. 
A straightforward calculation yields
\begin{equation}
b(N,\mu^2)=\frac{\gamma^{(0)}_{qg}(N)}{\gamma^{(0)}_{qq}(N)}
\left\{\left[1+\frac{\as(m_b^2)}{2\pi}\beta_0
\log\frac{\mu^2}{m_b^2}\right]
^{\frac{\gamma^{(0)}_{qq}(N)}{\beta_0}}-1\right\}g(N,m_b^2),
\end{equation}
where $\beta_0=11-\frac{2}{3} n_f$. This solution can be expanded in powers of
$\as(m_b^2)$. Using the binomial expansion
\begin{equation}
(1+x)^a=\sum_{k=0}^\infty\frac{\Gamma(a+1)}{\Gamma(a-k+1)}\frac{x^k}{k!}
\end{equation}
and standard properties of the $\Gamma$ function we find
\begin{equation}
b(N,\mu^2)=\gamma^{(0)}_{qg}(N)g(N,m_b^2)
\Bigg\{\frac{\as(m_b^2)}{2\pi}\log\frac{\mu^2}{m_b^2}
+\sum_{k=2}^\infty A_k(N)
\frac{1}{k!}\left[\frac{\as(m_b^2)}{2\pi}
\log\frac{\mu^2}{m_b^2}\right]^k\Bigg\},
\label{eq:b-expansion}
\end{equation}
where
\begin{equation}
A_k(N)=\left[\gamma^{(0)}_{qq}(N)-\beta_0\right]
\left[\gamma^{(0)}_{qq}(N)-2\beta_0\right]
\ldots
\left[\gamma^{(0)}_{qq}(N)-(k-1)\beta_0\right].
\end{equation}
Clearly, the resummation terms of order $\as^2$ and higher
become relevant when the factorization scale $\mu$, typically of the order
of the hard scale ${\cal Q}$ of the process, is much larger than $m_b$.
We note, however, that the coefficients $A_k(N)$ behave as $\log^k N$
at large $N$, due to the large-$N$ behavior $|\gamma^{(0)}_{qq}|\sim\log N$,
originated by the soft pole in the Altarelli-Parisi
splitting function. We therefore expect the effects of resummation
to be more important as the momentum fraction $z$ gets close to 1,
which corresponds to the large-$N$ region in Mellin space.

As we will see in the following, this simple derivation nicely reproduces the qualitative
behaviour of the evolution of the $b$-PDF at medium to large $x$ both at LO and NLO.
However, the approximation of no-evolution for the gluon pdf is untenable at small $x$ and therefore 
it cannot be used in this regime. In fact, as we will show in the following, even the exact evolution at LO
is not  reliable at small $x$ and the NLO accuracy is needed.


If no intrinsic bottom content of the proton is envisaged, then the
bottom parton density which appears in the 5F scheme is generated at
threshold (by setting $b(x, m_b^2)=0$ as a boundary condition) by gluon
and light parton distributions. As a result the
bottom PDF is not an independent quantity: it depends on the
four-light flavour ($u$,$d$,$c$,$s$) and the gluon ($g$) densities in
the 4F scheme. Following the notation of Ref.~\cite{Buza:matchPDF} the bottom PDF can
be written as a function of the light flavor singlet and the gluon up
to $\mathcal{O}(\as^2)$ as
\begin{eqnarray}
b^{\rm (p)}(x,\mu^{2}) &\equiv& 
\tilde{b}^{(p)}(x,\mu^{2}) + \mathcal{O}(\as^{p+1}(\mu))\nn\\
&=&\sum_{i=q,g}\int_{x}^{1}\frac{dz}{z}f_i^{4F,(p)}
\left(\frac{x}{z},\mu^{2}\right)\,A_{i,b}^{(p)}
\left(z,\frac{\mu^{2}}{m_{b}^{2}}\right),
\end{eqnarray}
where by $\tilde{b}^{(p)}$ we indicate the approximated $b$
distribution that one obtains when truncating the perturbative
expansion of the bottom PDF evolution at $\mathcal{O}(\as^p)$.  The
operator matrix elements $A_{i,k}^{(p)}$ have been computed in the
Mellin $N-$ space in the asymptotic limit $Q^2\gg m_b^2$
up to $\mathcal{O}(\as^3)$ and in the physical
$x-$space up to $\mathcal{O}(\as^2)$ in
Refs.~\cite{Buza:matchPDF,Chuvakin:2001ge,Bierenbaum:2009zt,Buza:1995ie,
Bierenbaum:2007qe,Bierenbaum:2008yu,Bierenbaum:2009mv}.
In the physical $x-$space up to $\mathcal{O}(\as^2)$, they read
\begin{eqnarray}
\tilde{b}^{(2)}(x,\mu^{2}) &=&\int_{x}^{1}\frac{dz}{z}\Sigma^{4F,(2)}\left(\frac{x}{z},\mu^{2}\right)\,\left(\frac{\as}{4\pi}\right)^{2}a_{\Sigma,b}^{(2)}(z,\mu^{2}/m_{b}^{2})\\
&+& \int_{x}^{1}\frac{dz}{z}g^{4F,(2)}\left(\frac{x}{z},\mu^{2}\right)\,\left[\left(\frac{\as}{4\pi}\right)a_{g,b}^{(1)}(z,\mu^{2}/m_{b}^{2})+\left(\frac{\as}{4\pi}\right)^{2}a_{g,b}^{(2)}(z,\mu^{2}/m_{b}^{2})\right]\nn,
\end{eqnarray}
where $\Sigma = \sum_{i=1}^{n_l}\,(q_{i}+\bar{q}_{i})$ and
$\as\equiv\as(\mu^{2}_{R})$. The explicit coefficients, taken from Ref.~\cite{Buza:matchPDF}, 
are given in Appendix C. Note that in the above formula, the $^{(1)}$ and $^{(2)}$
superscripts indicate the perturbative order for both the perturbative
matching coefficient and the PDFs. At LO, it is easy to check that
$\tilde{b}^{(1)}$ is the first term in Eq.~\eqref{eq:b-expansion},
namely
\begin{equation}
\tilde{b}^{(1)}(x,\mu^{2})=\frac{\as}{2\pi} \log\frac{\mu^2}{m_b^2}\int_x^1
\frac{dz}{z}\,P_{qg}(z)g\left(\frac{x}{z},\mu^2\right).
\end{equation} 

\begin{figure}[h]
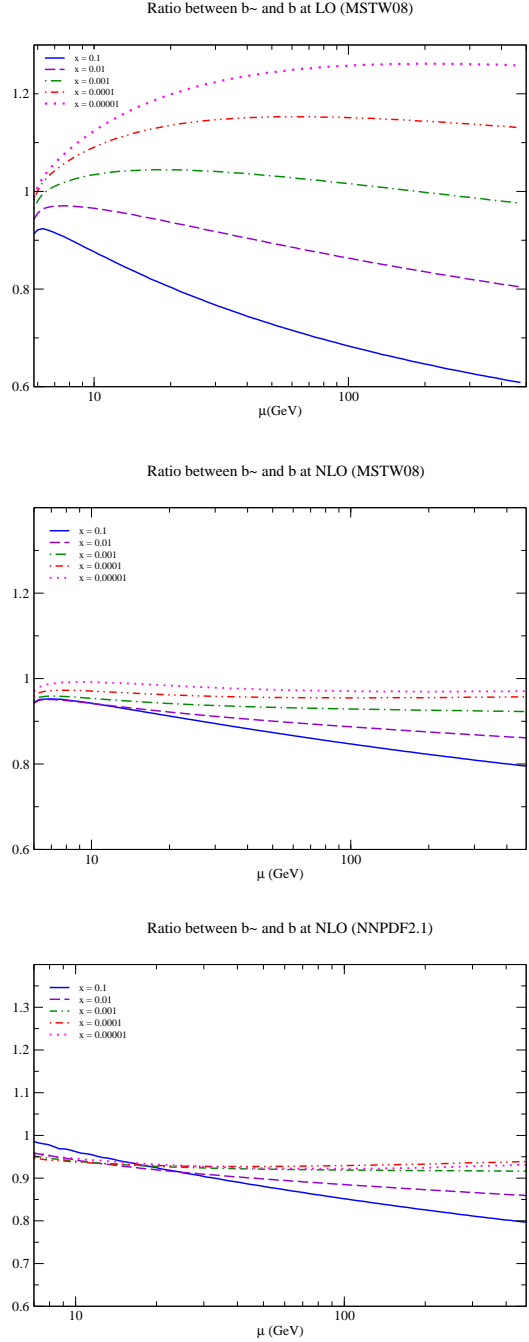

\begin{center}
\includegraphics[width=0.45\textwidth]{ratio-mstw-lo.eps}\\
\vspace*{0.5cm}
\includegraphics[width=0.45\textwidth]{ratio-mstw-nlo.eps}\\
\vspace*{0.5cm}
\includegraphics[width=0.45\textwidth]{ratio-nnpdf-nlo.eps}
\end{center}
\caption{Ratio $\tilde{b}/b$ for several values of $x$ as a function of the scale $\mu$. The 4F-FFNS and GM-VFNS are associated to the $\tilde{b}$ and $b$ PDF computations respectively at LO (top) and NLO order (centre and bottom) for the {\tt MSTW08}~\cite{MSTW08} and {\tt NNPDF21}~\cite{NNPDF21} parton sets. 
\label{fig:btilde-ratio}}
 \end{figure}

In order to assess the accuracy of the $\mathcal{O}(\as^1)$ and $\mathcal{O}(\as^2)$
approximation, we plot in Fig.~\ref{fig:btilde-ratio} the ratio
$\tilde{b}(x_0,\mu)/b(x_0,\mu)$ as a function of
$\mu$ between $m_b$ and $10^2\times m_b$, for several values of $x_0$. 
Deviations from 1 of these curves are an indication of the size of terms of
$\mathcal{O}(\as^2)$ ($\mathcal{O}(\as^3)$) and higher which have been resummed in the QCD
evolution of the heavy quark PDFs. 

We observe that at LO the higher order logarithms  are important and that the $\tilde{b}^{(1)}$ approximation is up to 40\% lower than the $b$ distribution at large $x$. At small $x$, on the other hand, the ratio becomes larger than one, clearly suggesting the inadequacy of the LO approximation to perform a meaningful resummation. 

The NLO ratio, displayed in the middle and bottom plots of Fig.~\ref{fig:btilde-ratio}, clearly shows that the explicit collinear logs present in  a 4F calculation at NLO already provide a rather accurate approximation of
the whole resummed result at NLL, significant effects of order up to 20\% appearing predominantly at large Bjorken $x$. 
At small $x$, the resummation at NLO is well behaved and predicts very  small differences between $\tilde b$ and $b$.
It would be interesting to explore and understand more in detail the small $x$ evolution of the $b$-PDF and in particular
the origin of such a large difference between LO and NLO results. While going somewhat beyond the scope of this paper, it is possible to qualitatively understand why the LO approximation appears too crude in this
region. At small $x$, i.e. at small $N$, the evolution is dominated by the rightmost pole of the anomalous dimensions, which is located at $N=1$ because of the singularity at $x=1$ in the Altarelli-Parisi splitting functions. This singularity appears
in $P_{qq}$ and $P_{qg}$ only at NLO (while it is present in $P_{gq}$ and $P_{gg}$ already
at leading order), and therefore its impact is missed by leading-order evolution.

Our results are consistent with previous findings
in the context of the $c$-PDF~\cite{Buza:matchPDF} even though for the charm 
PDF the effect of the resummation is more important, as it is also clearly 
shown in Ref.~\cite{Alekhin:2009ni}.
They also  provide a very simple
 explanation to the somewhat counterintuitive fact that for $b$-initiated
 processes, such as for example single-top~\cite{maltoni:stop1} and
 $Hbb$~\cite{Campbell:2004pu}, total cross sections calculated in the 5F
 scheme differ with respect to the 4F ones more at the Tevatron than
 at the LHC.  Naively, one could either expect the size of the logs to
 be independent of the collision centre-of-mass energy or were this
 not the case (as we will argue in the next section) to have more
 phase space available at higher energy to 'develop' large logs and
 therefore effects of resummation to be more important the higher the
 collider energy. The  dependence on the Bjorken $x$ of the
 resummation and in particular the larger effects present at higher
 $x$, on the contrary, do point to more important differences at small
 collider energies, as effectively seen.


\section{Heavy quark production in lepton-hadron collisions}
\label{sec:dis}

The production of heavy quarks in lepton--hadron scattering is one of
the first processes for which the choice of the heavy flavor schemes
has been extensively analysed from both the theoretical and the
experimental points of view. For a broad overview see Refs.~\cite{Olness:1994zn,
Buza:matchPDF,vanNeerven:2001tb,Tung:2001mv,acot,ForteNason,Tung:2006tb,
Olness:2008px,mstwHQ,Thorne:2008xf,Thorne:2012qh} 
and references therein and the experimental analyses performed by
the H1 and ZEUS collaborations~\cite{Aaron:2009af,Aaron:2009jy,
Aaron:2010ib,Aaron:2011gp,Chekanov:2004tk,Chekanov:2009kj,Abramowicz:2011kj}.

\begin{figure}[htb]
\begin{center}
\includegraphics[width=0.8\textwidth]{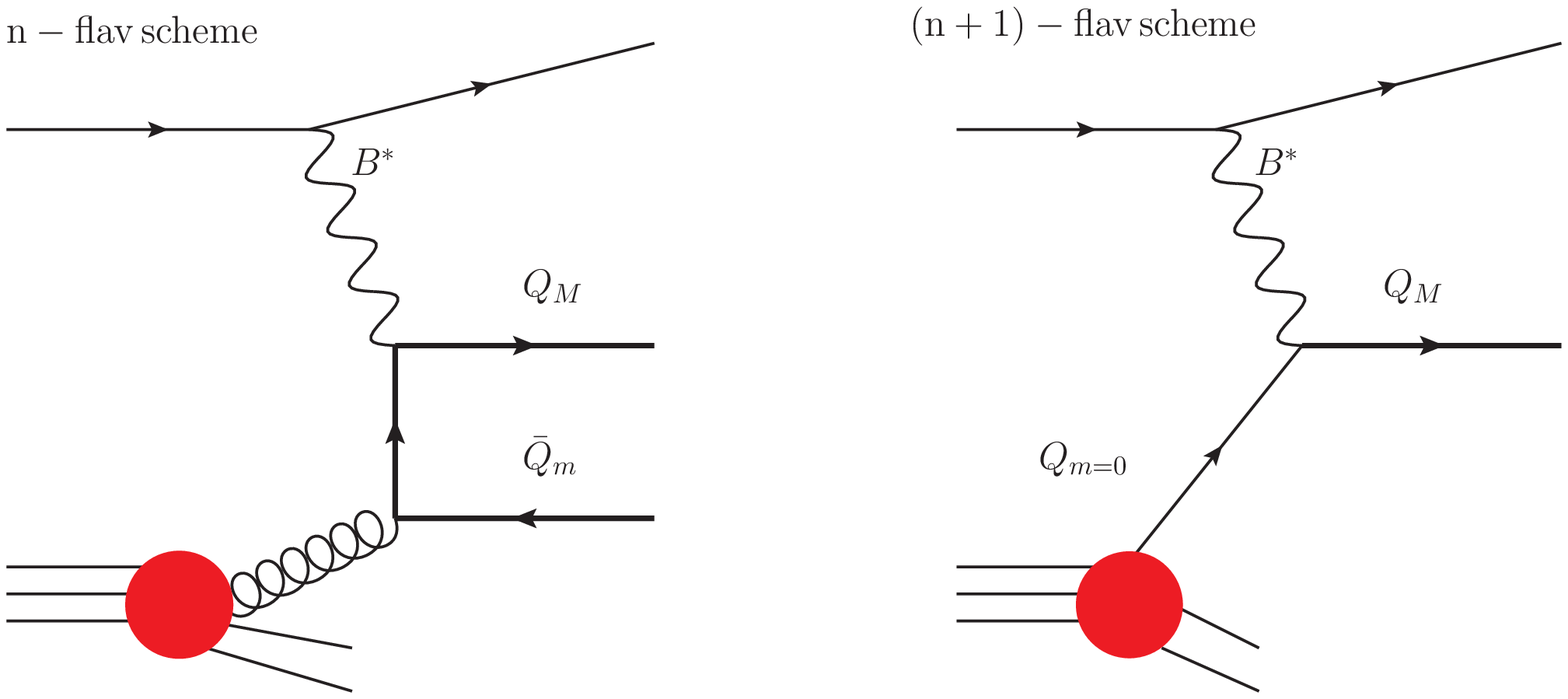}
\end{center}
\caption{Leading--order diagrams for heavy quark production in a $n$-flavor 
(left) and $(n+1)$-flavor (right) schemes.}
\label{fig:hHQprod}
\end{figure}
The relevant cross section can be expressed in terms of the structure
functions usually defined in the context of deep-inelastic scattering.
The contribution of heavy--quark final states to DIS structure
functions has been historically computed in the massive scheme, where
a heavy quark--antiquark pair is produced via vector boson--gluon fusion, as
shown in the left diagrams of Fig.~\ref{fig:hHQprod}.

Alternatively, one can adopt a variable flavor number scheme, whose
leading--order contribution is shown in the right diagram of
Fig.~\ref{fig:hHQprod}.  The main difference between the two
production mechanisms can be attributed to the fact that for massive
heavy quark production the quark and the anti--quark are produced in
pairs, while in the second approach only one heavy quark is produced
at the leading--order. This difference reflects in the transverse
momentum distribution of the bottom quark. At NLO, however, these
striking differences are milder as, for instance, the gluon splitting
process appears also in the massless scheme. 



\subsection{Bottom--Antibottom production}

The total cross section for the process of heavy quark production
in DIS can be written
\begin{eqnarray}
\sigma_b(\mu^2) &=& \int_{y_{{\rm min}}}^{y_{{\rm max}}}
\,dy\int_{Q^2_{{\rm min}}}^{Q^2_{{\rm max}}}\,dQ^2 \,
 \frac{2\pi\alpha_l\alpha_h}{y(M^2+Q^2)^2}
\Bigg\lbrace[1+(1-y)^2]F^b_2(x,Q^2,m_{b}^2)\nn\\
&&\hspace{1cm}-\,y^2F^b_L(x,Q^2,m_b^2)+\, [1-(1-y)^2]F^b_3(x,Q^2,m_{b}^2)
\Bigg\rbrace,
\label{eq:s2:sigmaDIS}
\end{eqnarray}
where $\alpha_l,\alpha_h$ are the lepton and quark coupling
constants to the exchanged vector boson, $M$ its mass,
$Q^2$ its virtuality, and 
\begin{equation}
y=\frac{Q^2}{xS}, 
\end{equation}
where $S$ is the centre--of--mass energy squared.  The functions
$F^b_{2,L,3}$ are the contributions to deep-inelastic scattering
structure functions coming from bottom quark production.  A
dependence of the total cross section on renormalization and
factorization scales arises because the calculation is performed at a
finite order in the QCD perturbative expansion.

We start our analysis by comparing the scale dependence of the 4F and
5F calculations. For simplicity, renormalization and factorization
scales are taken to be equal. We choose
\begin{equation}
\mu_F=\mu_R=k\sqrt{Q^2+4m_b^2}
\end{equation}
and we let $k$ vary between 0.5 and 2.
We have studied the scale dependence of both 4F and 5F results for two different
experimental configurations: the HERA Run II and one of the possible
future LHeC scenarios. At HERA the beam energies are $E_p= 920$~GeV 
and $E_e = 27.5$~GeV, and the kinematical cuts on the
outgoing electron are set to
\begin{equation*}
Q^2 \ge 20 \, {\rm GeV}^2, \,\,\, 0.05\le y\le 0.7
\end{equation*}
as in Ref.~\cite{Chekanov:2009kj}
%
\begin{figure}[htb]
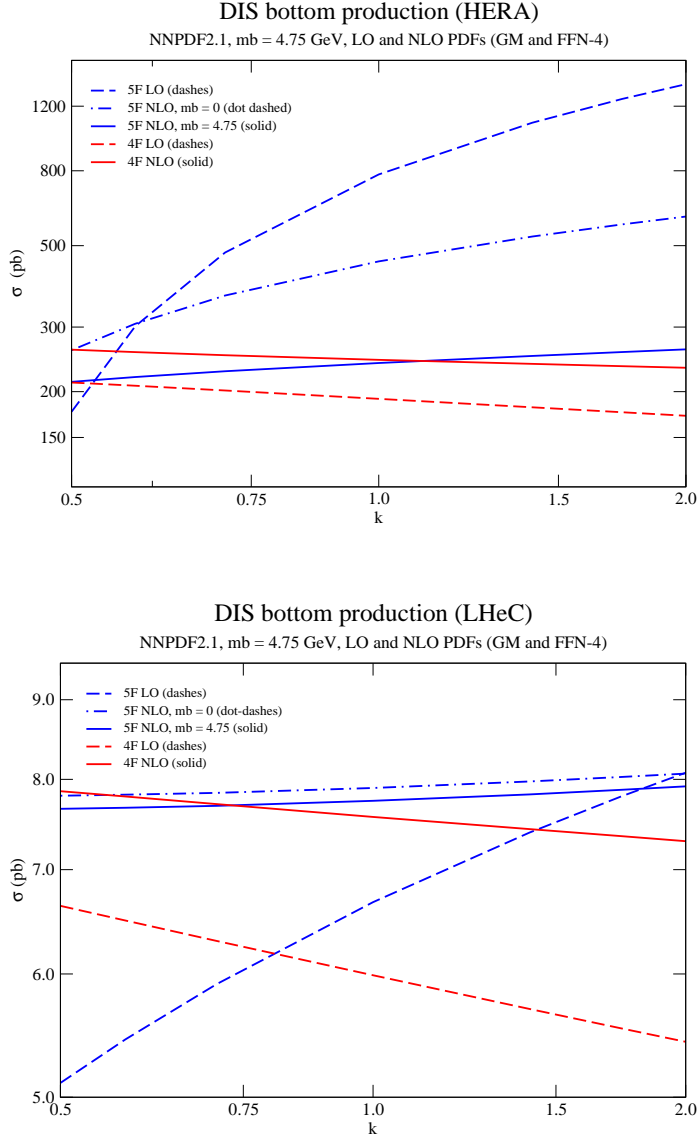

\begin{center}
\includegraphics[width=0.6\textwidth]{bbar-hera-dynscale.eps}

\vspace*{1cm} 

\includegraphics[width=0.6\textwidth]{bbar-lhec-dynscale.eps}
\end{center}
\caption{Comparison between 4F and 5F production for $b\bar{b}$
  production at HERA run II (top) and LHeC (bottom). Input PDFs:
  {\tt NNPDF2.1\_FFN\_NF4} (LO and NLO) for the 4F scheme (run with HVQDIS) and
  {\tt NNPDF2.1} (LO and NLO) for the 5F scheme (run with internal code). 
  Parameters: $m_b$ = 4.75 GeV, $k = \mu/\sqrt{Q^2+4m_b^2}$, 
  with $\mu = \mu_F = \mu_R$.\label{fig:scaledep-disbb}}
 \end{figure} 
In the LHeC scenario the energy of the proton beam has been set to
$E_p=7$ TeV and that of the electron beam to $E_e=50$ GeV. The
cuts in $y$ and $Q^2$ are set to
\begin{equation*}
Q^{2} \ge 2000 \, {\rm GeV}^2, \,\,\, 0.1\le y\le 0.9
\end{equation*}
The results are shown in Fig.~\ref{fig:scaledep-disbb}. 
The 4F scheme curves are obtained by running the code HVQDIS~\cite{HVQDIS},
which is based on the calculation of Ref.~\cite{Laenen:1992zk}.
The 5F curves have been computed using the results presented in
Ref.~\cite{ForteNason}.

In the HERA~II configuration the flavor excitation (the LO 2$\to$2
process) is considerably larger than the other predictions, being
completely driven by the $b$ quark PDF, which does not provide any
transverse momentum for the produced bottom quark. At NLO, the Zero
Mass prediction (where $m_b$ is set to 0 everywhere
except in the boundary conditions for the evolution
of the $b$ PDF) displays a sizeable scale
dependence, even though it is milder than in the pure flavor
excitation.  The matched prediction, computed according to the FONLL-A
scheme, where the collinear logarithms are resummed and mass
effects are also taken into account, shows a mild dependence on the
scale and it is close to the predictions of the 4F scheme. This
indicates that, in the low $Q^2$ region covered in this experimental
configuration, finite-$m_b$ effects dominate over the effects due
to the resummation of the logarithms, which we expect to be very
small. 

In the LHeC configuration, where much larger $Q^2$ scales are probed,
the scale dependence of the 4F LO prediction displays a scale
dependence which is comparable to the one of the 5F LO prediction, the
former being driven by the $\as$ scaling and the second being driven
by the bottom PDF evolution, which go in opposite directions. At NLO
the ZM and the FONLL predictions in the 5F scheme are very close to
each other, thereby showing that the size of finite-$m_b$ effects
are much smaller at such high scales. On the other hand, as compared to
the 4F NLO prediction, they display a milder scale dependence. This
tells us that in this kinematic region the resummation of the collinear
logarithms dominates over the finite $m_b$ mass effects.

\begin{figure}[h]
\begin{center}
\includegraphics[width=0.45\textwidth]{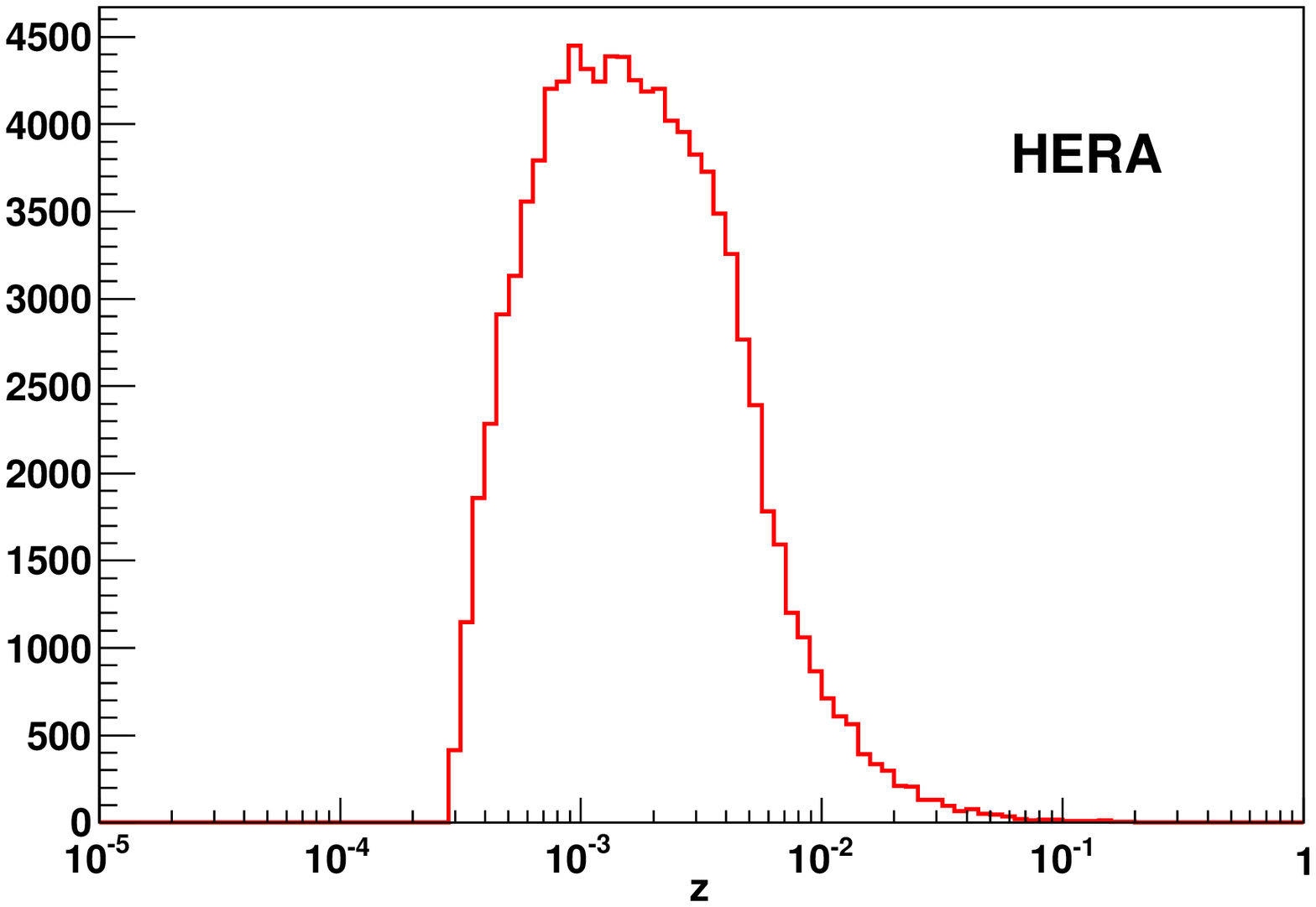}
\includegraphics[width=0.45\textwidth]{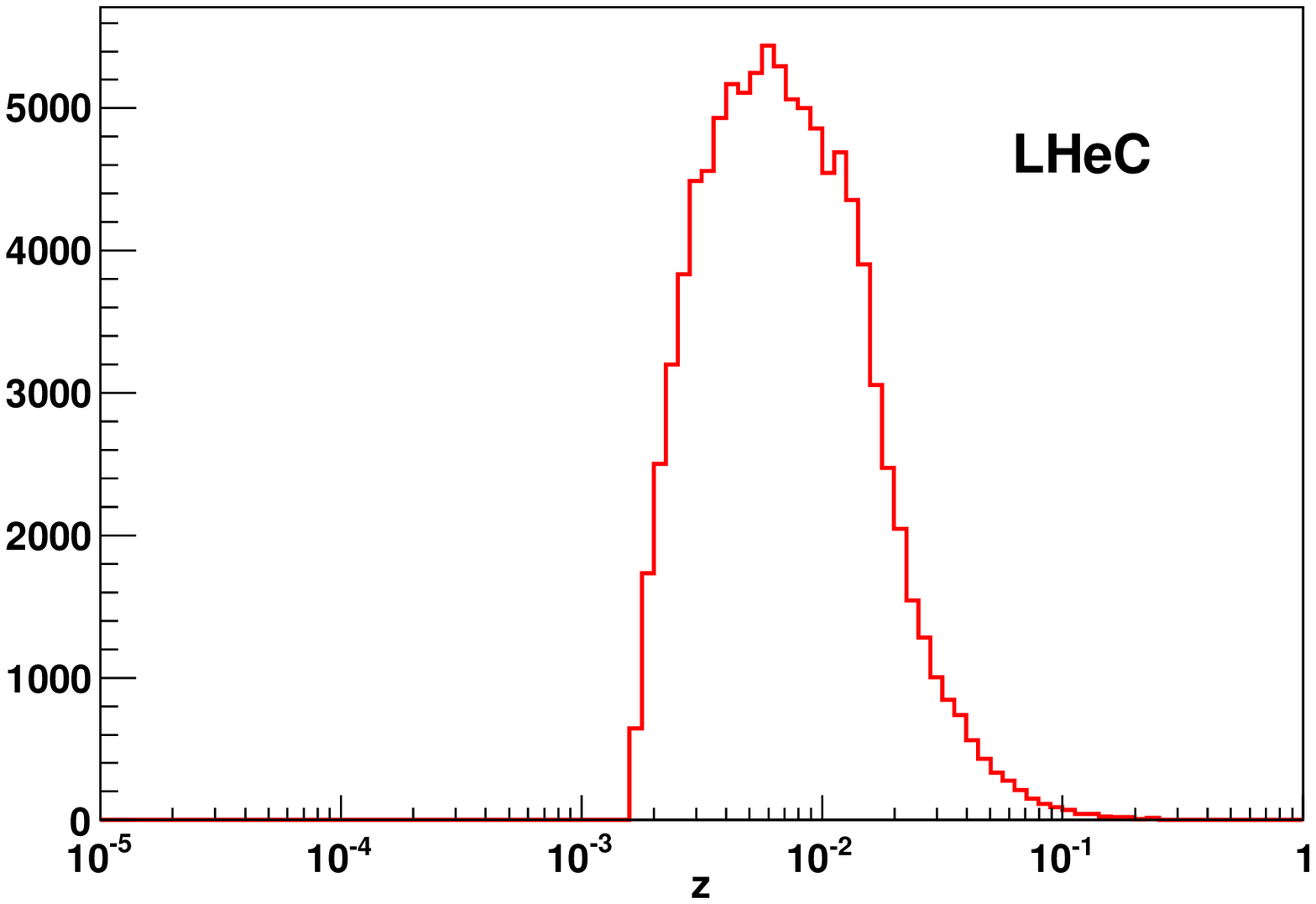}
\caption{\label{x-bbar}Distributions of events as a function of
  momentum fraction carried by the bottom quark in the LO 5F 
  $eb\to eb$ process, in the HERA (left) and LHeC (right)
configurations. Input PDF: {\tt NNPDF21} (LO)}
\end{center}
\end{figure}
In Fig.~\ref{x-bbar} we plot the distribution of the momentum fraction
$z$ carried by the bottom quark in the 5F process. We see that the
distribution has a maximum at $z\sim 10^{-3}$ for the HERA~II
kinematics, while the peak shifts to larger values, $z\sim 10^{-2}$,
in the LHeC case. Comparing to the plots in
Fig.~\ref{fig:btilde-ratio}, we conclude that indeed at the LHeC the
$\tilde{b}$ approximation works much worse than at HERA, not only because
of the larger energy scales probed in the latter kinematic region, but also
because the relevant values of the parton fraction $z$ are comparatively
larger than in the case of HERA~II.

We now turn to the problem of the scale choice for collinear emission.
The explicit expression of the LO partonic cross--sections for
$b\bar{b}$ production in DIS is reported in Appendix~\ref{hq}.  The
partonic cross--sections $d\hat\sigma_2$, relevant for the structure
function $F_2^b$, displays logarithmic contributions both in the $t-$
and $u-$channels, which become large as $m_b\to 0$. In order to see
this explicitly, and to compare the results in the 4F and 5F schemes,
it is useful to rewrite Eq.~(\ref{app:sigmadis}) as a Laurent
expansion around $t-m_b^2=0$ in order to isolate the collinear
singularity, and further neglect terms which are suppressed by
powers of $m_b^2/Q^2$. We obtain
\begin{equation}
\frac{d\hat\sigma_2}{dt} = \frac{\pi\alpha_ee_b^2\as C_F}{16} 
\Bigg[-\frac{4z}{Q^2(t-m_b^2)}\frac{z^2+(1-z)^2}{2}
+\text{non--singular terms}
\Bigg],
\label{sigmahat2bb}
\end{equation}
where $z = Q^2/(s+Q^2)$. The rhs of Eq.~\eqref{sigmahat2bb} is immediately
recognized to be proportional to the Altarelli--Parisi splitting function $P_{qg}(z)$
as reported in Eq.~\eqref{eq:apqg}.  The contribution of the collinear
region to the cross-section takes the form
dictated by the factorization theorem:
\begin{align}
\int_{t_-}^{t_+}dt\, \frac{d\hat\sigma_2}{dt} 
&=\frac{\pi\alpha_e e_b^2\as C_F }{4 Q^2}\, 
zP_{qg}(z)\log\frac{1+\beta}{1-\beta} \nonumber\\
&=\left(\frac{\pi^2\alpha_e e_b^2C_F}{2 Q^2}\right)\,\frac{\as}{2\pi}
zP_{qg}(z)\left[\log\frac{m_b^2}{s}+O\left(\frac{m_b^2}{s}\right)\right], 
\end{align}
where we have used
\begin{equation}
t_\pm=m_b^2-\frac{s+Q^2}{2}(1\pm\beta);
\qquad 
\beta=\sqrt{1-\frac{4m_b^2}{s}}.
\end{equation}
Thus, as expected, the contribution of the collinear
region is proportional to
\begin{equation}
\frac{\as}{2\pi}P_{qg}(z)\log\frac{s}{m_b^2}
=\frac{\as}{2\pi}P_{qg}(z)\log\left[\frac{Q^2}{m_b^2}\frac{1-z}{z}\right]
\label{eq:c7:scaledis}
\end{equation}
In a leading-order 5F computation, only this logarithmically divergent
contribution is taken into account, and resummed to all orders in the
evolution of the $b$ parton density. The argument of the collinear logarithm 
in the 5F scheme is $\frac{\mu_F^2}{m_b^2}$, and the factorization scale
$\mu_F^2$ is chosen to be  $Q^2$.
We now note that in the full massive calculation, Eq.~\eqref{eq:c7:scaledis},
the scale is not simply proportional to $Q^2$, but rather to 
the invariant mass of the produced $b\bar{b}$ pair
\begin{equation}
L_{\rm DIS} 
\equiv \log\left[\frac{Q^2}{m_b^2}\frac{1-z}{z}\right]
= \log \frac{M_{b\bar{b}}^2}{m_b^2}.
\end{equation}
The scale of the logarithm in Eq.~\eqref{eq:c7:scaledis} is a
dynamical scale which changes on an event--by--event basis depending
on the momentum fraction carried by the gluon and on the kinematic
invariants $Q^{2}$ and $z$. (This is well known for example 
in the context of Sudakov resummation~\cite{Forte:2002ni},
where $Q^2(1-z)$ is identified as the 
characteristic energy scale of soft emission.)

Therefore, to make any consideration
about its size, for a given collider energy and acceptance, one has to
check what is the distribution of $(1-z)/z$. In the case this
distribution is peaked in a region which is much smaller than one,
then the logarithms of $(M_{b\bar{b}}^2/m_b^2)$ are not in fact large, even when
$Q^2\gg m_b^2$. If, on the other hand, it is close or even larger than
one, logarithms should be resummed.

\begin{figure}[h]
\begin{center}
\includegraphics[width=0.31\textwidth]{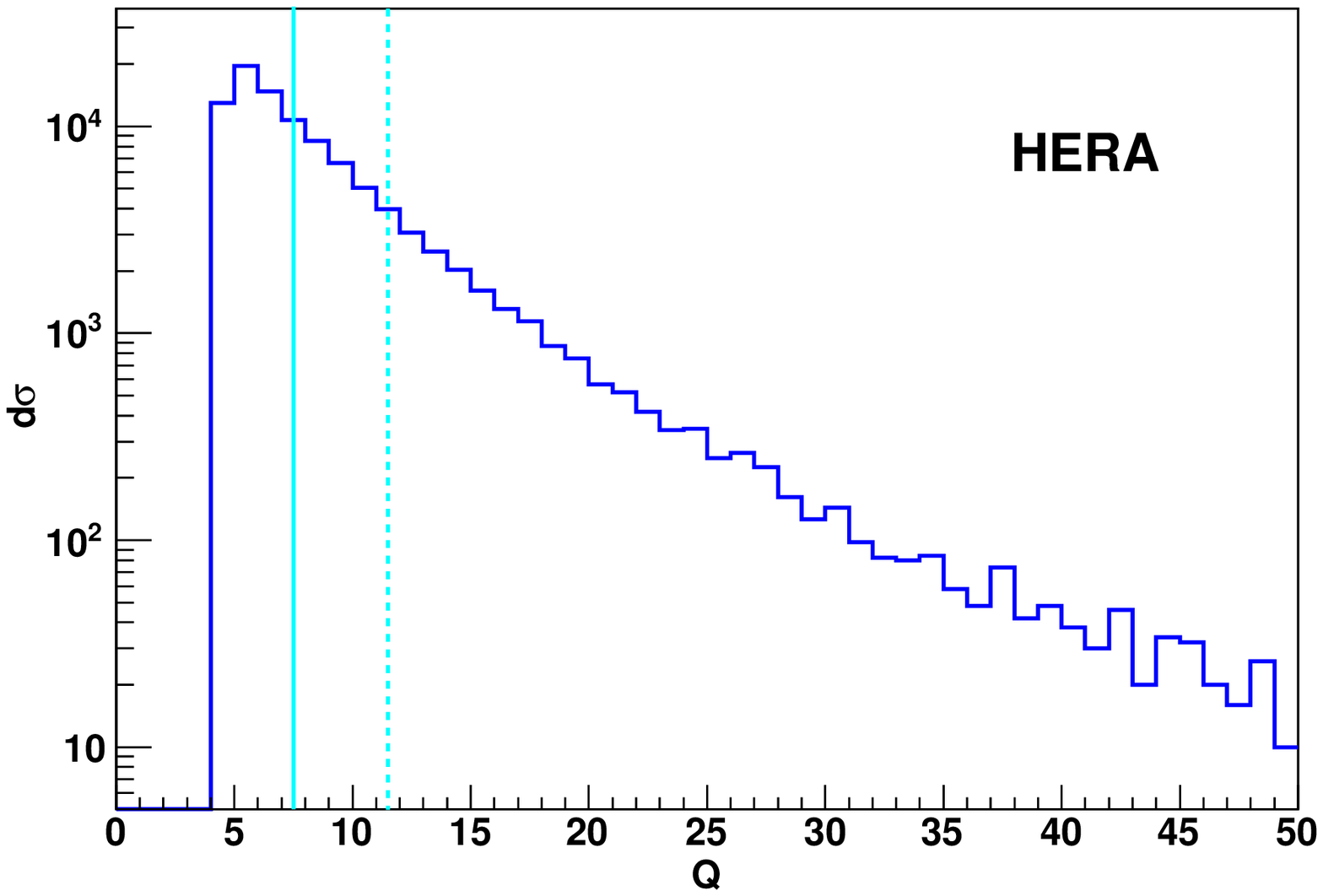}
\includegraphics[width=0.31\textwidth]{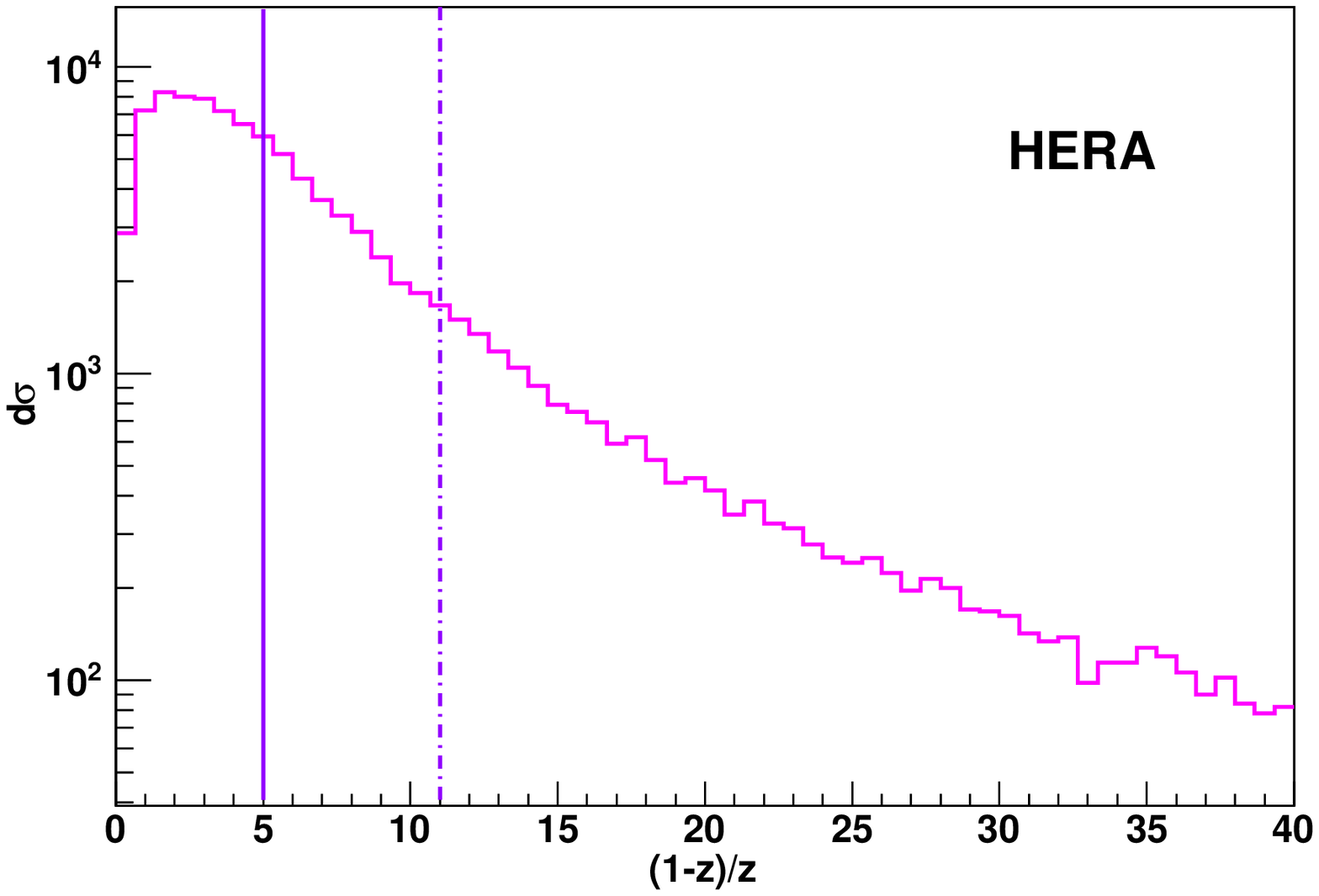}
\includegraphics[width=0.31\textwidth]{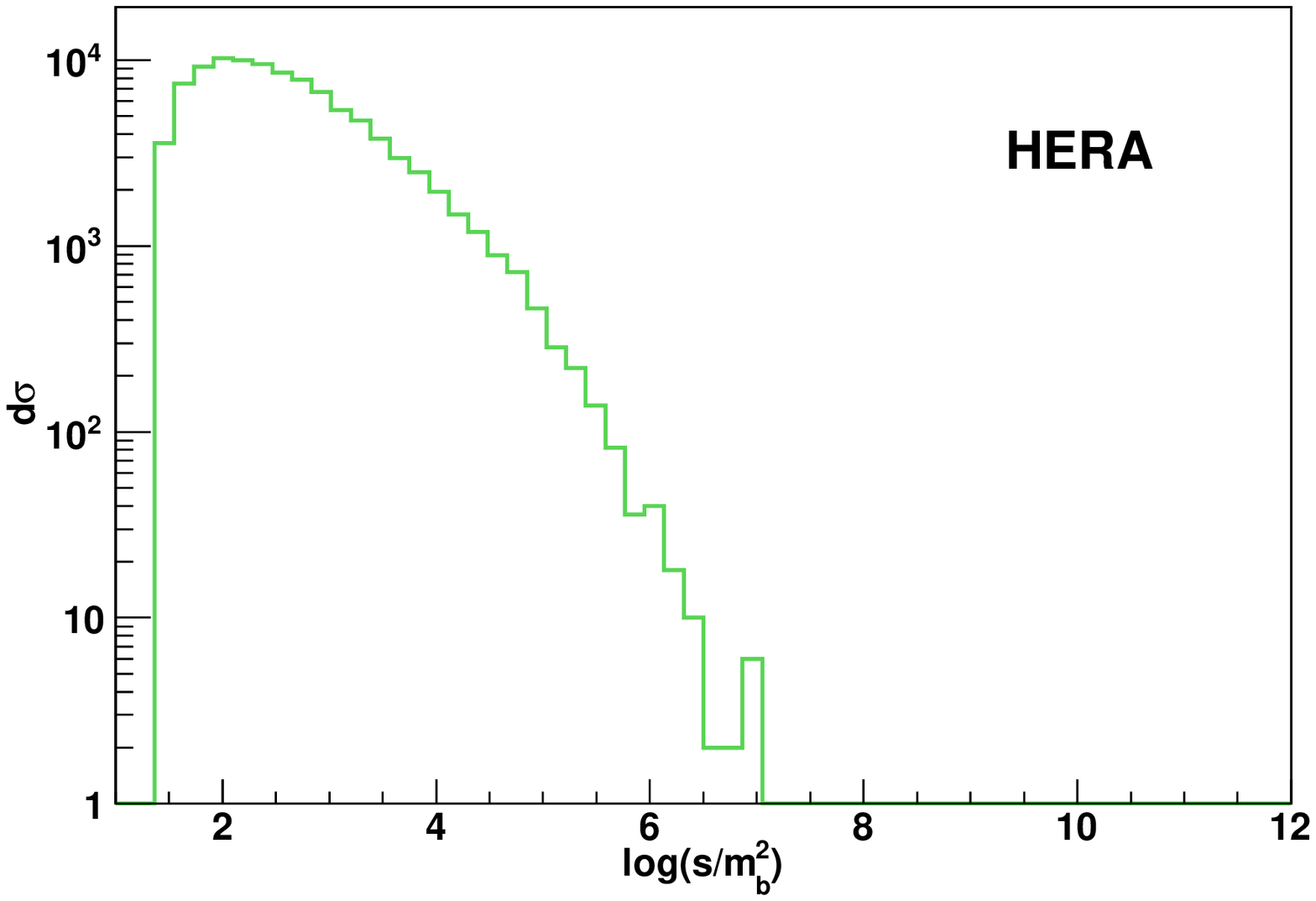}\\
\includegraphics[width=0.31\textwidth]{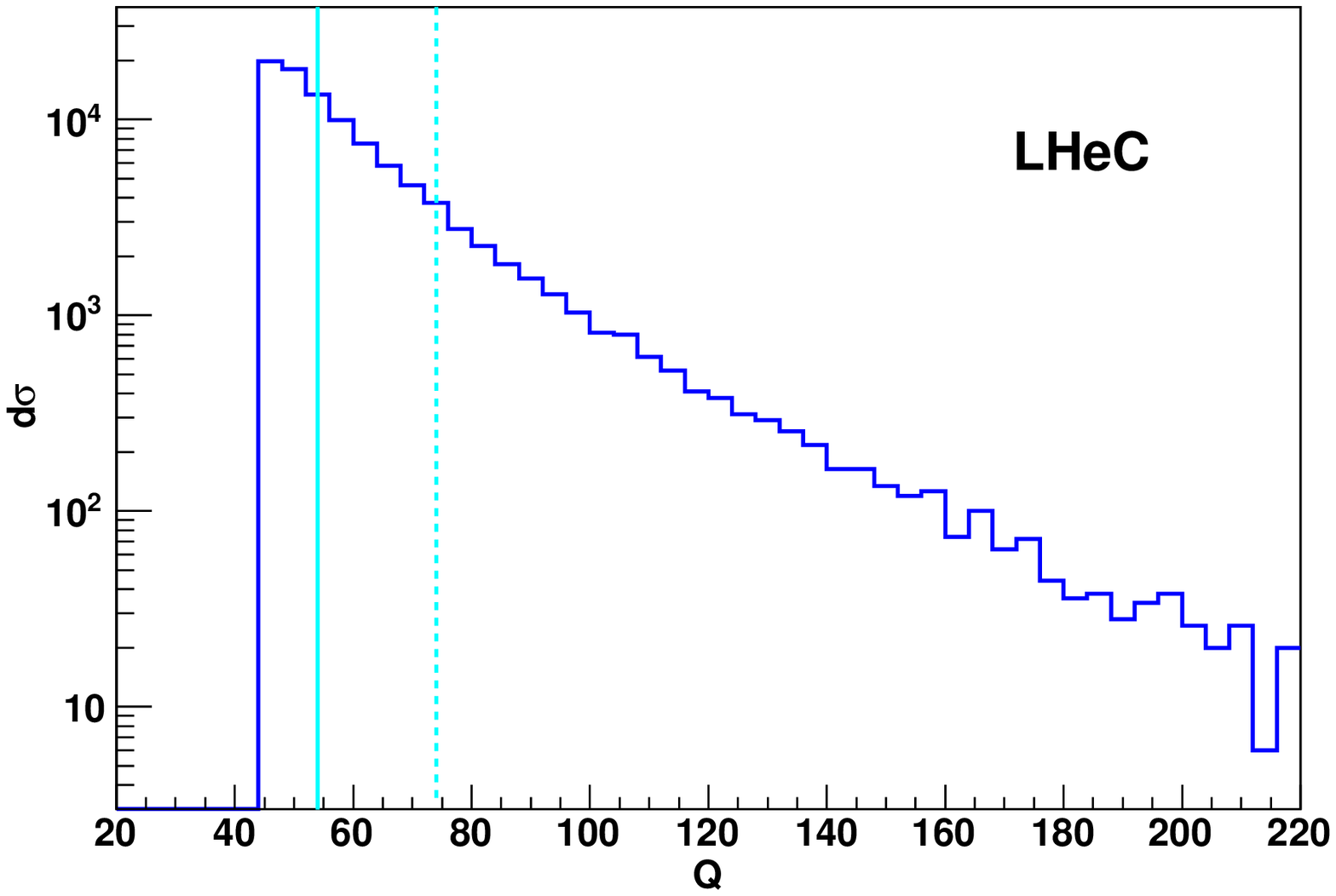}
\includegraphics[width=0.31\textwidth]{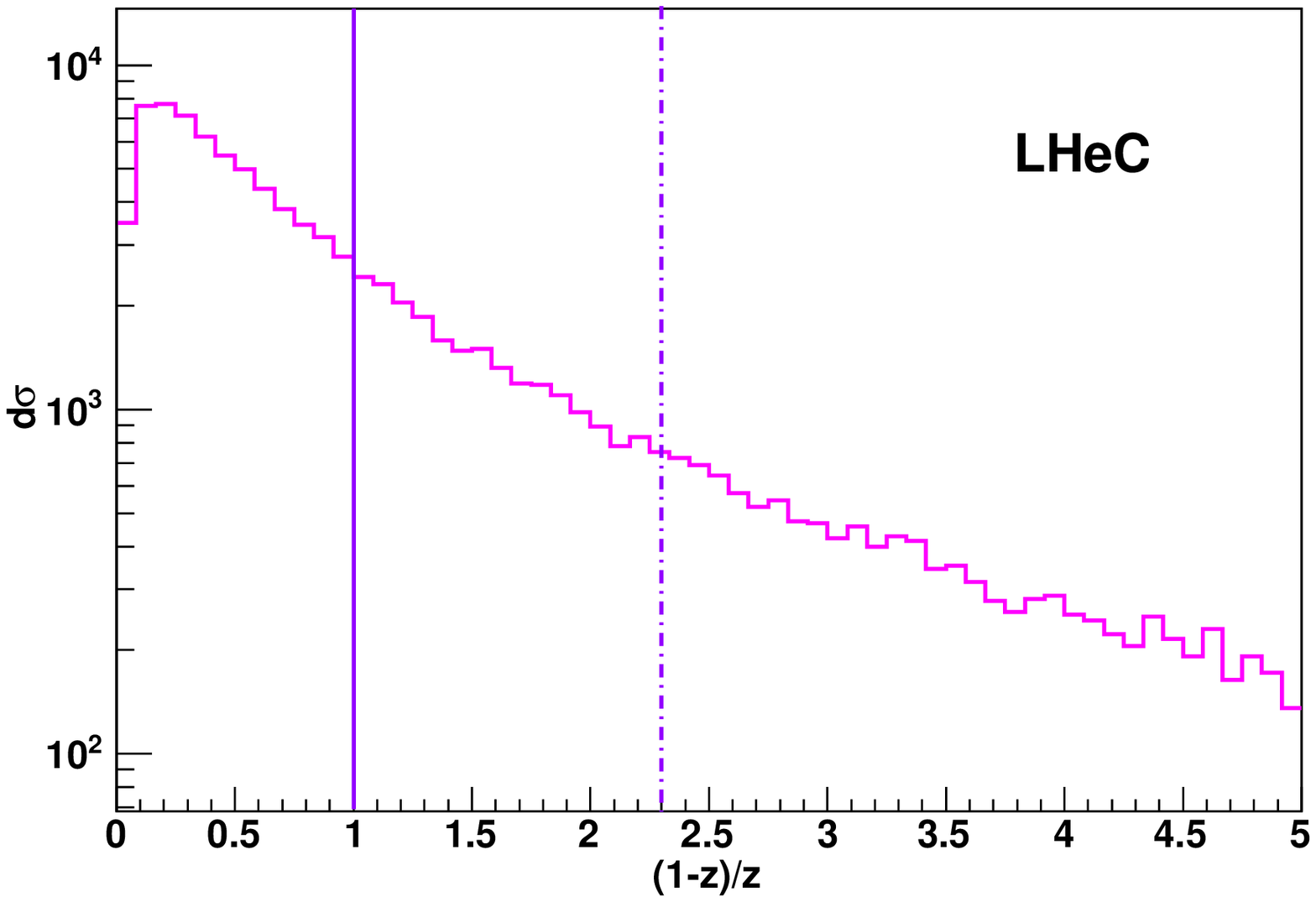}
\includegraphics[width=0.31\textwidth]{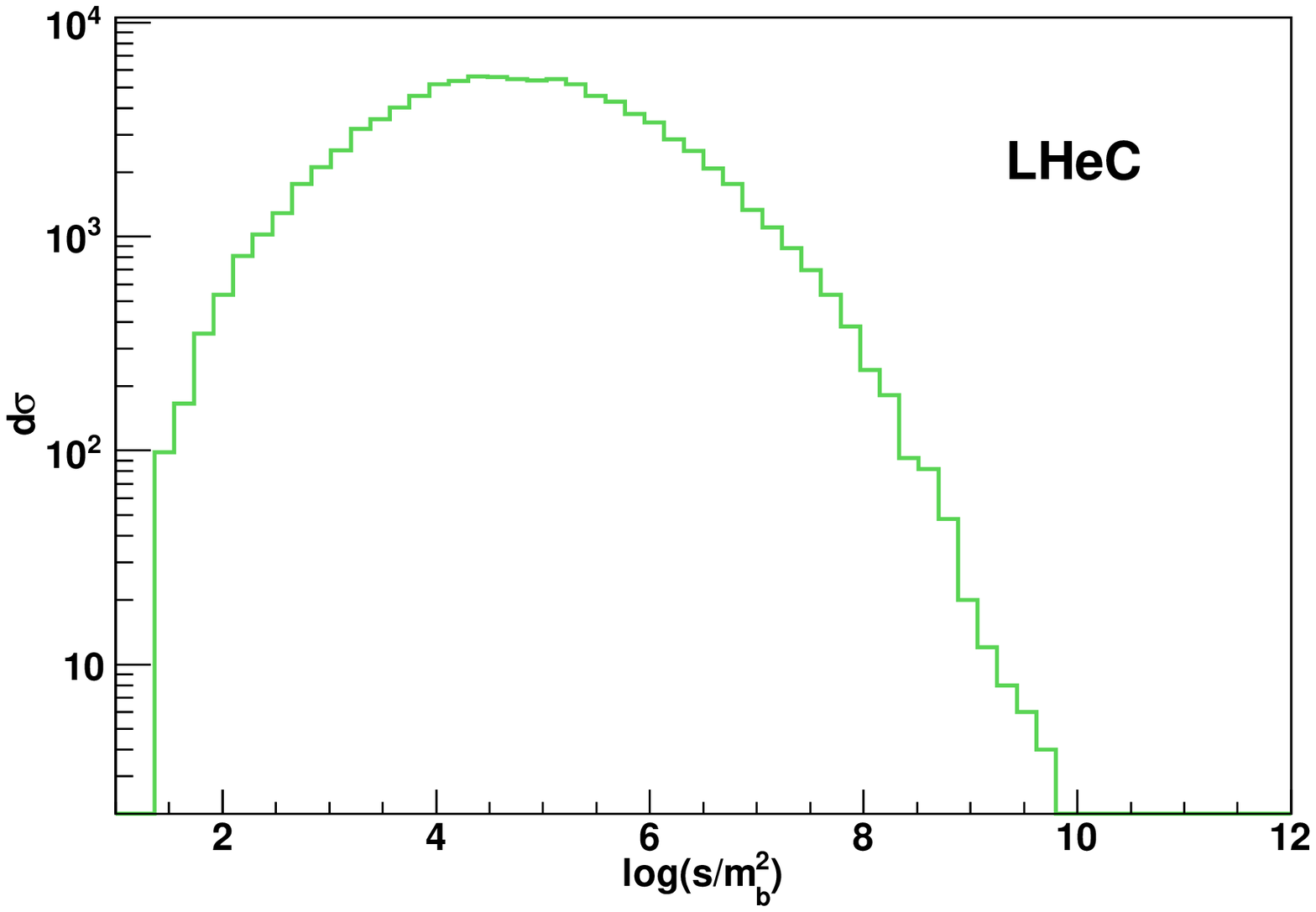}
\caption{\label{dists-bbar}Distributions of events as a function of
  $Q$ (left), $\frac{1-z}{z}$ (centre) and $\log(s/m_b^2)$ (right) 
  for HERA~II (top) and LHeC (bottom). In left and centre plots,
  50\% (80\%) of events are to the left of the
  vertical solid (dot-dashed) line.
  Input PDF: {\tt NNPDF21\_FFN\_NF4} (LO).
}
\end{center}
\end{figure}

In Fig.~\ref{dists-bbar} we plot the distributions of
$Q$, $\frac{1-z}{z}$ and the collinear
logarithm. From the plot of the $Q$ distribution we see that the two
experimental configurations we are considering 
do really explore two complementary regions: for
HERA~II the bulk of events lies in the region $Q\gsim m_b$, while
at LHeC $Q$ is typically much larget than $m_b$. 
The event distribution in the kinematic factor
$\frac{1-z}{z}$ is also rather different in the two cases.
For HERA~II, 80\% of events are in the region $(1-z)/z\lsim 10$.
This means that the scale of the logarithm is sizeably larger than
$Q^2/m_b^2$ for a large fraction of events. 
Therefore, even if the experimental cuts are such
that $Q^2$ lies in a region where $\log(Q^2/m_b^2)$ is very small,
the effect of these logarithms is enhanced by a scale which is
effectively up to ten times larger than $Q^2$. As a result,
in this case
the logarithms are not as negligible as one would expect
from the typical size of $Q^2$.
On the other hand, in the LHeC kinematics the prefactor
$\frac{1-z}{z}$ is peaked around a value smaller than one. This means that in
the LHeC configuration the scale associated to the collinear splitting is
slighlty smaller than $Q^2$ but not dramatically and that therefore
the logarithms which are resummed in the bottom PDF are significant as
one would expect by looking at the $Q^2$ distribution.

\section{Hadron--hadron collisions}
\label{sec:pp}

In this section we discuss $b$-initiated processes at hadron
colliders.  Such processes fall in two wide categories depending
whether they can be described by one or two $b$'s in the initial
state. Associated production of a $b$ plus a jet, 
a $W$ or $Z$ boson, a Higgs boson, $t$-channel and
associated $tW$ and $tH^+$ single top production are examples of the
first class, while $Zbb$ and $Hbb$ production of the second.  For the
sake of the argument, in this work we mainly focus on one-$b$ initiated
processes. In fact, as we will argue in the final discussion, two-$b$
initiated processes can be analysed following the same lines
providing, for instance, a useful framework to decipher well-known
apparent discrepancies between 4F and 5F predictions for $pp \to Hbb$
production~\cite{Campbell:2004pu}.
 
We will consider specifically $W$ production initiated by a bottom quark, and
single top production. The large energy scale associated
to these processes is the invariant mass of the final state, rather
than the high virtuality of the incoming particles, at variance with
DIS. As we will prove in the following this makes a difference in the
phase space available for the spectator $b$ quark or extra radiation
in general. In addition, single top provides already all the
kinematical complexity needed to obtain a general expression for the
initial-state collinear logarithm, a form that can then be used for
any process of this type, including those involving two $b$'s in the
initial state.

We first analyse the associated production of a heavy quark with a
gauge boson or a charged scalar.  In the 5F scheme these processes
start with the simplest possible topology, {\it i.e.}, $2\to1$ at Born
level, and feature a $b$ quark as one of the initial-state
partons. While not really of prime phenomenological importance as for
themselves, 4F and 5F scheme predictions can be easily compared at NLO,
and provide a very simple baseline for studying more complex
final states.

We then turn to the analysis of $t$-channel single top
production. This probes directly the charged--current interaction of
the top quark and it is therefore sensitive to possible new
physics associated to the charged--current weak interaction of the top
quark. A full calculation of the NLO corrections to $W$--gluon fusion
has been already presented in the
literature~\cite{Willenbrock:1986cr,Yuan:1989tc,Ellis:1992yw}, and several
schemes have been used. The calculation was originally
performed~\cite{Bordes:1994ki,Stelzer:1997ns} in the 5F scheme. More
recently~\cite{maltoni:stop1,maltoni:stop2} the 4F scheme was adopted
in a NLO calculation where a first comparison between 4F and 5F
results was performed.

In the last subsection we generalise the results obtained by analysing
$t$-channel single-top and present formulae that can be applied to any
$b$-initiated process, with either one or two $b$'s in the initial
state.

\subsection{Associated $W$ and bottom production}

Associated production of a $W$ with heavy quarks, such as, charm and
bottom, is of great phenomenological importance at the Tevatron as
well as the LHC. The production of a $W$ boson and a charm quark, for
example, can provide very useful information on the strange content of
the proton; it has been subject of measurements at the
Tevatron~\cite{:2007dm} and it is being explored now at 
the LHC~\cite{CMSnote}. $Wb$
associated production is also of phenomenological interest, yet its
main production mechanism at hadron colliders does not involve a $b$
in the initial state, since at leading order it amounts to $W$ Drell-Yan
production with an extra gluon splitting into a $b \bar b$ pair.

In this subsection, we consider the simple $2 \to 1$ processes
$u\,\bar{b}\rightarrow W^+$ and $\bar{u}\,b \rightarrow W^-$.  This is
to be considered as a simple alternative, yet fully analogous to, the
sample process, $\bar b t \to H^+$ considered in the seminal work by
Olness and Tung~\cite{Olness:1987ep}. The main difference here is that
we focus on the $b$ quark and consider a valence--like PDF in the
other initial-state leg, ignoring the suppression originated by the
small CKM coupling. In this way, our analysis is not affected by the
presence of a sea quark in the initial state.

\begin{figure}[h!]
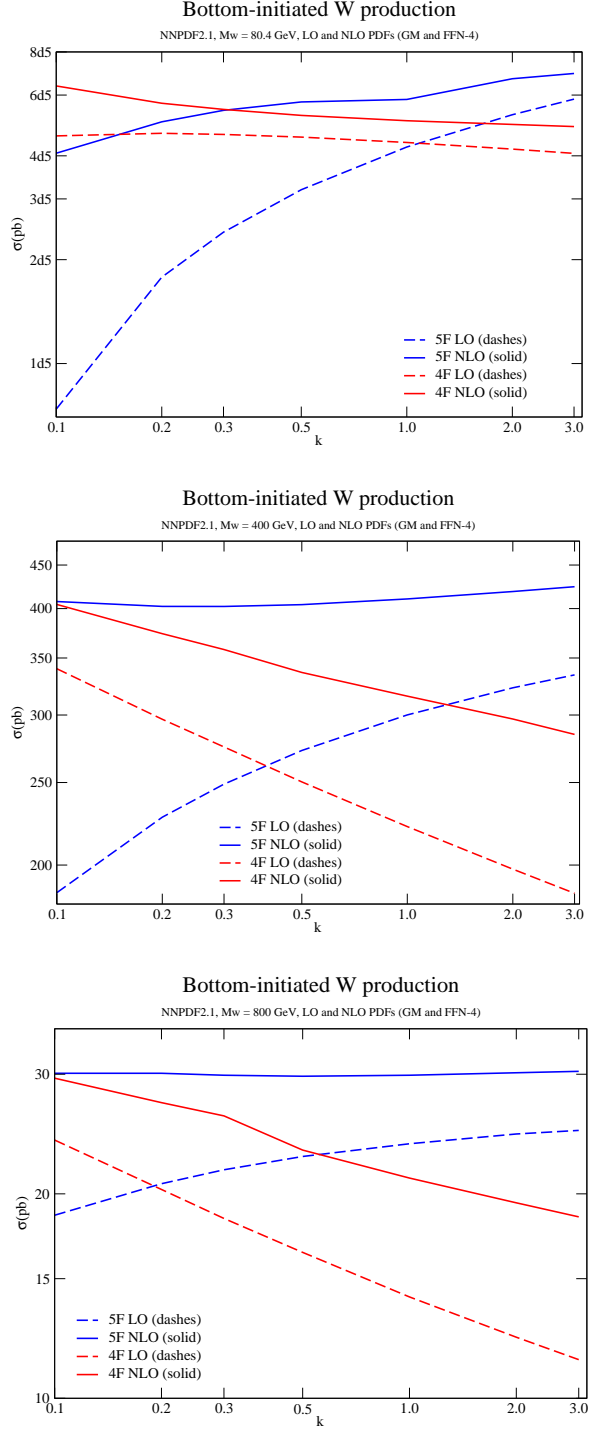

\begin{center}
 \includegraphics[width=0.5\textwidth]{wb-80-nnpdf.eps}
\vspace*{0.5cm}

 \includegraphics[width=0.5\textwidth]{wb-400-nnpdf.eps}
\vspace*{0.5cm}

 \includegraphics[width=0.5\textwidth]{wb-800-nnpdf.eps}
 \end{center}
\caption{Comparison between 4F and 5F production for $b$-initiated $W$
   production at LHC 14 TeV, as a function of $k=\mu/M_W$, with 
   $\mu=\mu_F=\mu_R$, for $M_W=80.4$ GeV (top), $M_W=400$ GeV (middle),
   $M_W=800$ GeV (bottom) and $m_b$=4.75
   GeV. Input PDFs: {\tt NNPDF21\_FFN\_NF4} and  {\tt NNPDF21}
   for 4F and 5F respectively. 
   \label{fig:wb-comp-nnpdf}}
\end{figure}

We first compare the size and the scale dependence of
the 4F and 5F predictions at leading and next-to-leading order. Our
results are shown in Fig.~\ref{fig:wb-comp-nnpdf}.
The 4F curves are obtained using the MCFM code~\cite{Campbell:2003hd},
while we have computed the 5F cross sections. We have used the
NNPDF2.1 family of parton distribution functions. In particular the
default GM-VFN scheme-based set was used for the 5F calculation, and
the 4-FFN scheme set was used for the 4F computations. We have considered
both the case of the production of a real $W$ boson (first plot in
Fig.~\ref{fig:wb-comp-nnpdf}), and that of a virtual one, for two
different values of the invariant mass of its decay products, namely
400~GeV (second plot) and 800~GeV (third plot).  We take the
renormalization and factorization scales to be equal to $k M_W$, with
$0.1\leq k\leq 3$.

We observe that, for large values of the invariant mass of the $W$
decay products, the 5F cross sections is typically larger than the 4F
prediction at the central value $k=1$.  The difference decreases with
decreasing $k$, and the crossing point is around $k\sim 0.1$.

\begin{figure}[htb]
\begin{center}
\includegraphics[width=0.32\textwidth]{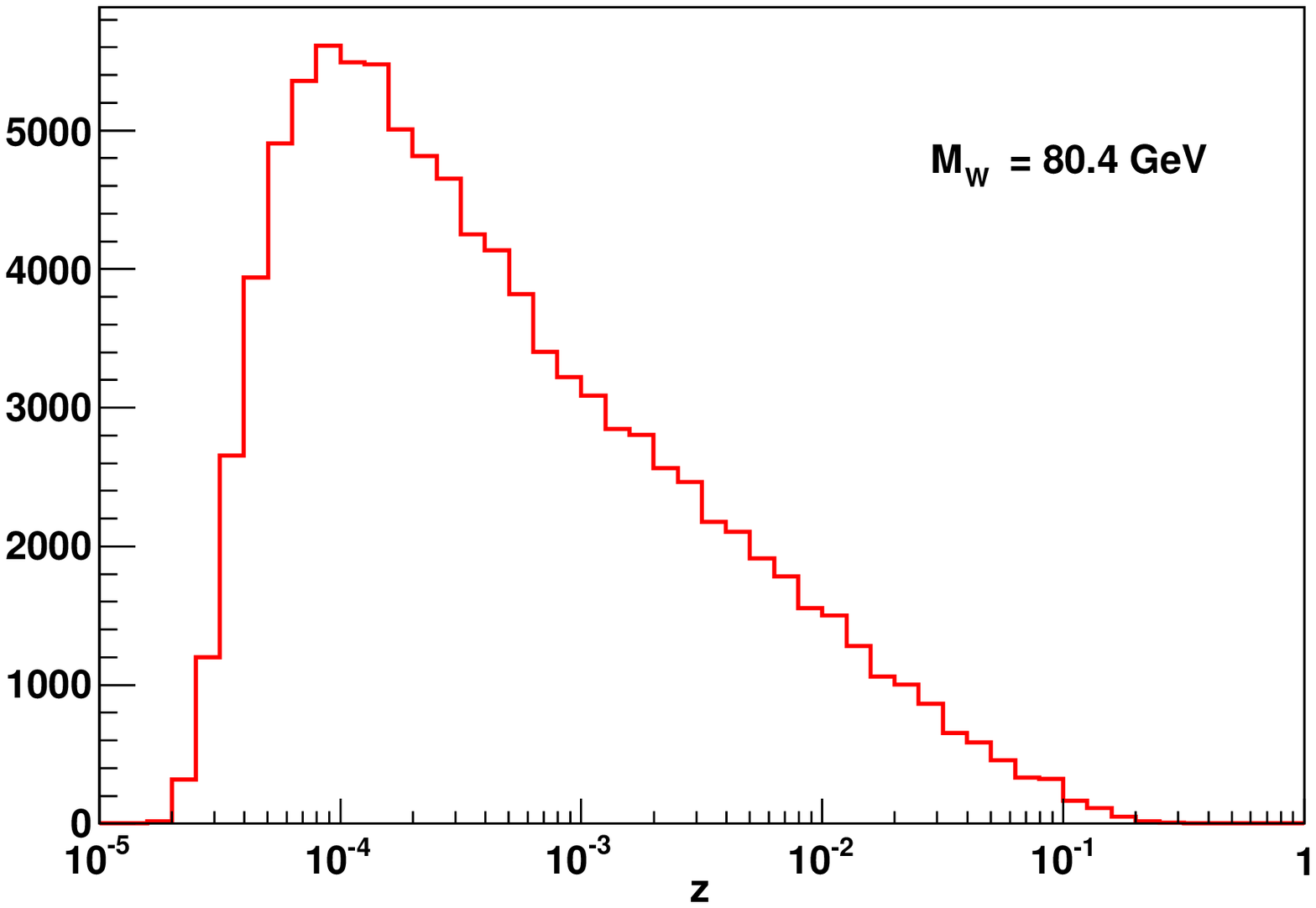}
\includegraphics[width=0.32\textwidth]{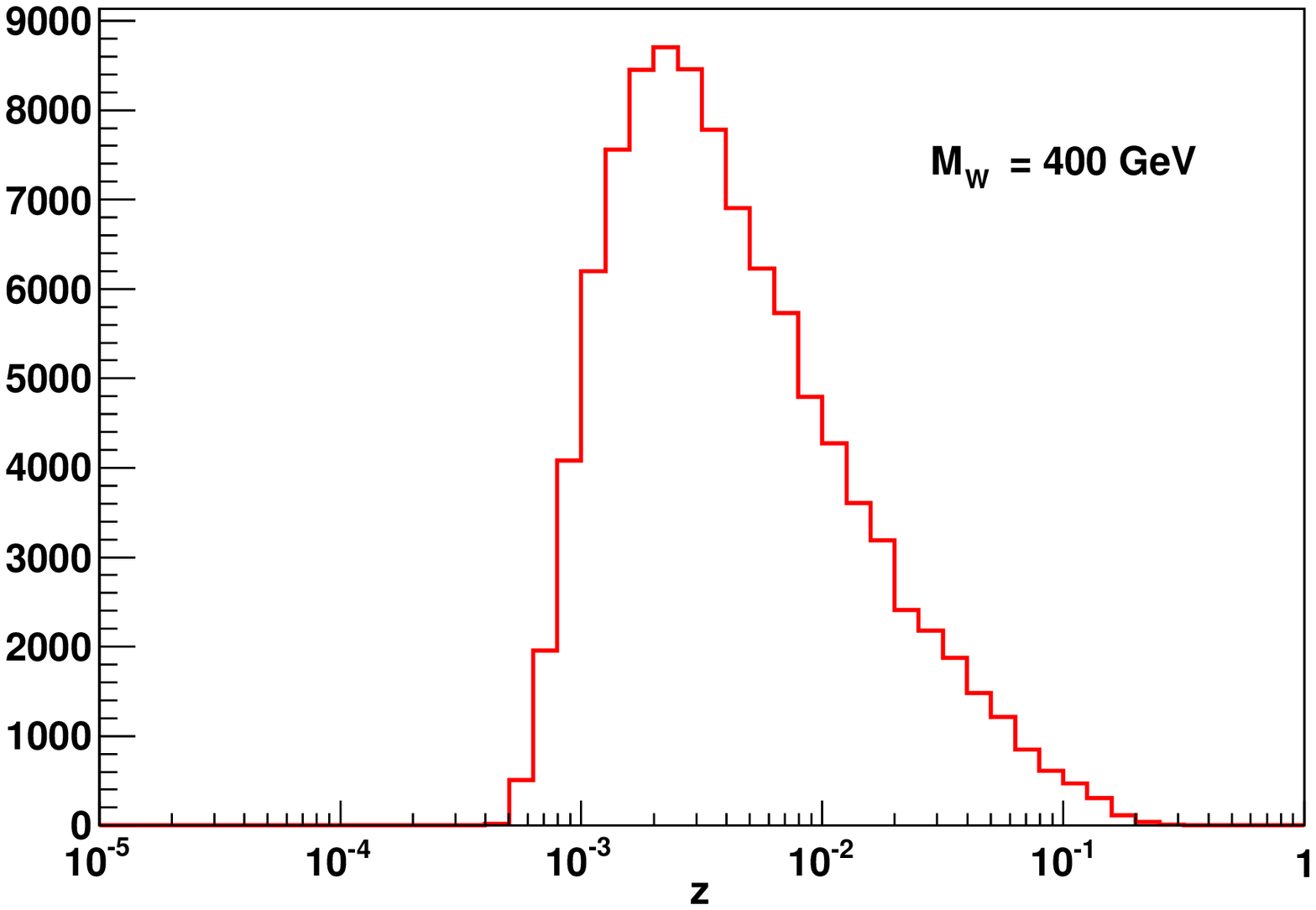}
\includegraphics[width=0.32\textwidth]{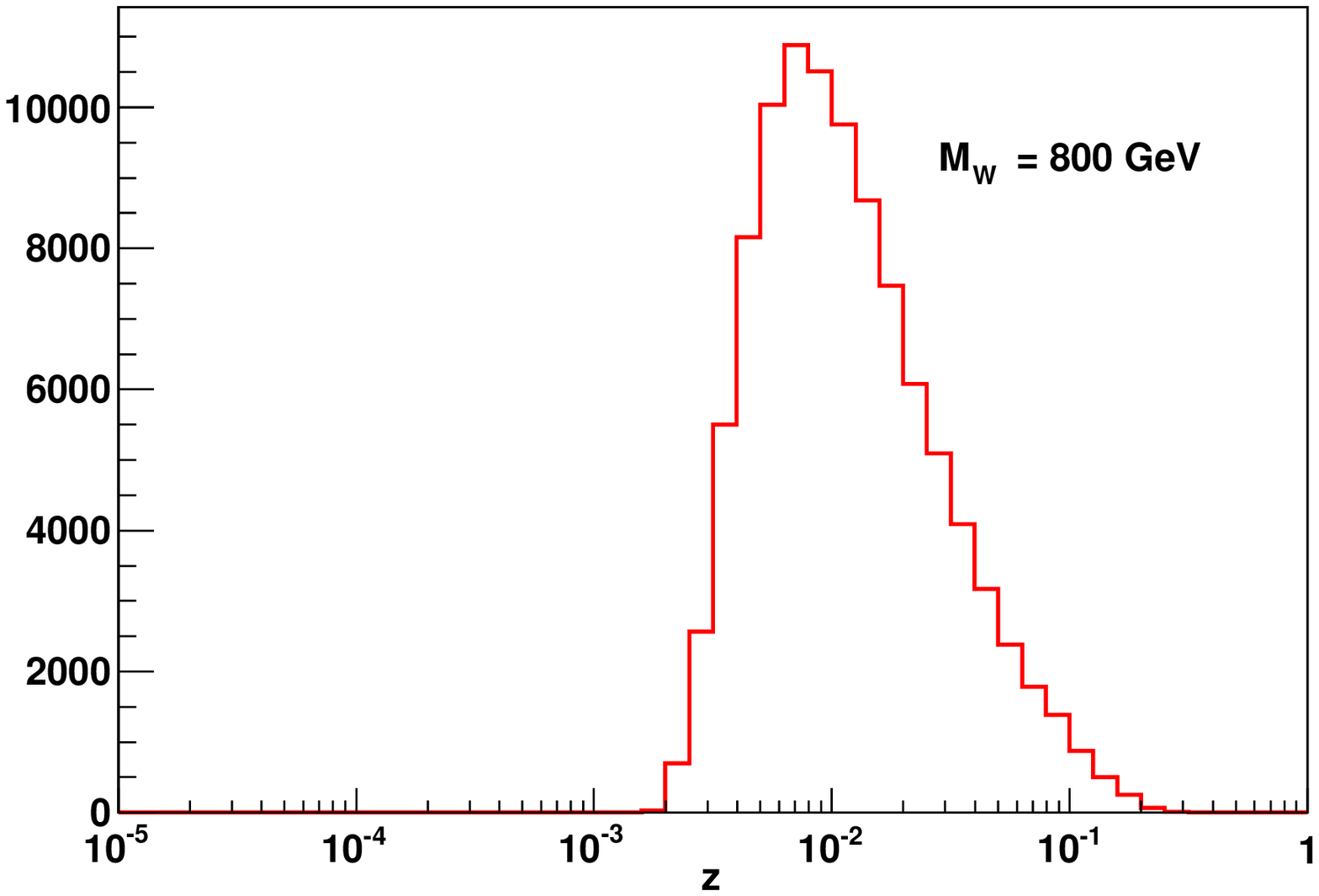}
\end{center}
\caption{Distribution of the momentum fraction of momentum carried by the
   $b$ quark(antiquark) in $W$ production initiated by bottom at
   LHC 14 TeV, LO 5F scheme, for $M_W$ = 80.4 GeV (left), $M_W$ = 400 GeV
   (centre), $M_W$ = 800 GeV (right). Input PDFs: {\tt NNPDF2.1} (LO) 
\label{fig:wb-pdfs}}
\end{figure}

These features can be qualitatively understood by studying the
explicit expressions of the leading-order cross sections in the two
schemes.  At leading-order, the relevant partonic subprocesses for the
5F calculation are
\begin{eqnarray}
b(k_1) + u(k_2)&\longrightarrow& W(k)
\end{eqnarray}
The corresponding cross-section is
\begin{equation}
\sigma^{\rm 5F}(\tau) =
\frac{\pi\sqrt{2}}{3}G_F\tau\,\mathcal{L}_{ub}(\tau,\mu_F^2).
\end{equation}
where $\tau= M_W^2/S$, $S$ is the hadronic centre-of-mass energy squared,
and we have set $V_{ub}=1$.
The parton luminosity is given by 
\begin{equation}
\mathcal{L}_{ub}(\tau,\mu_F^2) 
= 2\int_\tau^1\,\frac{dz}{z}u^+\left(\frac{\tau}{z},\mu_F^2\right)\,
b(z,\mu_F^2),
\label{lumdef}
\end{equation}
where $u^+=u+\bar{u}$. As explained in Section 3, the heavy quark
parton distribution function $b(z,\mu_F^2)$ arises dynamically as a
consequence of Altarelli-Parisi evolution, which resums the leading
powers of $\log(\mu_F^2/m_b^2)$ to all orders in perturbation theory. If
only the first order of this resummation is retained, one obtains an
approximate form for the 5F cross section:
\begin{equation}
\sigma^{5F}(\tau)
= \left(\frac{\pi\sqrt{2}}{3}G_F\tau\right)
\int_\tau^1\frac{dz}{z}\,{\cal L}_{ug}\left(\frac{\tau}{z},\mu_F^2\right)
\frac{\as}{2\pi}P_{qg}(z)\log\frac{\mu_F^2}{m_b^2}+{\cal O} 
\left(\as^2\log^2\frac{\mu_F^2}{m_b^2}\right).
\label{eq:1}
\end{equation}
The choice of factorization scale will be discussed later in this
Section. Here we recall from Section 3 that, for fixed $\mu_F$, the
size of the neglected terms in Eq.~\eqref{eq:1} is smaller, the
smaller is the momentum fraction carried by the $b$ quark. It is
therefore interesting to study the distribution in this variable, for
different values of the virtuality of the produced $W$ boson. This
distribution is displayed in Fig.~\ref{fig:wb-pdfs}.
We note that the dominant contribution to the convolution integral in
Eq.~\eqref{lumdef} comes from very small values of the momentum
fraction, ranging from $10^{-4}$ to $10^{-2}$ in an order of
increasing values of $M_W$.  On the basis of the discussion in
Section~3, we conclude that the impact of resummation from
Altarelli-Parisi evolution is in this case rather small indeed, 
apart from very high virtuality of the $W$.

The factorization scale in the 5F calculation is typically chosen to
be  the hard scale characteristic of the process, in
our case $M_W$. This is nothing but an order-of-magnitude indication. A
less vague indication is given by the comparison with the 4F computation,
which we now illustrate. In analogy with the previous case, we
consider the subprocesses
\begin{eqnarray}
g(p_1) + u(p_{2})&\longrightarrow& b(p_{3})+W(p_{4})
\end{eqnarray}
The total cross-section is given by
\begin{equation}
\frac{\sigma^{\rm 4F}(\tau)}{\tau} = \int_{\tau}^{1} \frac{dz}{z}\, 
\mathcal{L}_{ug}\left(\frac{\tau}{z},\mu_F^2\right)
\frac{\hat\sigma^{4F}(z)}{z}\,,
\label{eq:s4FWb}
\end{equation}
where $\tau=\frac{(m_b+M_W)^2}{S}$, $z=\frac{(m_b+M_W)^2}{s}$,
$s=(p_1+p_2)^2$ and
\begin{equation}
\hat\sigma^{4F}(z)=
\int_{t_-}^{t_+} dt\, \frac{d\hat\sigma}{dt}(s,t,\as).
\label{total-xsec}
\end{equation}
The variable $t=(p_1-p_3)^2$ represents the collinearity of the bottom
pair produced by gluon splitting, and the kinematical bounds are
given in Eq.~\eqref{eq:t-extr}.  The ${\cal L}_{ug}$ parton luminosity
is defined in Eq.~\eqref{lumdef}.

\begin{figure}[ht!]
\begin{center}
\vspace*{3mm}
\includegraphics[width=0.8\textwidth]{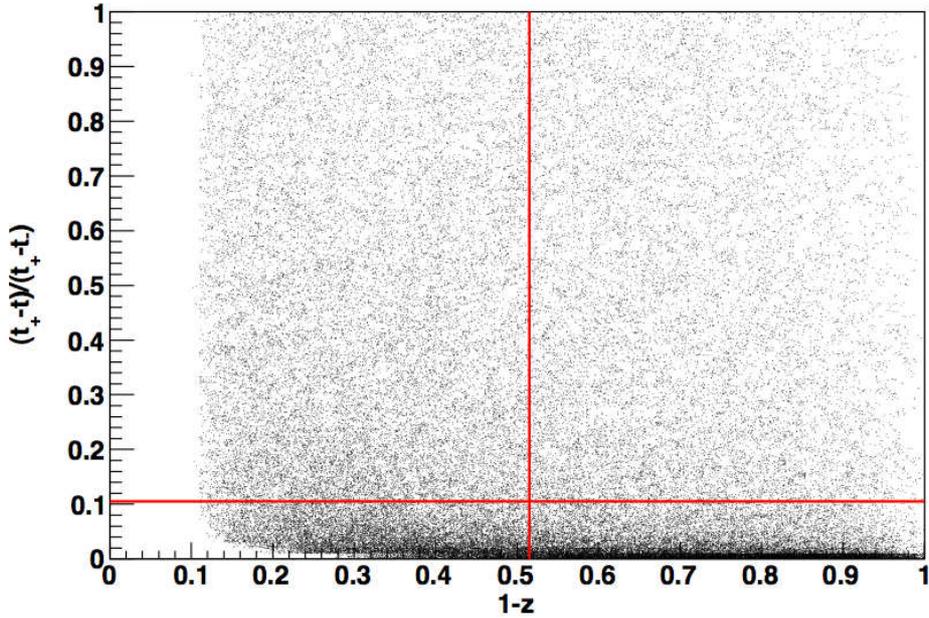}
\caption{Density plots of the hadronic cross-section of the leading
  order 4F $Wb$ production at LHC 14 TeV as a function of $1-z$ and
  $(t_+-t)/(t_+-t_-)$. The straight lines represent the median of
  the distributions in $x$ and $y$, respectively.}
\label{fig:density-80-wb}
\end{center}
\end{figure}

The partonic differential cross section is given in Eq.~\eqref{m0wb}.
In order to compare the results in the 4F and 5F schemes,
it is actually useful to rewrite Eq.~\eqref{m0wb} as a Laurent expansion
around $t-m_b^2=0$ in order to isolate the collinear singularity,
and further neglect terms which are suppressed by powers
of $m_b^2/M_W^2$.
We find
\beq
\frac{d\hat\sigma}{dt}
= \frac{\as G_{F}}{6\sqrt{2}s^3}\frac{-2M_W^6+2M_W^4s- M_W^2s^2}{t-m_b^2}
+ \mathcal{O}\left(\frac{m_b^2}{M_W^2}\right)
+\text{non-singular terms}\,.
\label{eq:coll-limit}
\eeq
Since, in this limit,
\beq
t_-=-(s-M_W^2);\qquad
t_+=-\frac{m_b^2 s}{s-M_W^2}\,,
\eeq
the total partonic cross section is
\begin{align}
\hat\sigma^{\rm 4F}(z) 
&=\frac{\as G_F}{6\sqrt{2}s^3}(2M_W^6-2M_W^4s+ M_W^2s^2)
\log\left[\frac{s}{m_b^2}\left(1-\frac{M_W^2}{s}\right)^2\right]
\nonumber\\
&=\frac{\as}{2\pi}
\left(\pi\frac{\sqrt{2}}{3}G_F\right)
z\frac{z^2+(1-z)^2}{2}
\log\left[\frac{M_W^2}{m_b^2}\frac{(1-z)^2}{z}\right]+\mathcal{O}(m_b^0),
\end{align}
where $\mathcal{O}(m_b^0)$ stands for constant or vanishing terms 
in the limit $m_b\to 0$.
Recalling Eq.~\eqref{eq:s4FWb}, we can rewrite it as 
\begin{equation}
\sigma^{\rm 4F} (\tau)=
\left(\pi\frac{\sqrt{2}}{3}G_F\tau\right)
\int_\tau^1\frac{dz}{z}\,{\cal L}_{ug}\left(\frac{\tau}{z}\right)
\frac{\as}{2\pi}
P_{qg}(z) \, L_{\rm DY}+\mathcal{O}(m_b^0)
\label{eq:2}
\end{equation}
with
\begin{equation}
L_{\rm DY} \equiv \log \left[\frac{M_W^2}{m_b^2}\frac{(1-z)^2}{z}\right].
\label{eq:calQ0}
\end{equation}
In order to assess the impact of neglecting non singular
terms and the $b$ mass, we have computed the total cross section both
with the exact and approximated partonic cross section. The results
are $\sigma^{\rm tot}= 43.7$~nb and $\sigma^{\rm tot}=37.6$~nb
respectively.  In Fig.~\ref{fig:density-80-wb} the density plot of the
cross-section is displayed as a function of $1-z$ and
$\frac{t_+-t}{t_+-t_-}$. It shows that indeed most of the cross section
lies in the small-$t$ region, while the points are uniformely distributed in
$1-z$. In particular they are not concentrated in the threshold
region $z\sim 1$, clearly showing that collinear splitting  $g\to b\bar b$ is
not a soft effect.
 
\begin{figure}[htb]
\begin{center}
\includegraphics[width=0.32\textwidth]{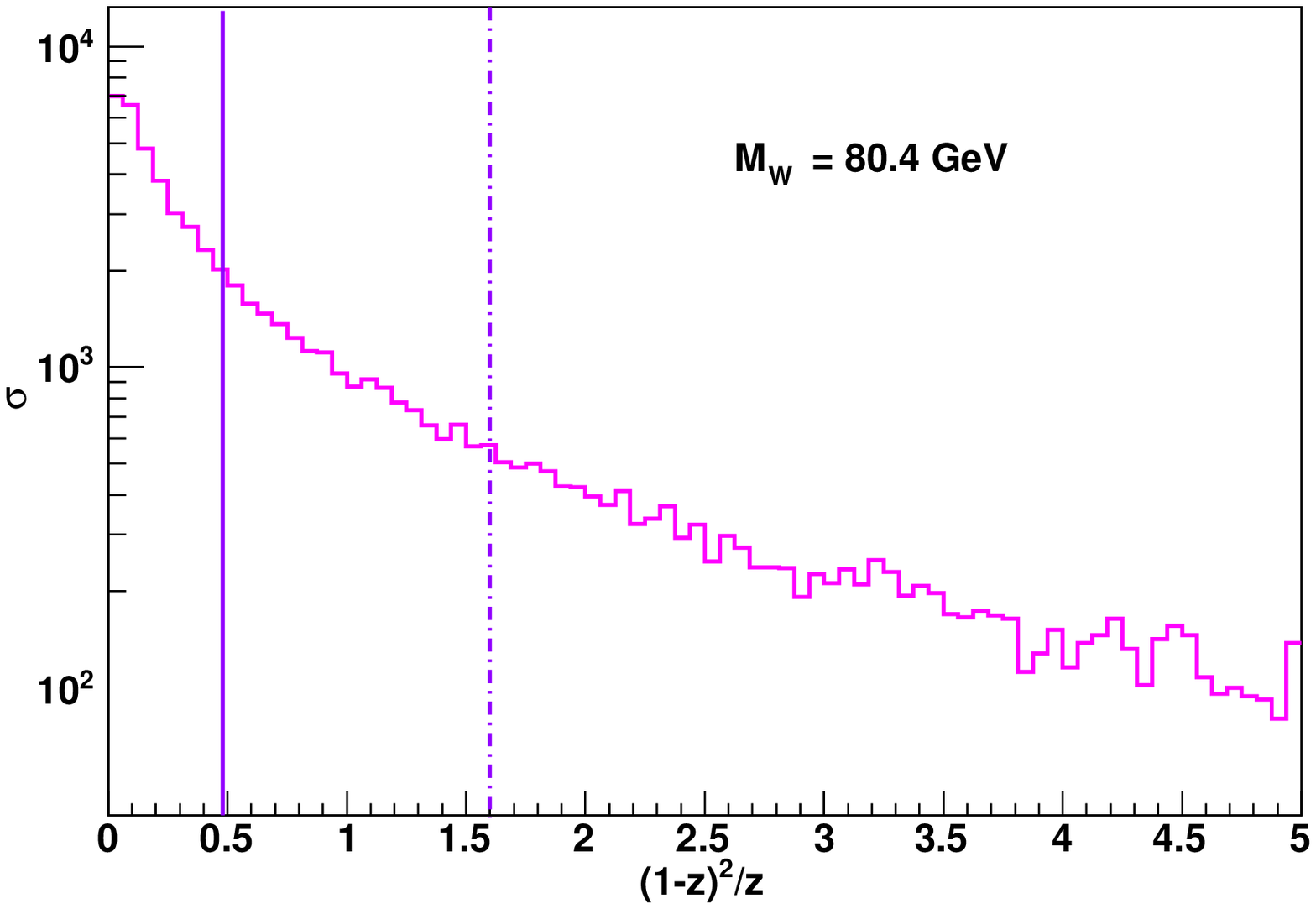}
\includegraphics[width=0.32\textwidth]{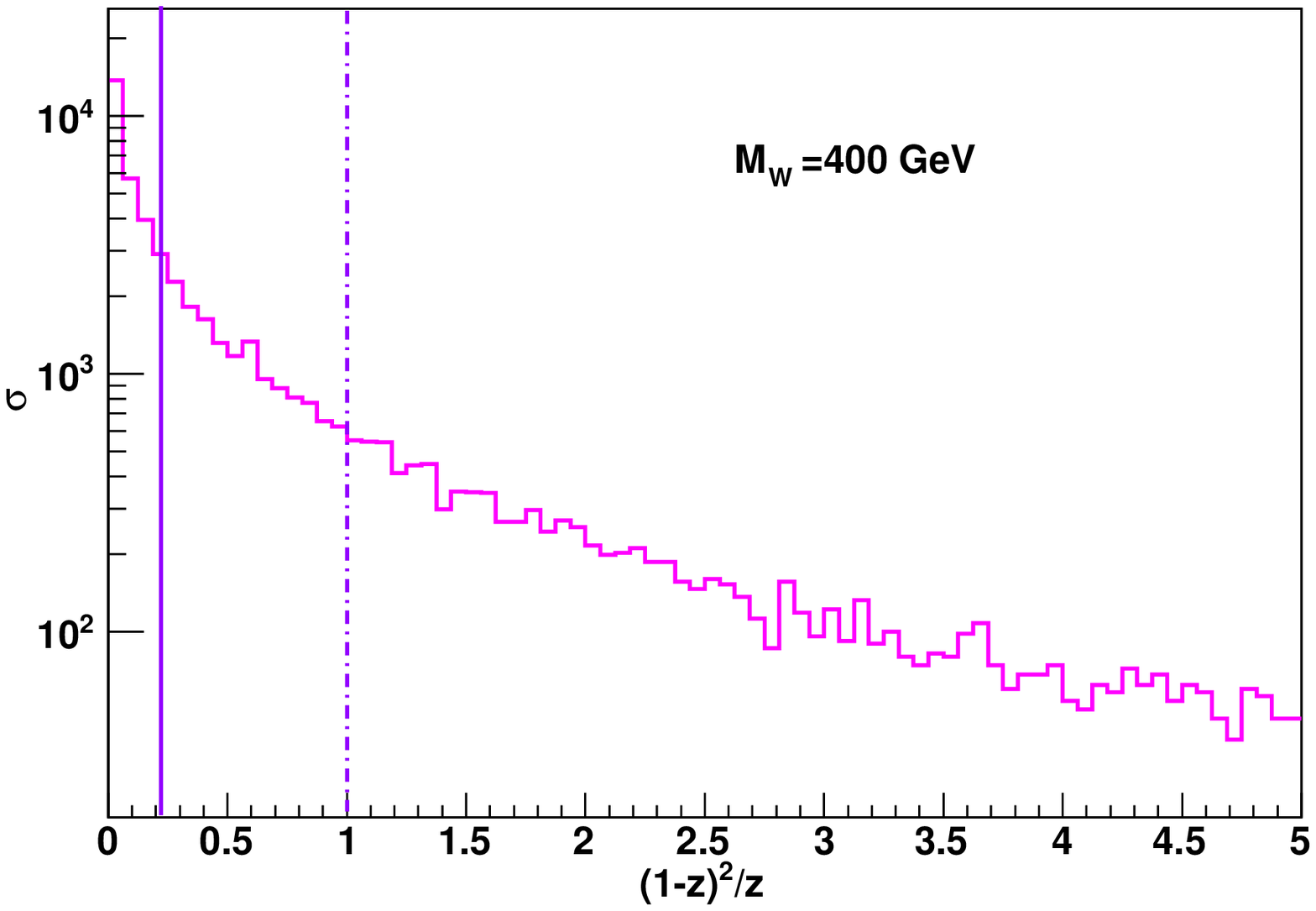}
\includegraphics[width=0.32\textwidth]{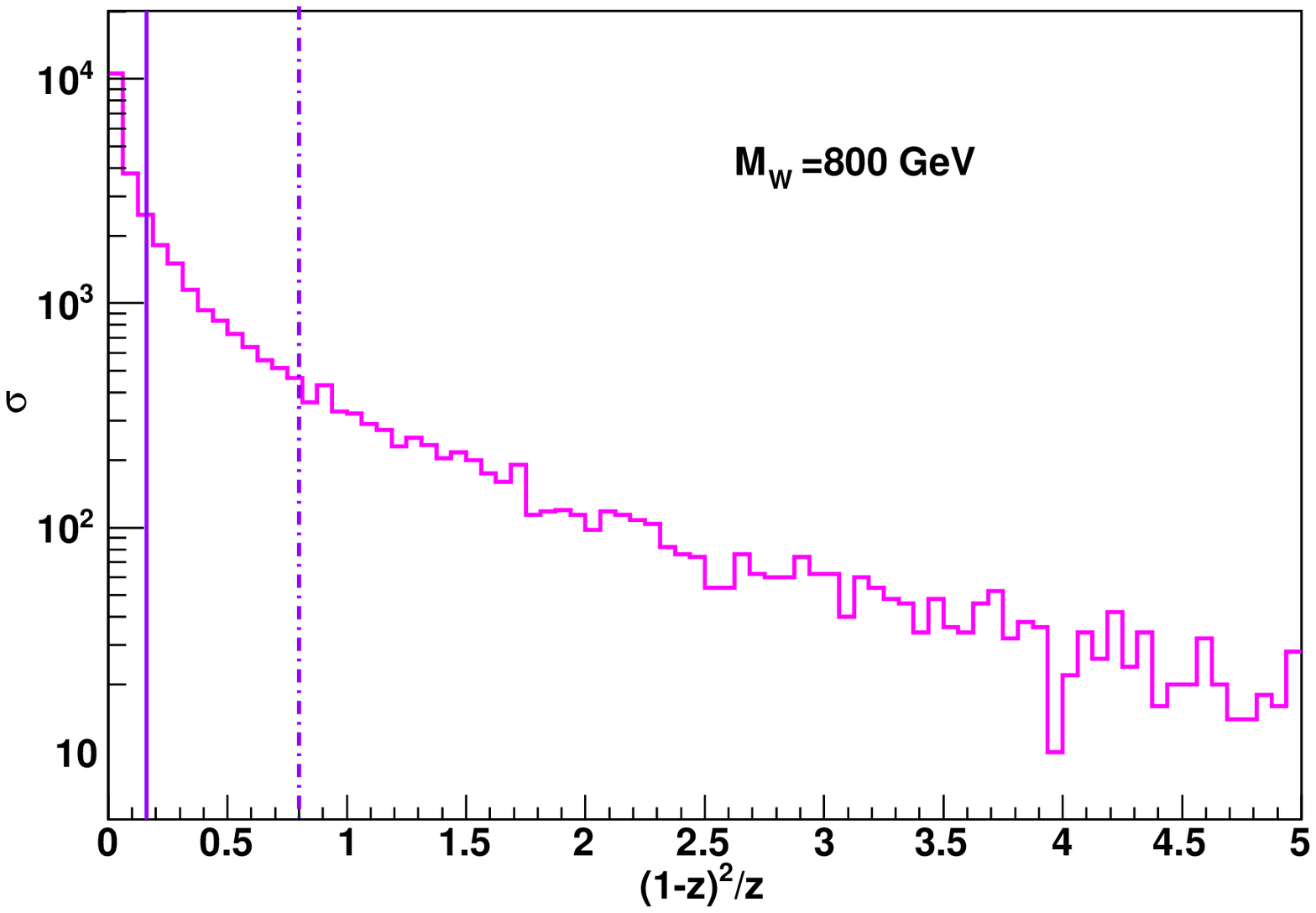}\\
\includegraphics[width=0.32\textwidth]{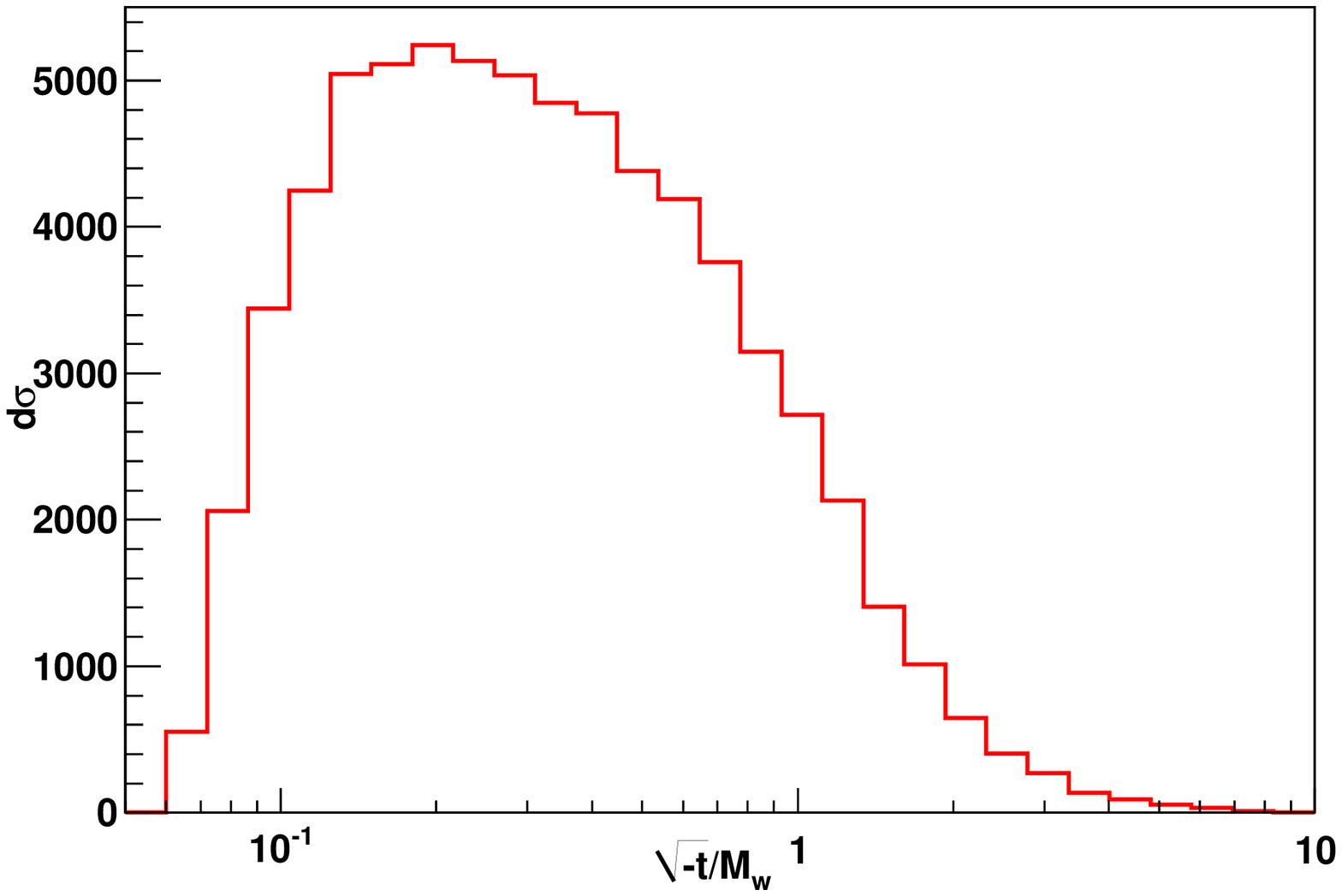}
\includegraphics[width=0.32\textwidth]{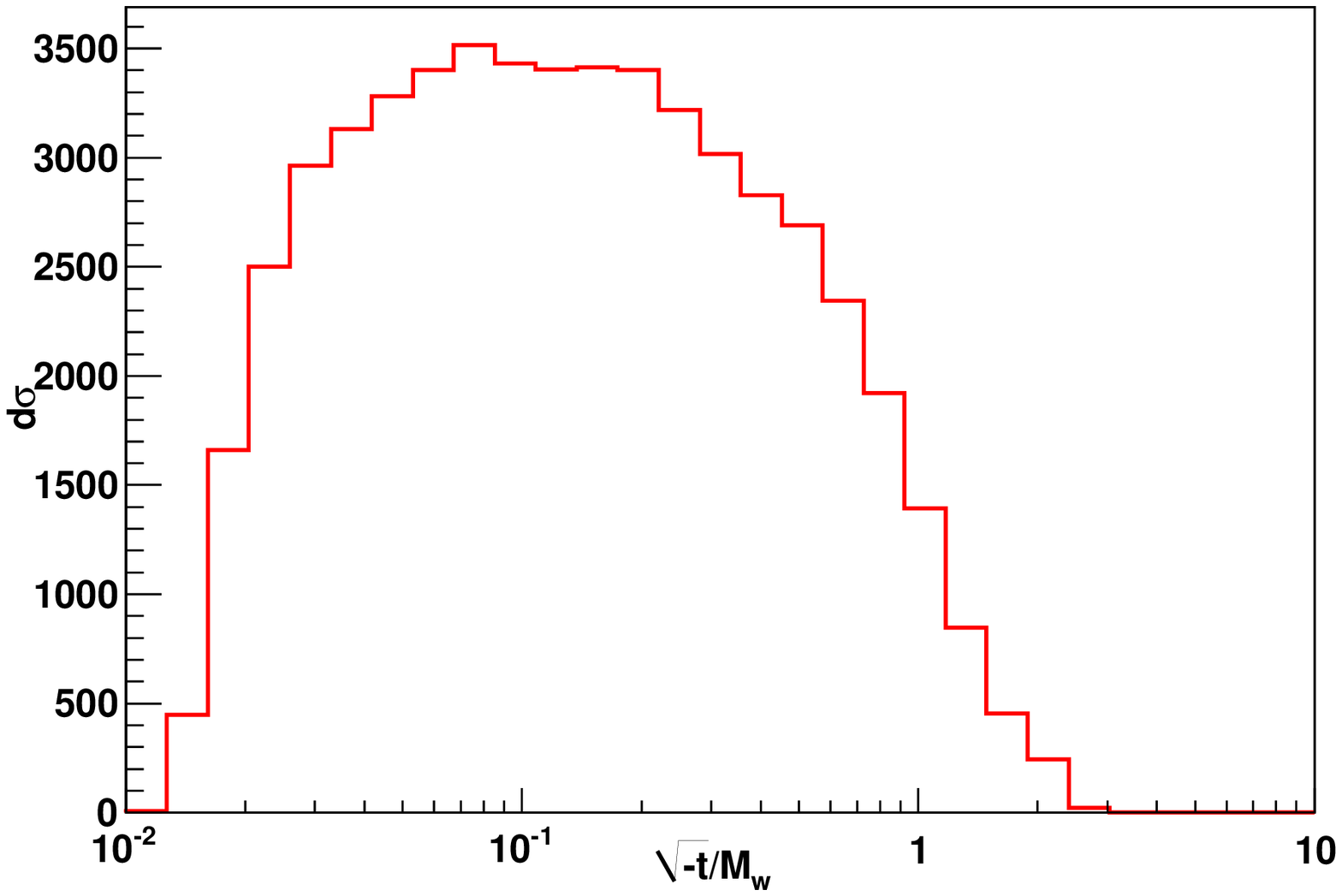}
\includegraphics[width=0.32\textwidth]{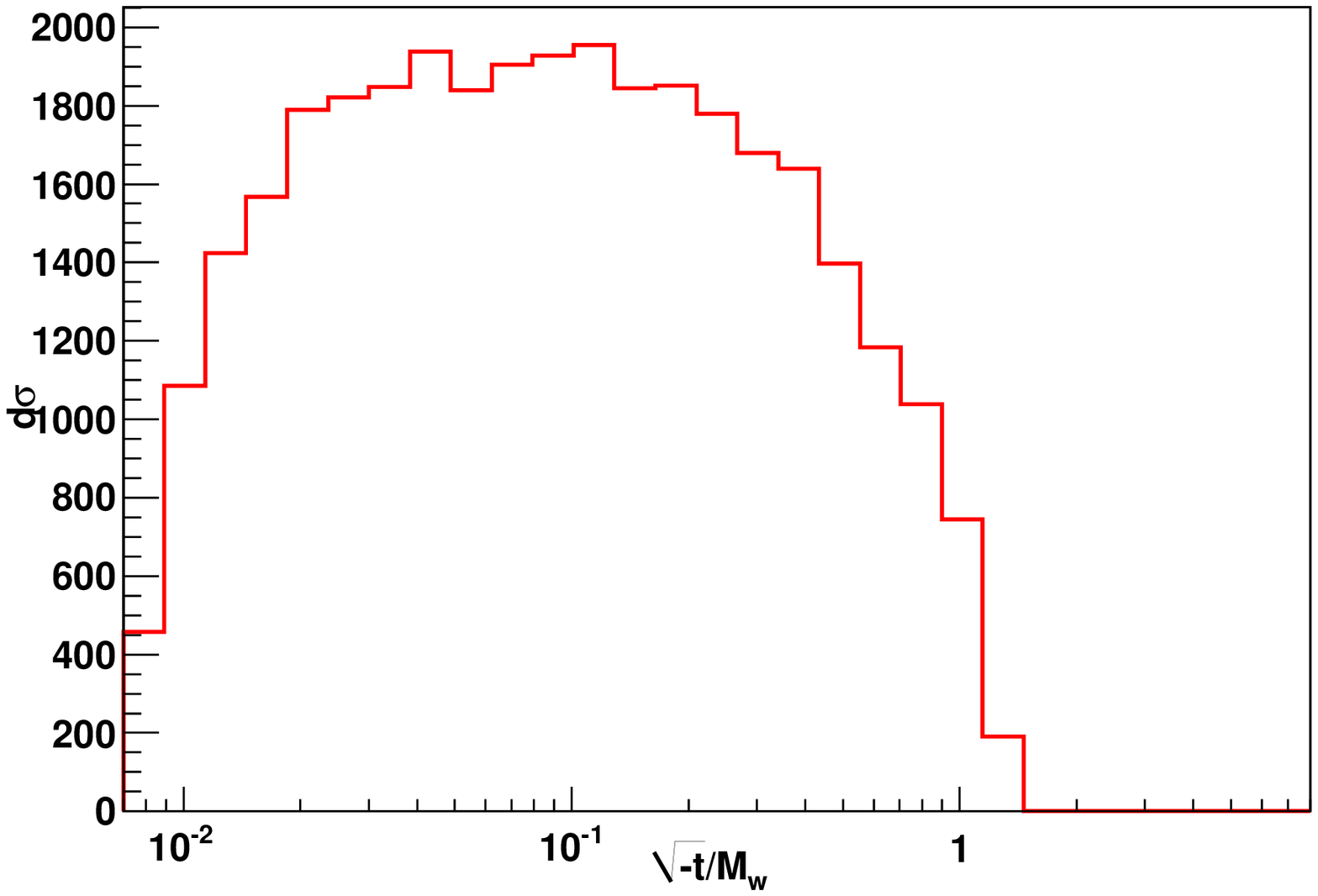}
\caption{\label{dists-wb}$b$-initiated $W$ production at the LHC 14
  TeV. $M_W=80.4$ (left), $M_W=400$ GeV (centre) and $M_W=800$ GeV
  (right).  {\bf Upper plots}: Distribution of events (in pb/bin) as a
  function of ${\cal Q}^2_{\rm DY}(z)/M_W^2=(1-z)^2/z$. 50\% of the
  events lie on the left of the vertical solid line, 80\% of the
  events on the left of the vertical dot--dashed line.  {\bf Lower
    plots}: Distribution of events (in pb/bin) as a function of
   $\log \sqrt{|t|}/M_W$.}
\end{center}
\end{figure}

It is now easy to compare Eq.~\eqref{eq:2} with the 5F computation in
the approximation of Eq.~\eqref{eq:1}.  Equations~\eqref{eq:1} and
\eqref{eq:2} have the same structure; they only differ in the argument
of the collinear logarithm, which has a fixed value in the case of the
5F calculation, while it depends on the momentum fraction $z$ in the
4F result.
In order to estimate the size of the dynamical scale, we have computed
the distribution 
of $(1-z)^2/z$, which is the suppression factor in the
argument of the collinear logarithm with respect to $M_W/m_b$. 
The result is shown in Fig.~\ref{dists-wb}.
We see that this distribution is indeed peaked around values which
are sizably smaller than one, and that the peak is shifted to
smaller values with increasing $M_W$.

As a confirmation of the conclusion arising from inspection of 
Fig.~\ref{dists-wb}, we observe that the
two calculations are seen to give the same result for
$\mu_F=\tilde\mu_F$, such that
\begin{equation}
\log\frac{\tilde\mu_F^2}{m_b^2}
=\frac{\int_\tau^1\frac{dz}{z}\,{\cal L}_{ug}\left(\frac{\tau}{z}\right)
P_{qg}(z)\log\left[\frac{M_W^2}{m_b^2}\frac{(1-z)^2}{z}\right]}
{\int_\tau^1\frac{dz}{z}\,{\cal L}_{ug}\left(\frac{\tau}{z}\right)
P_{qg}(z)}.
\label{mutilde}
\end{equation}
For  $\sqrt{S}=14$~TeV and $M_W=80.4$~GeV we find
\begin{equation}
\tilde\mu_F =33\; {\rm GeV} 
\quad (\tilde\mu_F = 42\; {\rm GeV}),
\end{equation}
where the result in parenthesis is obtained by using the exact
partonic cross section for the 4F calculation, i.e., keeping the exact
$m_b$ and $t$ dependence.  Both results point to a value remarkably
smaller than $M_W$, i.e. in the range $\mu_F \simeq [0.4,0.5] \, M_W$.
This reduction is even more pronounced at larger values of the $W$
virtuality:
\beq
\begin{array}{ll}
M_W=400\, {\rm GeV} &, \qquad  \tilde\mu_F \simeq   [0.3,0.4]\,M_W\ \\
M_W=800\, {\rm GeV} &, \qquad  \tilde\mu_F \simeq   [0.25,0.35]\,M_W.
\end{array}
\eeq
The above results clearly suggest that "fair" comparisons between 4F
and 5F calculations should be performed at factorization scales that
are in general smaller than a na\"{\i}ve choice. This fact had been
noticed previously in several studies concerning $b$-initiated
processes, see for
instance~\cite{Boos:2003yi,Maltoni:2003pn,Campbell:2004pu,Maltoni:2005wd,
Maltoni:2007tc}.

Finally, it is instructive to plot the differential cross section as a
function of $\log \sqrt{|t|}/M_W$ (lower plots of
Fig.~\ref{dists-wb}), i.e., $t\, d\sigma/dt$.  For a massless quark
one expects a plateau to develop at small $t$ corresponding to the
collinear pole $d\sigma/dt \propto 1/t$. For finite $b$ quark the
collinear divergence is regulated by the quark mass and the
distribution goes to zero (a.k.a. dead cone).  The drop of the plateau
at larger $t$ can be taken as signaling the end of the collinear
enhanced region, i.e., where the $2 \to 2 $ hard scattering dominates.
We observe that in general the collinear plateau drops at scales
smaller than $M_W$ for the production of on-shell $W$ bosons, in line
with the results found above for the factorization scale.  The scale
associated to the splitting of the gluon is softer that the scale
associated to the hard $W$ production process, which is naturally
given by $M_W$. This implies that the factorization scale that should
be set for the bottom PDFs in the four--flavor scheme is smaller than
$M_W$, as suggested in several earlier studies. The considerations in
this subsection help to understand the dynamical origin of this fact.

\subsection{Single top production}

We now consider the production cross-section for events with a single
top quark in the final states in hadron-hadron collisions.  It was
shown in Refs.~\cite{maltoni:stop1,maltoni:stop2} that the central
values of the cross section predicted at next--to--leading order
according to the 4F and 5F schemes differ by 5\% or less, both at the
Tevatron and at the LHC, by setting the factorization scale around the
mass of the top quark (or the invariant mass of its decay products).
At the Tevatron, the difference is well within the combined
uncertainty from higher orders and PDFs, while at the LHC (10 TeV) the
consistency was found to be marginal. For larger masses, \ie, for
$t^\prime$ production, the differences were found to be much
larger. For a $t'$ of mass of 1 TeV, the $2 \to 2$ prediction 
is almost twice as large at the Tevatron and
$20\%$ larger at the LHC. Therefore, for such large top masses it
could well be that the logarithm that is implicitly resummed in the
bottom quark distribution function might become relevant, or that an
even smaller factorization scale should be used. Here we further
investigate this process by using the same procedure adopted in the
previous section.

\begin{figure}[h!]
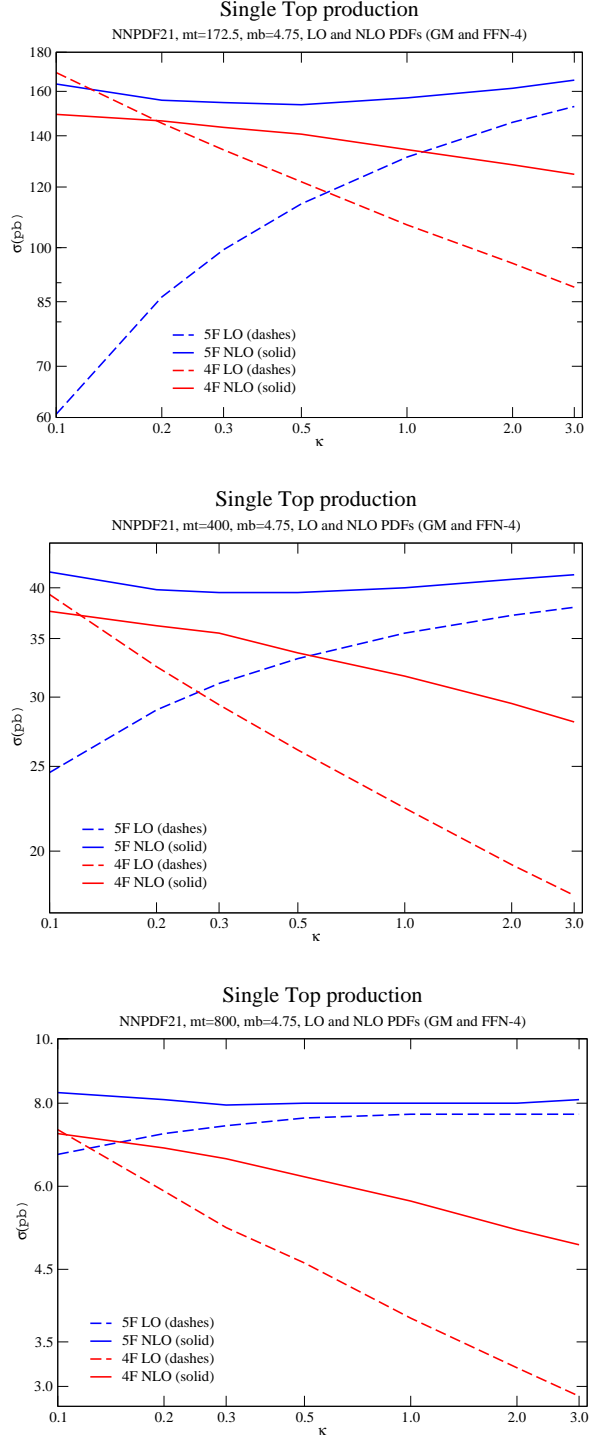

\begin{center}
 \includegraphics[width=0.5\textwidth]{singletop-172-nnpdf.eps}
\vspace*{0.5cm}

 \includegraphics[width=0.5\textwidth]{singletop-400-nnpdf.eps}
\vspace*{0.5cm}

 \includegraphics[width=0.5\textwidth]{singletop-800-nnpdf.eps}
 \end{center}
 \caption{Comparison between 4F and 5F production for single top at LHC 14 TeV as a function of 
$k=\mu/M_t$, with $\mu = \mu_F = \mu_R$, for $M_t$ =172.5 GeV (top), $M_t$ = 400 GeV (middle), 
$M_t$ = 800 GeV (bottom) and $m_b$ = 4.75 GeV. Input PDFs: {\tt NNPDF21\_FFN\_NF4} and 
{\tt NNPDF21} (LO and NLO) for 4F and 5F respectively.\label{fig:stop-comp-nnpdf}}
\end{figure}
 
In Fig.~\ref{fig:stop-comp-nnpdf} the scale dependence of the total
cross section for the single top production in the $t$-channel is
displayed for both the leading and the next-to-leading order
computations. For this process the scales are varied about the mass of
the top. On top of the on-shell top quark case $M_t=172.5$ GeV, we consider
also two cases where the top has a higher virtuality: $M_t=400$
GeV and 800 GeV. All curves have been produced by running the MCFM
code. In particular, the ACOT computation for the 5F scheme is
obtained by matching the LO massive computation with the NLO
massless. The input PDFs belong to the NNPDF2.1
family~\cite{NNPDF21}. The set associated to the 5F scheme is the
default NNPDF2.1 NLO set, based on a FONLL-A scheme which at NLO is
equivalent to the ACOT scheme~\cite{ForteNason}. The set associated to
the 4F scheme is the {\tt NNPDF2.1\_FFN\_NF4} set which is based on a
4F fixed flavor number scheme.

In both figures we observe that the LO curves have an opposite
behavior, being the 5F curve driven by the scale dependence of the
bottom PDFs at the relevant $z$ where the fraction of the momentum
is peaked, while the 4F one is driven by the running of $\as$. We
further observe that both the leading order and the next-to-leading
curves obtained in the two schemes are closer to each others for a
value of the factorization and renormalization scales which is about
half of the mass of the top quark.  For heavier top, the scale
dependence of the 4F computation deteriorates and the scale at which
the two computations get closer decreases as increasing the value of
the mass of the top quark. 

\begin{figure}[htb]
\begin{center}
\includegraphics[width=0.32\textwidth]{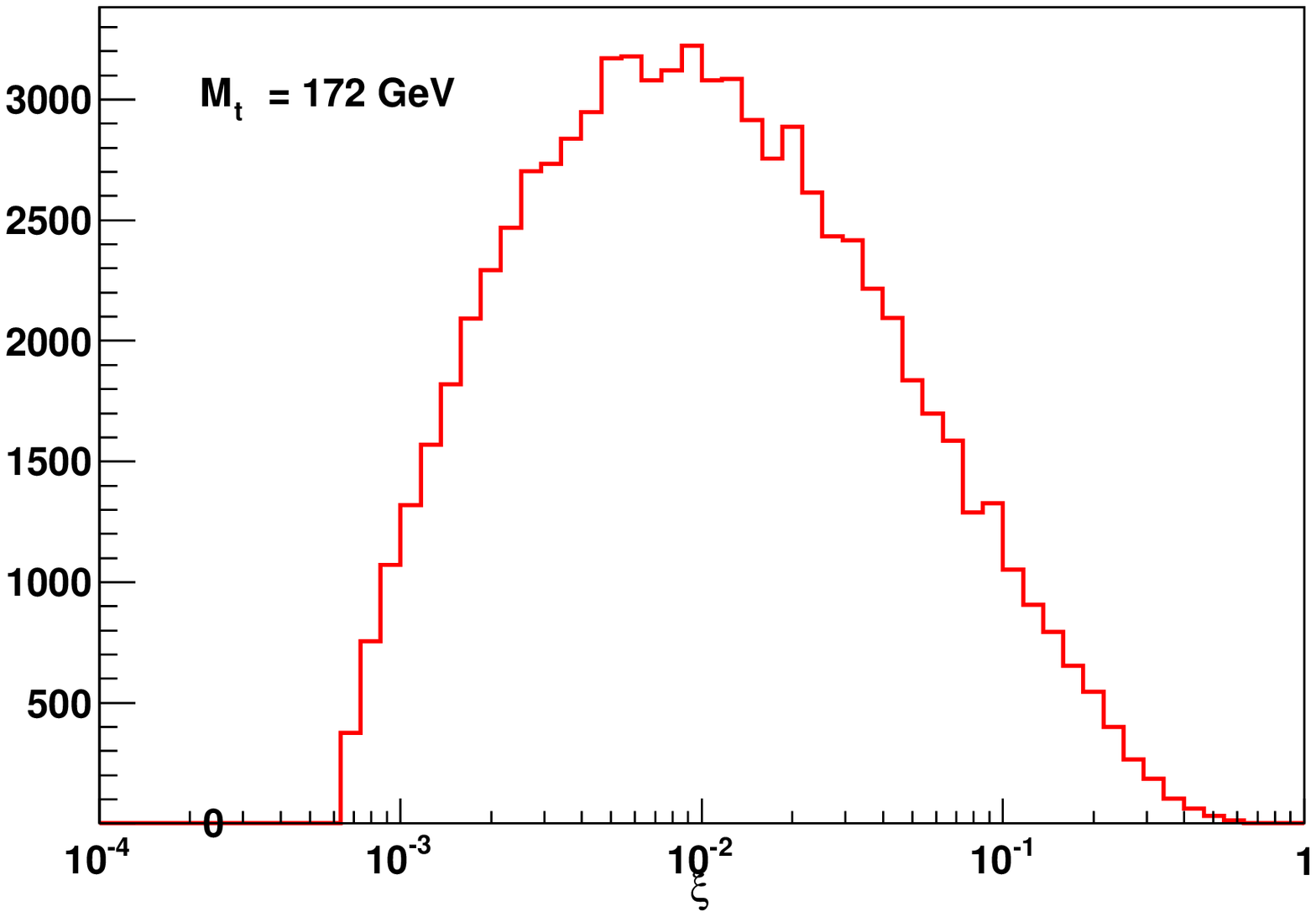}
\includegraphics[width=0.32\textwidth]{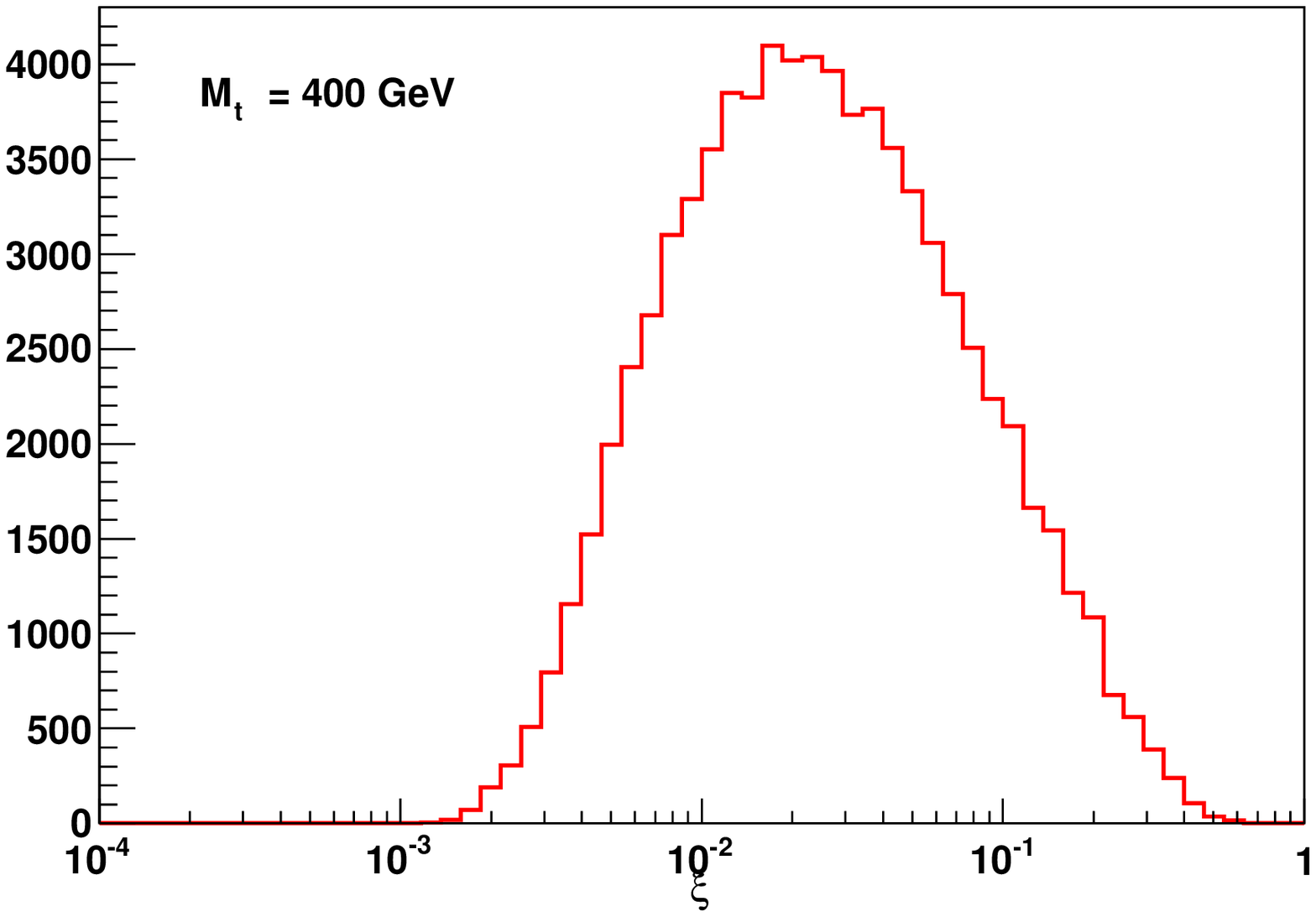}
\includegraphics[width=0.32\textwidth]{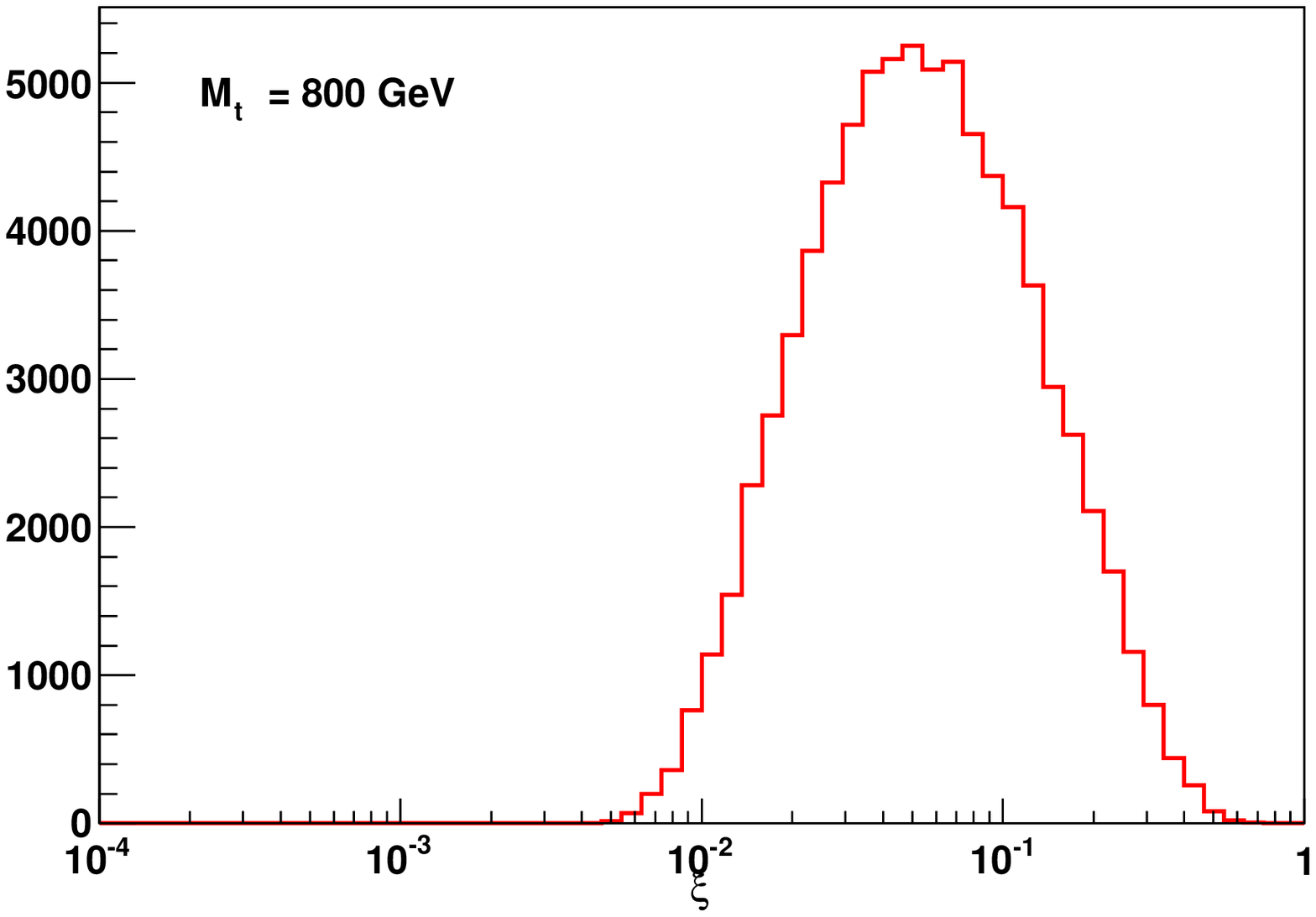}
\caption{\label{dists-x-stop}Distributions of events as a function of
  the proton momentum carried by the incoming $b$ quark in single
  top production at LHC in the 5F scheme, leading order. 
$\sqrt{S}=14$ TeV and $M_t=172.5$ (left), $M_t=400$ GeV (centre) and
  $M_t=800$ GeV (right). Input PDF: {\tt NNPDF21} (LO).}
\end{center}
\end{figure}

We first consider the leading-order computation in the 5F scheme.
The relevant partonic subprocess is
\begin{equation}
W^*(q)+b(p)\to t(k),
\end{equation}
and the corresponding amplitude is easily computed:
\beq
\mathcal{M}_\mu=\frac{g_W}{\sqrt{2}}\bar u(k)\gamma^\mu P_L u(p).
\eeq
As explained in Appendix~\ref{hq}, the contribution to the structure function
$F_2$ is obtained by applying a suitable projection operator
on the squared amplitude, and integrating over the one-particle phase space:
\begin{align}
\hat\sigma_2^{5F}(z)&=\int d\Phi_1(p+q;k)
\left[-\frac{1}{2}g^{\mu\nu}+\frac{3}{2}\frac{4Q^2}{(s+Q^2)^2}p^\mu p^\nu
\right]\mathcal{M}_\mu\mathcal{M}^*_\nu
\nn\\
&=\pi g_W^2 \delta(1-z),
\end{align}
where we have defined, in analogy with deep-inelastic scattering,
\beq
z=\frac{M_t^2+Q^2}{s+Q^2};\qquad s=(p+q)^2;\qquad Q^2=-q^2\,,
\label{zdef}
\eeq
so that $0\leq z\leq 1$, and we have used
\beq
d\Phi_1(p+q;k)=\frac{d^3 k}{(2\pi)^32k^0}(2\pi)^4\delta(p+q-k)
=\frac{2\pi}{s+Q^2}\delta(1-z).
\eeq
Thus, at leading order the cross section (or better, the contribution
to $F_2$) from $t$ production is simply proportional to the
$b$ parton distribution function. From Fig.~\ref{dists-x-stop} we conclude that
the fraction of momentum carried by the $b$ quark is peaked at larger value as
the virtuality of the produced top increases.

\begin{figure}[htb]
\begin{center}
\includegraphics[width=0.32\textwidth]{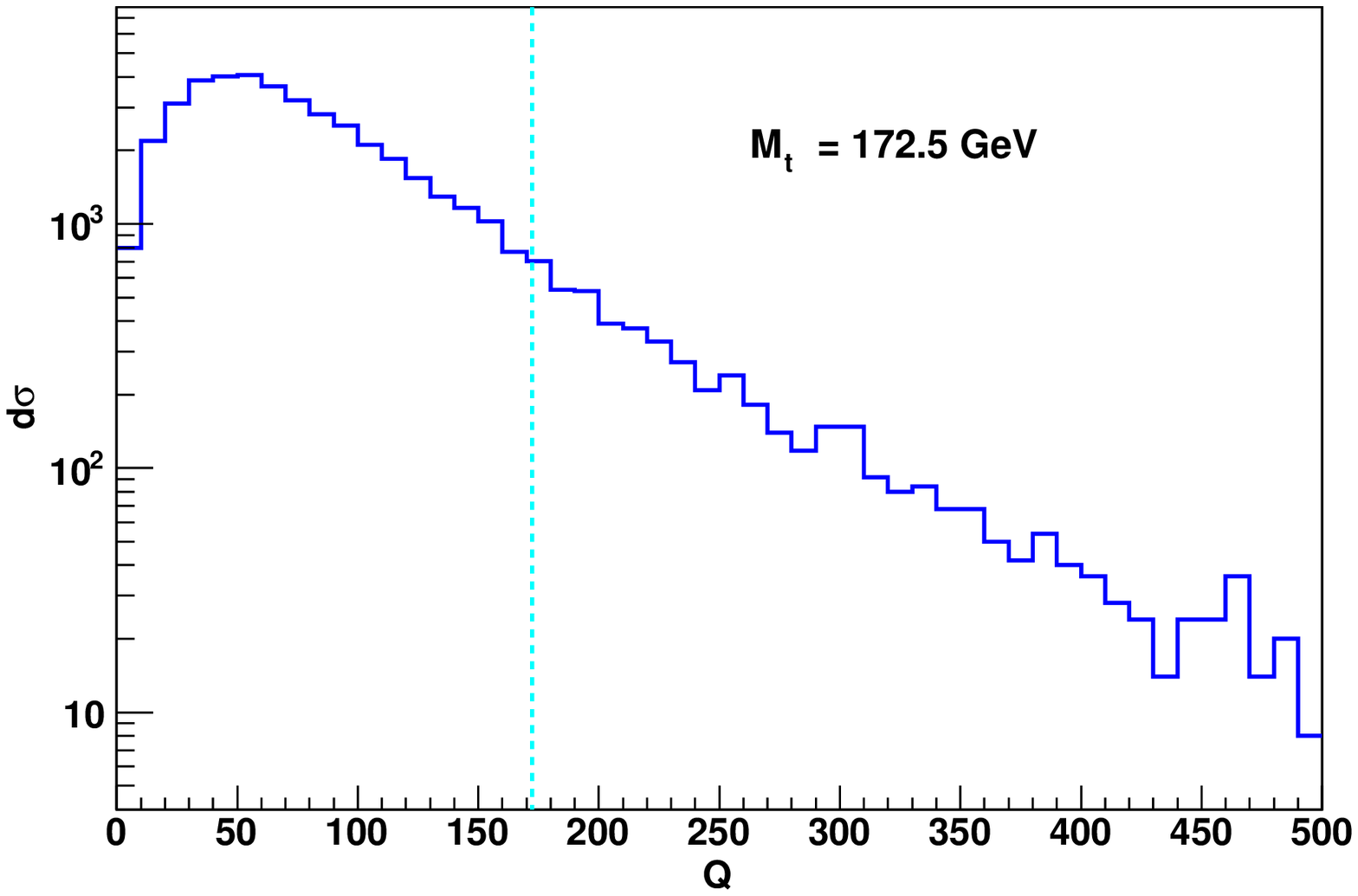}
\includegraphics[width=0.32\textwidth]{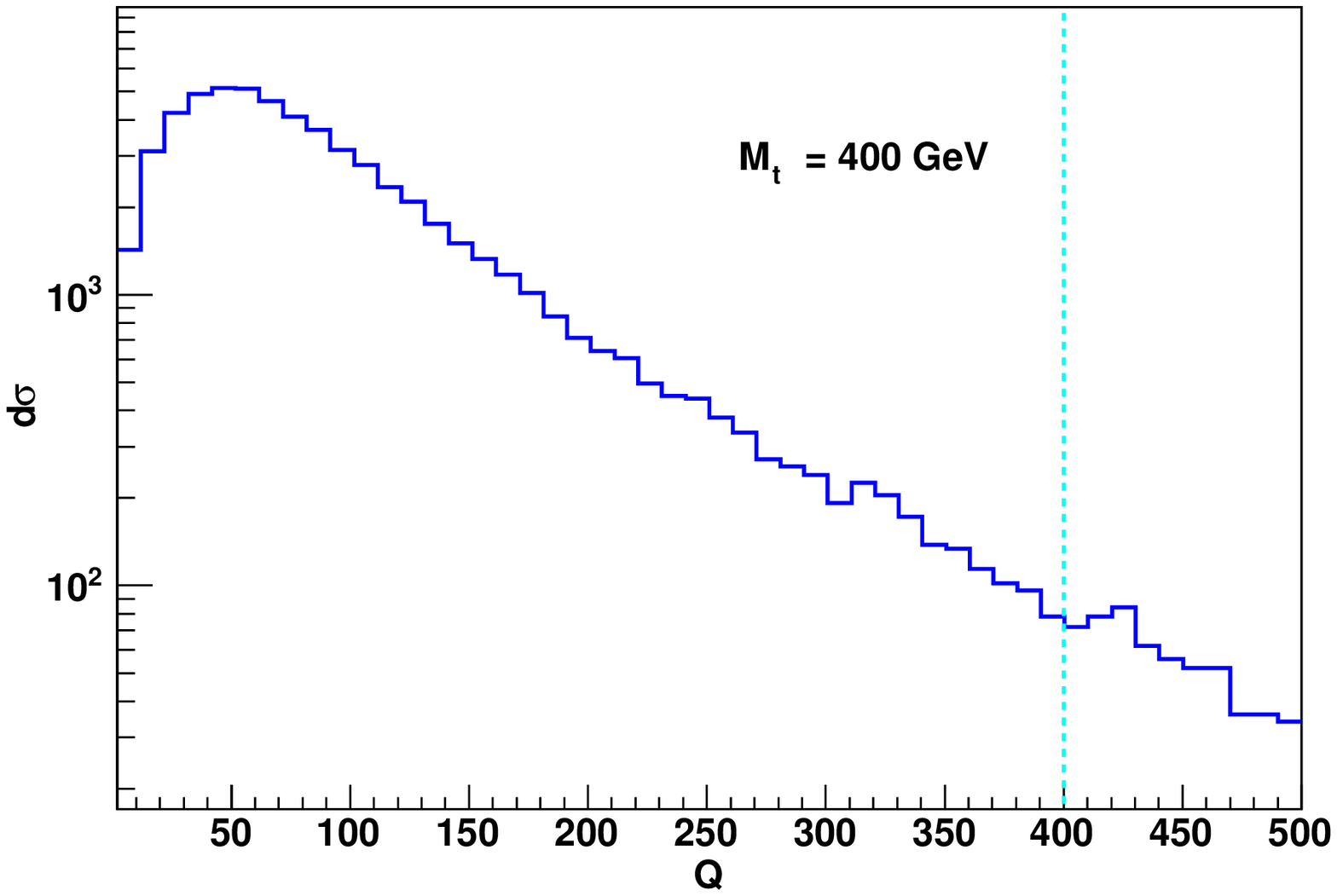}
\includegraphics[width=0.32\textwidth]{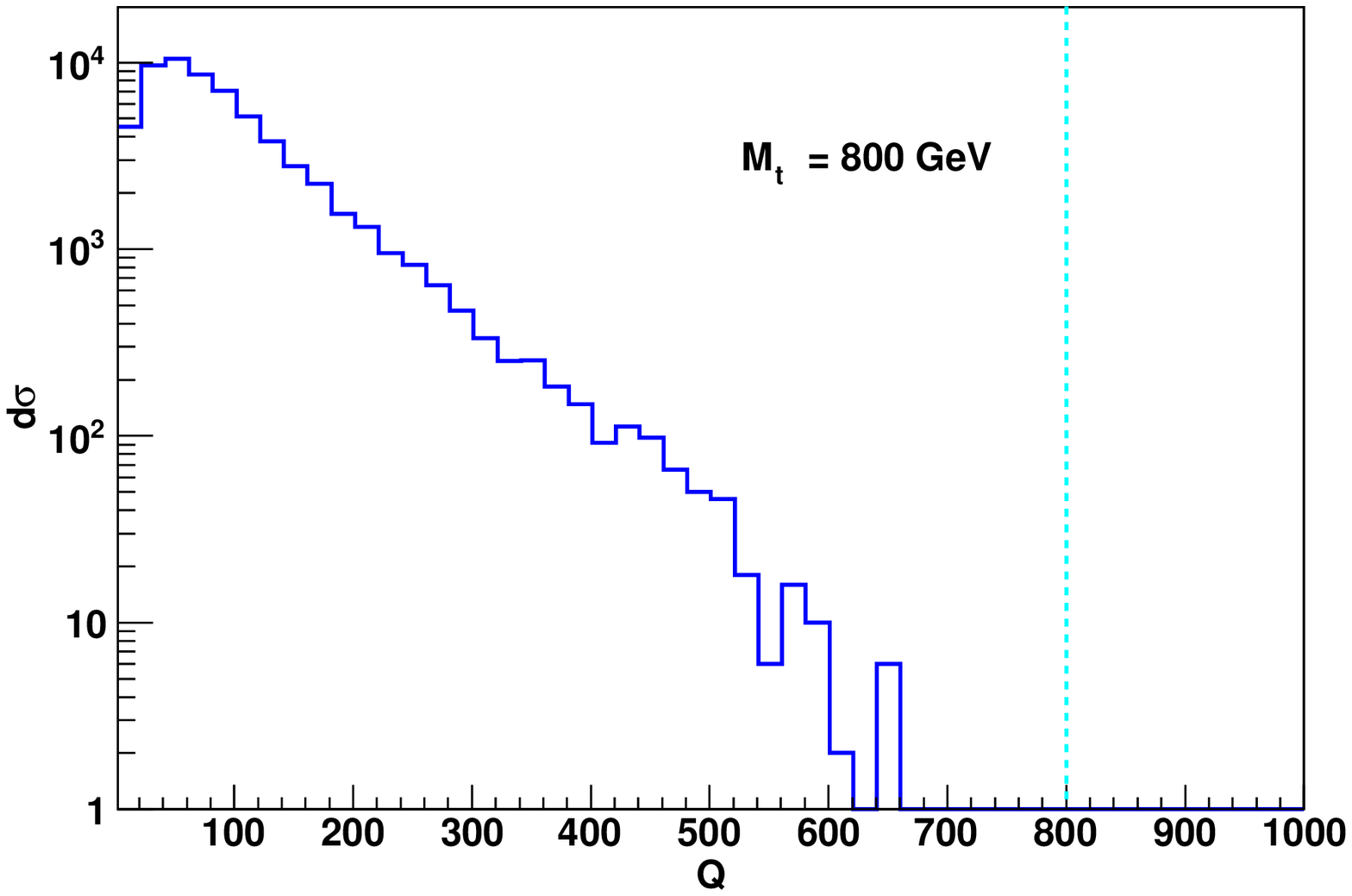}\\
\includegraphics[width=0.32\textwidth]{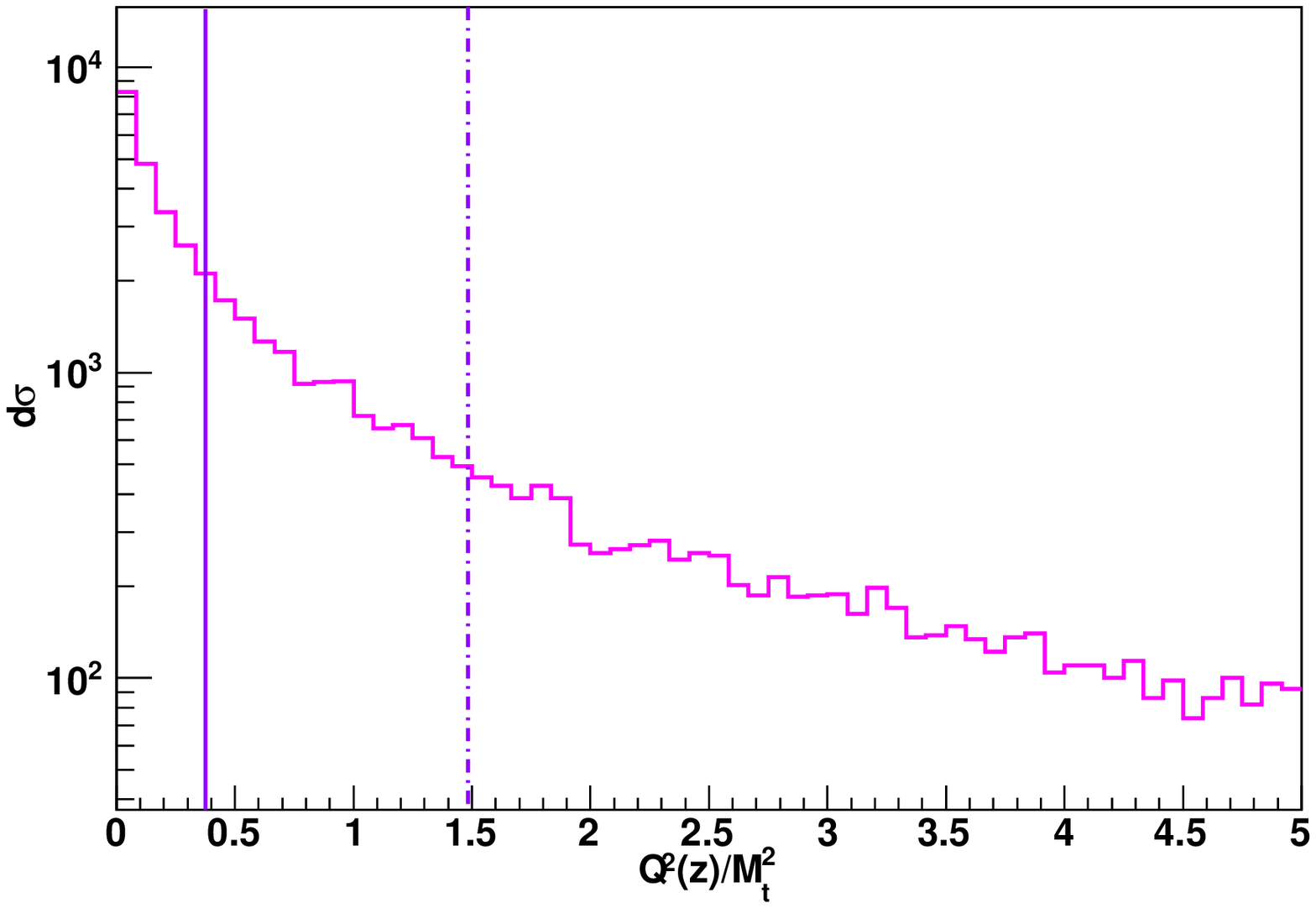}
\includegraphics[width=0.32\textwidth]{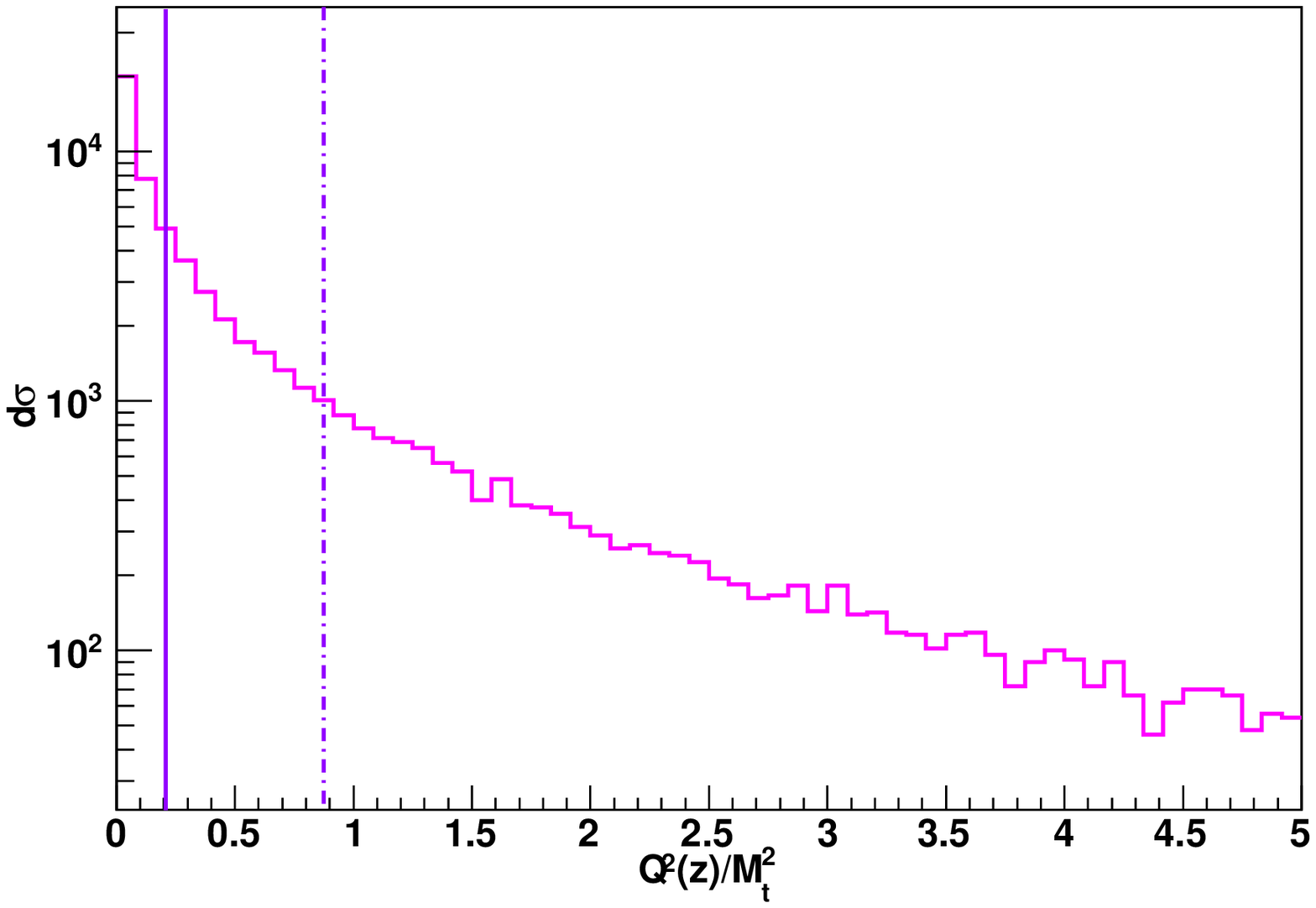}
\includegraphics[width=0.32\textwidth]{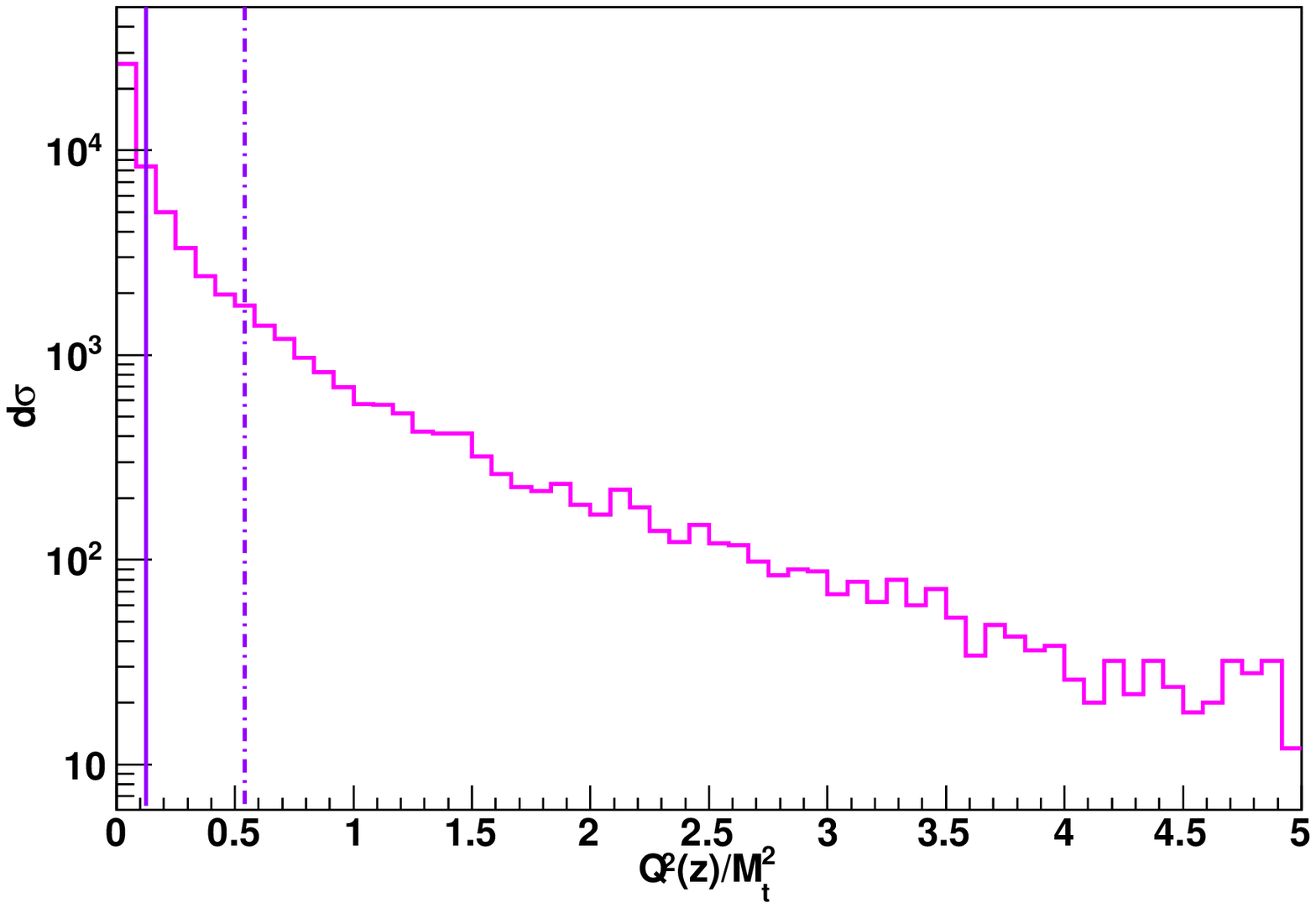}
\includegraphics[width=0.32\textwidth]{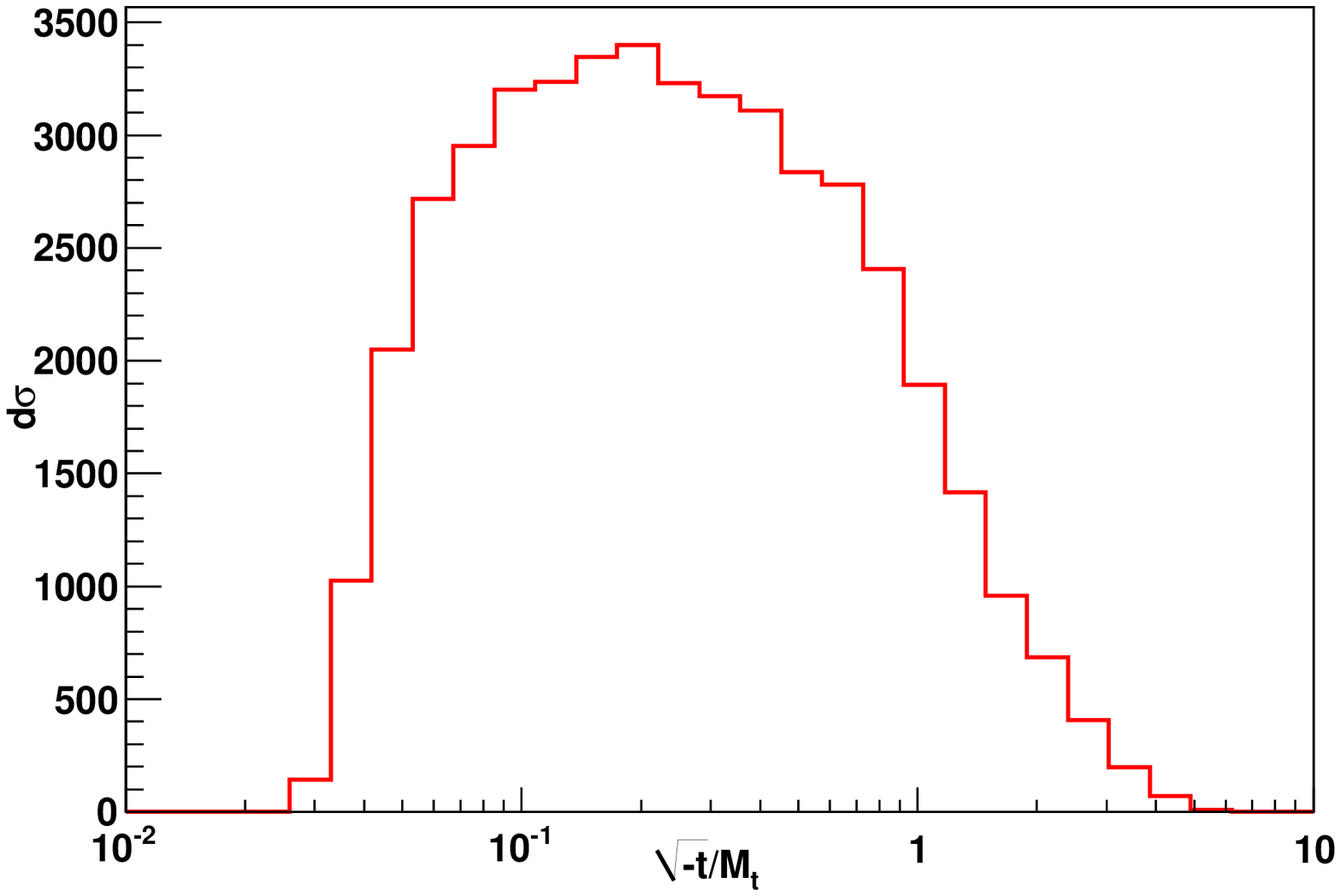}
\includegraphics[width=0.32\textwidth]{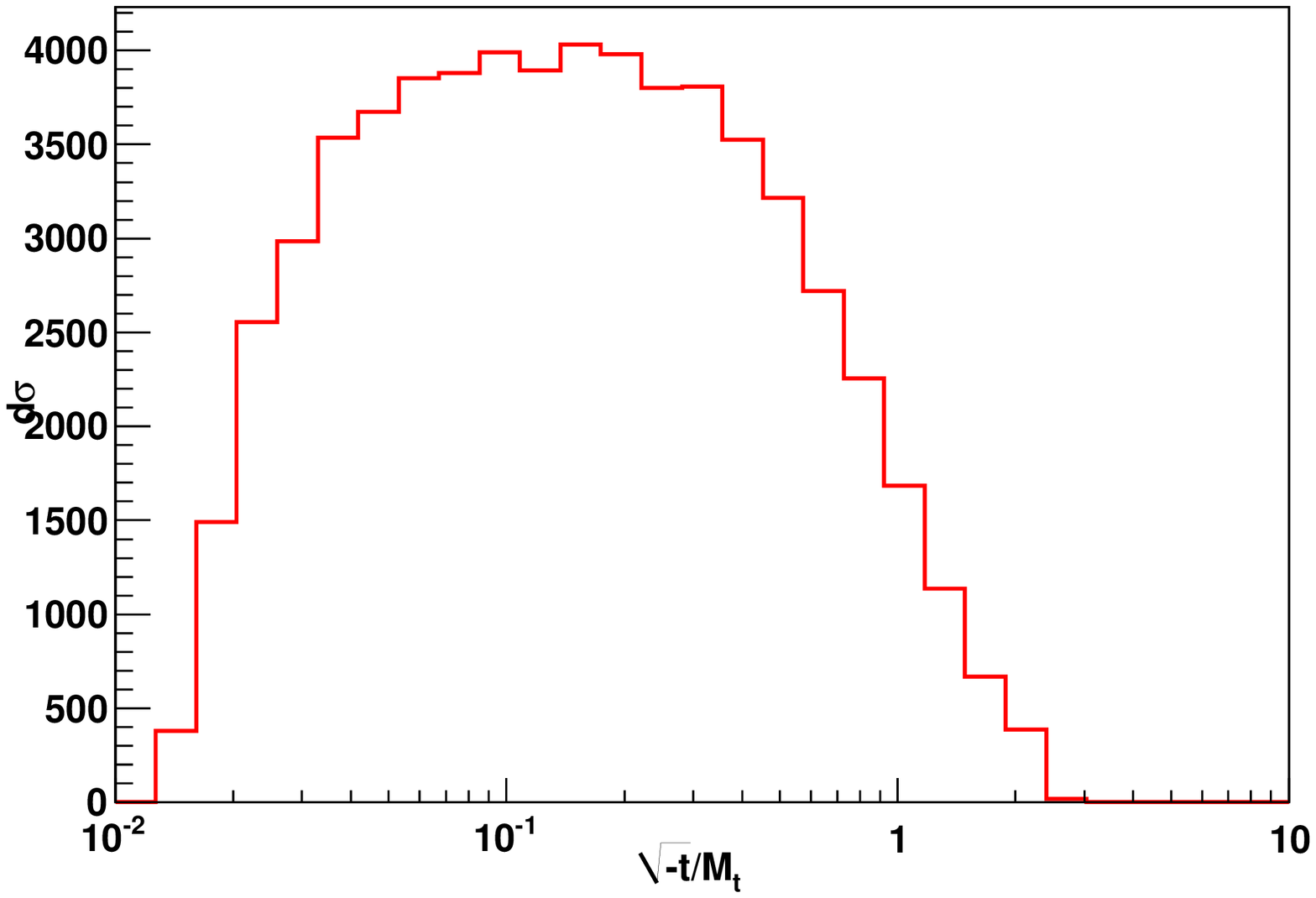}
\includegraphics[width=0.32\textwidth]{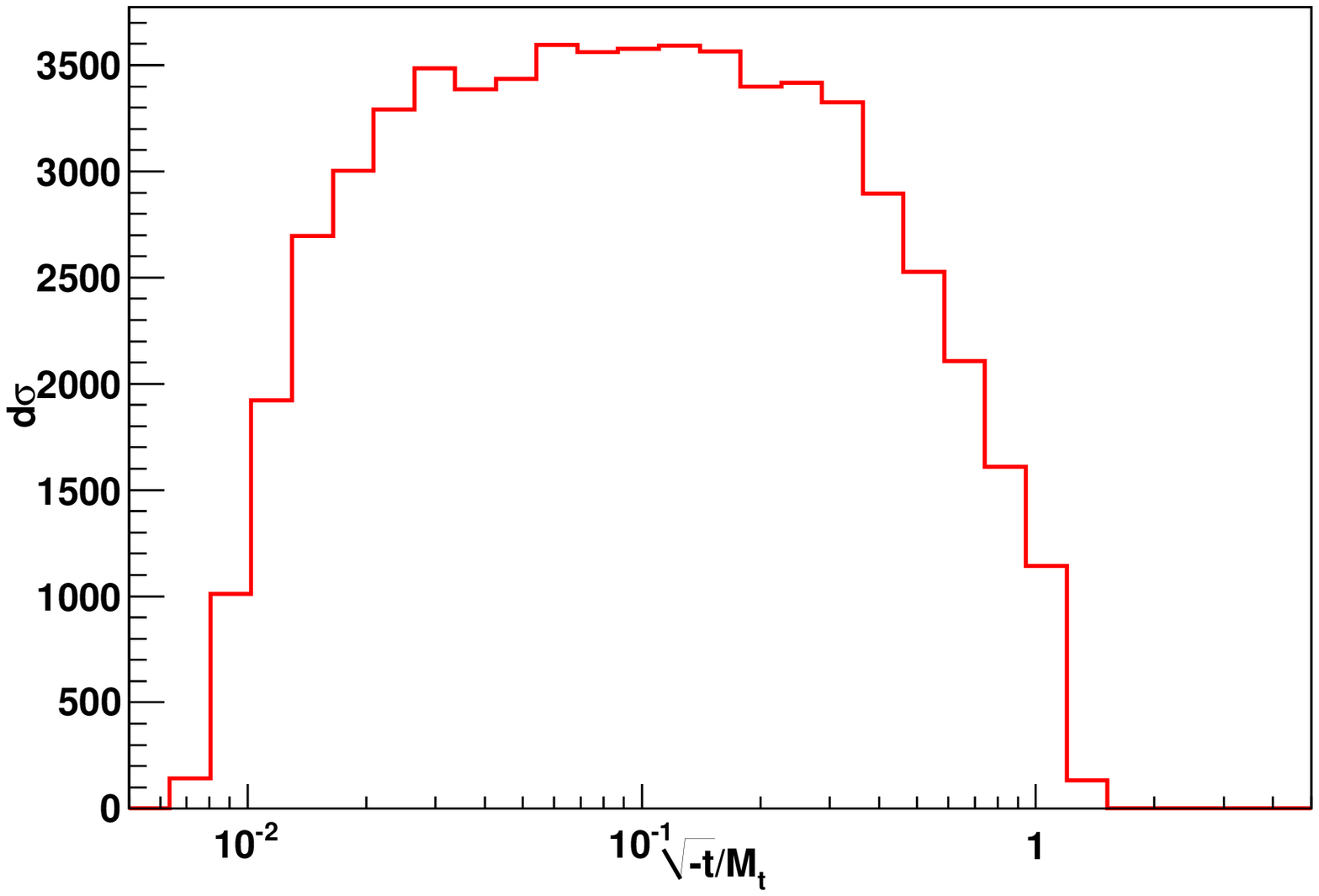}
\caption{\label{dists-s-top} Single--top production at the LHC 14 TeV. Input
  PDF: {\tt NNPDF21\_FFN\_NF4} (LO).  {\bf Top}: Distributions of events as
  a function of $\sqrt{Q^2}$; the vertical blue line corresponds to
  $Q=M_t$, for $M_t=172.5$ (left), $M_t=400$ GeV (centre) and
  $M_t=800$ GeV (right).  {\bf Middle}: Distributions of events as a
  function of ${\cal Q}^2(z)/M_t^2$, the vertical solid line
  corresponds to the median of the distribution and the dot--dashed
  line corresponds to the 80\% of the distribution, for $M_t=172.5$
  (left), $M_t=400$ GeV (centre) and $M_t=800$ GeV (right).  
{\bf Bottom}: Distributions of events as a function of
  $\log\frac{\sqrt{|t|}}{M_t}$ for $M_t=172.5$ (left), $M_t=400$ GeV
    (centre), $M_t=800$ GeV (right). }
\end{center}
\end{figure}

In order to interpret correctly the scale dependence of the curves in
Fig.~\ref{fig:stop-comp-nnpdf} we consider the collinear limit of the
massive leading--order computation. The full expression of the LO
partonic differential cross--section $d\hat\sigma_2^{\rm 4F}/dt$ is given by
Eqs.~(\ref{mgt},\ref{mgl},\ref{mg3}) with $M=M_{t}$, $m=m_b$, $g_R= 0$
and $g_L=g_{W}/\sqrt{2}$.
This specific term in the total cross sectio is the only one
which has a pole in $t=0$ in the small $m_b$ limit.
Taking the limit $m_b\to 0$ and keeping only terms which give rise
to the collinear singularity we find
\begin{align}
\frac{d\hat\sigma_2^{\rm 4F}}{dt}&=\frac{3 \as g_W^2 C_F}{64(s+Q^2)^3}
\left[
\frac{(M_t^2+Q^2)^2-M_t^2(s+Q^2)-Q^2(s+Q^2)+(s+Q^2)^2/2}{t-m_b^2}\right]
\nonumber\\
&+{\rm non\,singular\,terms}.
\end{align}
Hence, in this limit,
\begin{align}
\int_{t_-}^{t_+}dt\, \frac{d\hat\sigma_2^{\rm 4F}}{dt} &=
\frac{3 \as g_W^2 C_F}{128 (s+Q^2)^3} 
\nn\\
&\left[2M_t^4+4Q^2 M_t^2-2(s+Q^2)M_t^2+s^2+Q^4\right]
\log\left[\frac{s}{m_b^2}\left(1-\frac{M_t^2}{s}\right)^2\right].
\label{singletop}
\end{align}
With the definition Eq.~\eqref{zdef}, it is immediate to recognize that
Eq.~\eqref{singletop} can be written as
\begin{equation}
\int_{t_{{\rm min}}}^{t_{{\rm max}}}\,dt \frac{d\hat\sigma_2^{\rm 4F}}{dt} =
\frac{3 \as g_W^2 C_F}{64 (s+Q^2)}\frac{z^2+(1-z)^2}{2}
\log\frac{{\cal Q}^2(z)}{m_b^2},
\end{equation}
with
\begin{equation}
{\cal Q}^2(z)=\frac{(M_t^2+Q^2)^2}{M_t^2+(1-z) Q^2} \frac{(1-z)^2}{z}.
\label{eq:calQ}
\end{equation}
This result can be interpreted as the scale corresponding to the
collinear splitting not being $M_t$ or $\sqrt{M_t^2+Q^2}$, but rather the
dynamical scale ${\cal Q}(z)$.  As we will comment at length in the
next subsection the expression above actually encompasses all cases
discussed so far and paves the way towards the generalisation of our
analysis to any process involving one or two $b$ quarks in the initial
state.  The fact that the collinear logs in single-top production are
in fact not very large can be also drawn by looking at the plots in
Fig.~\ref{dists-s-top}, where cross section distributions are shown as
a function of $Q, {\cal Q}^2(z)/M_t^2$, and $\log\sqrt{|t|}/M_t$, in the
first, second and third lines respectively. 


\subsection{The general case}
\label{sec:general}

\begin{figure}[t]
\begin{center}
\includegraphics[width=0.4\textwidth]{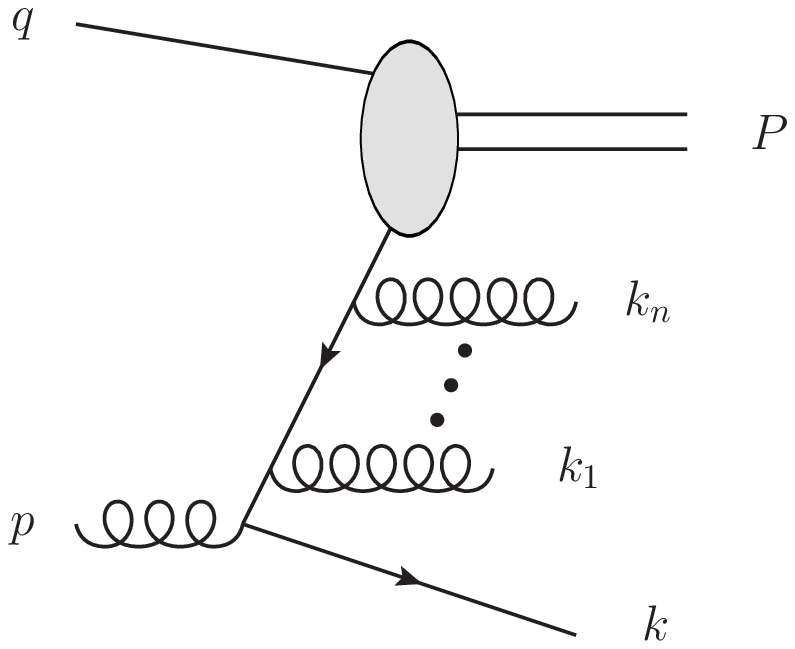}
\end{center}
\caption{Generic process involving gluon splitting into $b\bar b$ pair in the initial state and the production of an heavy system $P$. 
Multi-gluon emission from the quark line gives rise to the collinear logarithms at all orders  that  are resummed into
 the $b$ parton distribution function. \label{fig:ladder}}
\end{figure}

Single-top production at hadron collisions is not only interesting per
se, but also because it provides the simplest yet most general
kinematics for any process involving one gluon splitting in the
initial state:
\beq
I(q)+g(p)\to b(k)+g(k_1)+\ldots +g(k_n)+X(P),
\label{eq1:gen}
\eeq 
where $I$ is a generic initial state particle (whose virtuality
can also be space-like, such in the case of DIS) and $X$ is a generic
heavy system, possibly made out of several particle, such that
$P^2\geq M^2$.  The case of single-top production is obtained by
identifying $I$ with a virtual $W$ boson with virtuality $q^2=-Q^2$,
and $X$ with a top quark, so that $P^2\geq M_t^2$.  In the case of
$Wb$ production, $I$ is a light parton ($q^2=0$) and $X$ is a $W$, so
that $P^2\geq M_W^2$.  In both cases, no light parton is present in
the final state in the lowest order calculation of the 4F scheme; we
have included in Eq.~\eqref{eq1:gen} the possibility that the system
$X$ is produced in association with a collection of light partons with
momenta $k_1,\ldots,k_n$, which is obviously the case for higher-order
contributions to the cross-section. Multiple gluon emission from the
$b$ quark line gives rise to the collinear logs that are resummed in
the $b$-PDF's, Fig.~\ref{fig:ladder}.

In the previous sections we have reconsidered  the
limit $m_b^2\to 0$ of the lowest-order 4F cross-sections, which have a collinear
singularity due to the gluon splitting into $b\bar b$ pair in the initial state; in other words,
the $b$ quark mass acts as a regulator of the collinear
singularity.  Naively, one therefore expects  a logarithm of the type
\beq
L_{\rm naive} = \log \frac{M^2+Q^2}{m_b^2}
\eeq
i.e., of the hard scale to the collinear regulator, to develop in the integrated 
cross section. Large $M$ and/or $Q$  therefore motivate the resummation of such logs via 
the introduction of $b$-PDF's and the use of the DGLAP equations.

In fact, our analysis based on the lowest-order processes has shown that the 
logarithmically-enhanced contributions to the cross-sections are proportional to
\beq
L=\log\frac{{\cal Q}^2(z)}{m_b^2},
\label{eq1:general}
\eeq
where ${\cal Q}^2(z)$, defined in Eq.~(\ref{eq:calQ}), can be rewritten in the following revelatory form
\beq
{\cal Q}^2(z)=(M^2+Q^2)\frac{(1-z)^2}{z}\frac{1}{1-\frac{zQ^2}{M^2+Q^2}}
\qquad{\rm with}\qquad 
z=\frac{M^2+Q^2}{s+Q^2}, 
\label{eq1:mu2}
\eeq
which encompasses in one expression all interesting cases previously discussed.
Equation~(\ref{eq1:general}) reduces to the log 
in DIS $b\bar b $ production in the limit $M\to 0$:
\begin{equation}
L_{\rm DIS} =\log\left[\frac{Q^2}{m_b^2}\frac{1-z}{z}\right] \qquad  {\rm with} 
\qquad z=\frac{Q^2}{s+Q^2},
\label{eq1:DIS}
\end{equation}
and to the log in the hadronic $Wb$ associated
production in the limit $Q^2 \to 0$:
\begin{equation}
L_{\rm DY} =\log\left[\frac{M^2}{m_b^2}\frac{(1-z)^2}{z}\right]\qquad
{\rm with} \qquad z=\frac{M^2}{s}.
\label{eq1:W}
\end{equation}
In the case of single top production, both $M^2$ and $Q^2$
are in general non-negligible, and the collinear log keeps its
general form Eqs.~(\ref{eq1:general}-\ref{eq1:mu2}).

In this section we argue that the form Eqs.~(\ref{eq1:general}-\ref{eq1:mu2}) is not only
general enough to encompass all interesting processes at the lowest order, but  also it is valid at higher orders, 
where powers of collinear logarithms are generated by multi-parton emission
\beq
\log^n\frac{{\cal Q}_n^2(z)}{m_b^2}
\eeq
with ${\cal Q}_n^2(z)\leq {\cal Q}^2(z)$ and for processes with two initial-state gluon splitting into $b\bar b$.

A complete argument based on the ladder kinematics and valid at any order is detailed in Appendix~\ref{app:ho}. 
Here we provide an argument that while it has  the disadvantage of not providing 
the exact expression of the cross section in the collinear limit, yet it illustrates in a very simple way the 
origin and universality of the kinematic terms appearing in Eq.~(\ref{eq1:mu2}).
For the sake of the argument, we consider the case where $X$ is a real heavy particle,
such as a top quark or a $W$ boson. 

In the initial-state centre-of-mass system, the  four-momenta of the particles in the process
Eq.~\eqref{eq1:gen} can be parametrized as
\begin{align}
&p=\left(\frac{s+Q^2}{2\sqrt{s}},0,0,\frac{s+Q^2}{2\sqrt{s}}\right)
\label{p}
\\
&q=\left(\frac{s-Q^2}{2\sqrt{s}},0,0,-\frac{s+Q^2}{2\sqrt{s}}\right)
\label{q}
\\
&k_n=\left(\omega_n,0,\omega_n\sin\theta,\omega_n\cos\theta\right)
\\
&P_n=\left(\sqrt{\omega_n^2+M_n^2},0,-\omega_n\sin\theta,
-\omega_n\cos\theta\right)
\end{align}
where $P_n=P+k+k_1+\ldots+k_{n-1}$, $P_n^2=M_n^2$ and
\beq
\omega_n=\frac{s-M_n^2}{2\sqrt{s}}\leq 
\frac{\sqrt{s}}{2}\left(1-\frac{M^2}{s}\right)
\eeq
since
\beq
M_n^2\geq M^2.
\eeq
We may now define
\beq
z=\frac{M^2+Q^2}{s+Q^2}
\eeq
so that the upper bound on the energy of the gluon with momentum $k_n$ 
can be written
\beq
\omega_n\leq\frac{{\cal Q}(z)}{2}
\eeq
where ${\cal Q}(z)$ is given in Eq.~\eqref{eq1:mu2}.  As a consequence, the
transverse momentum of each of the $n$ gluons in the final state is
bound from above by the same quantity. We therefore expect at most a factor
of
\beq
\log\frac{{\cal Q}^2(z)}{m_b^2}
\eeq
for each emitted gluon. This covers all processes that involve one
gluon splitting in the initial state.  In fact, given that no use on the details
of the kinematics is made, the same argument can be equally applied to processes
such as $gg\to Zb\bar b$ or $gg\to Hb\bar b$ that involve two gluons splitting into $b\bar b$ pairs
in the initial state. A detailed proof in terms of $t$-channel kinematics is  provided 
for this case too in Appendix~\ref{app:ho}. 

We conclude this section by noting that  while the impact
 of the logarithms for processes with one or two $g\to b \bar b$ splittings 
in the initial state is similar, the correspondence between LO, NLO,... 4F calculations  with the LL, NLL,...accuracy of calculations in  5F schemes only applies to processes with only one splitting.
 This simple relation does not hold in the case of two splittings as for example in  $gg\to Hb\bar b$. At LO,  the 4F calculation 
displays the leading logs $\as^2 \log^2\frac{{\cal Q}^2(z)}{m_b^2}$ and part of the subleading ones,  
yet at NLO only one splitting at a time is correctly reproduced up to NLL the other being effectively only LL. In other words,
one would need to go up to NNLO in the 4F scheme to correctly account for all NLL terms of each splitting at that order. This is  easily achieved by the 5F calculation already at NLO. In this respect it is not surprising that  5F scheme calculations of total cross sections  for $b\bar b$ fusion into Higgs or $Z$  leads to very stable  results under scale variations~\cite{mstwHQ}. 
\section{Conclusions}

A quite important set of processes that feature gluon splitting into a
$b\bar b$ pair in the initial state are of phenomenological interest
at the LHC. Accurate predictions for such processes can be obtained
performing calculations in either the 4F and 5F schemes. Motivated
by the somewhat unexpected result that in most cases predictions in
the two schemes are found to substantially agree within uncertainties
if judicious scale choices are made, in this work we have addressed
the following two basic questions:
\begin{itemize}
\item What is the typical size of the effects of the resummation of
  initial-state collinear logs of the type $\log\frac{{\cal Q}^2}{m_b^2}$
  with respect to an approximation where only logs at a finite order
  in perturbation theory are kept?
\item What is the typical size of the $\log\frac{{\cal Q}^2}{m_b^2}$
  themselves in phenomenologically relevant processes at hadron
  colliders, and in particular at the LHC?
\end{itemize}

Our main conclusion is that unless the typical Bjorken $x$ probed by the process is large, the
effects of initial-state collinear $\log\frac{{\cal Q}^2}{m_b^2}$'s is always
modest, and that even though total cross-sections computed in 5-flavor
schemes may indeed display a smaller uncertainty, such logarithms do
not spoil the convergence of perturbation theory in 4-flavor scheme
calculations.
 
We have identified two main and different reasons, one of dynamical
and the other of kinematical nature.

The first is that the effects of the resummation of the 
$\log\frac{{\cal Q}^2}{m_b^2}$ 
universal terms is quite small and relevant mainly at large Bjorken $x$
and in general keeping only the explicit logs appearing at  NLO is in fact an excellent approximation.  This
observation accounts for previously noticed, and yet not well-understood,
behaviours, such as the more sizable differences between predictions
in the two schemes for single top and $b b \to H$ at the Tevatron
than at the LHC~\cite{maltoni:stop1,Campbell:2004pu}.

The second result of our study is that the effective scale
${\cal Q}$ which enters in the initial-state collinear logarithm while
proportional to hardest scale(s) in the process, turns out to be
modified by universal phase space factors that tend to reduce the size
of the logarithms for processes taking place at hadron colliders. In
our study we have provided a simple analytic formula, valid also in
the case of processes with two $b$ quarks in the initial state, that
can be used to quantitatively assess the size of such logs for any
process at the LHC. We have also provided a simple rule to choose the
factorization scale where to perform comparisons between calculations
in the two schemes. As a result, a consistent and quantitative
explanation is provided of the many examples where a substantial
agreement between total cross sections obtained at NLO (and beyond) in
the two schemes can be found within the expected uncertainties.

The main outcome of this study is that 4F and 5F schemes provide
complementary information and it strongly motivates having
calculations at higher orders available in both schemes for any given
process. (Improved) 5-flavor schemes, for example, can typically
provide quite accurate predictions for total rates and being simpler,
in some cases allow the calculations to be performed at NNLO, such as
those already available for $bb\to H,Z$ and foreseeable in the near
future, such as $t$-channel single-top production. On the other hand,
being often the effects of resummation very mild, 4-flavor
calculations can be also put to use. They can be useful to achieve
accurate fully exclusive predictions, such as those obtained from
Monte Carlo programs at NLO accuracy. Promoting a 4-flavor calculation
at NLO to an event generator is nowadays a fully automatic procedure
and kinematic effects due to the $b$ quark mass can be taken into
account from the start leading for example to a more accurate
description of the kinematics of the spectator $b$ quarks in all phase
space.

As we have argued, processes that can be described by two $b$ quarks
in the initial state, such as $pp\to Hbb $ and $pp \to Z bb$
entail a simple extension of our approach. The comparison between the
5F and 4F schemes can be performed at NNLO for the former and NLO
level for the latter and results could be compared with alternative
approaches, such as the proposal of Ref.~\cite{Harlander:2011aa}.  We
leave this to future work.  In addition, given  the absence of any phenomenological
motivation for an intrinsic  charm contribution to the proton, it would be certainly
interesting to investigate to which extent the results obtained here for $b$ quarks
can be applied to $c$ quarks.  On a more speculative level, one could also
wonder whether the approach followed in this work could be employed to
develop a criterium for an optimal central factorization scale choice
for processes at the LHC.  We have argued that initial state collinear
logs display a universal form valid at all orders which in fact is
independent of the details of splitting as well as of the hard
scattering, but only due to phase space.  This suggests that for
arbitrary processes with light partons in the initial state collinear
logs are of the form $\log \frac{{\cal Q}^2(z)}{\Lambda^2_{\rm QCD}}$
and therefore factorization scales chosen on event-by-event basis as
$\mu_{\rm F} \simeq {\cal Q}(z)$ would lead to an improvement of the
perturbative expansion. Developments in this direction would be
certainly welcome.

\section*{Acknowledgements}

We would like to thank Scott Willenbrock  and Stefano Forte for many useful discussions on this topic and for comments on this work. F.M. is thankful  for the  always lively  discussions on heavy-quark PDF's to Francesco Tramontano, Fred Olness, John Cambpell, Michael Kraemer, Michael Spira, Michelangelo Mangano,  Paolo Nason, and Robert Harlander.
F.M. and M.U.'s work is partially  supported by the Belgian IAP Program, BELSPO P6/11-P and IISN conventions. M.U. is supported by the Bundesministerium f\"ur
Bildung and Forschung (BmBF) of the Federal Republic of Germany
(project code 05H09PAE).

\appendix

\section{LO matrix elements for heavy quark production}
\label{hq}

In this Appendix we give the explicit expressions for leading order partonic
cross sections relevant for $Wb$ production and $tb$ production.

\subsection{Heavy quark production (DIS and single top)}
\label{app:qmqM}
\begin{figure}
\includegraphics[width=0.9\textwidth]{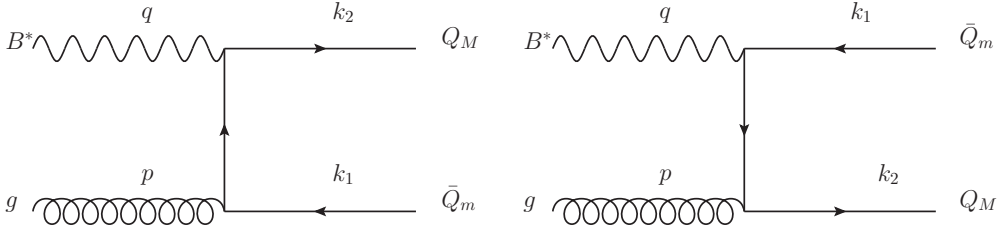}
\caption{\label{diag1}LO Feynman diagrams associated to the $Q_M\,\bar{Q}_m$ production}
\end{figure}

We first consider the production of a heavy quark $Q_m$ of
mass $m$ in association with another heavy quark $Q_M$ of mass $M$ (with
$M\ge m$). In the 4F scheme the heavy
quark of mass $m$ is treated as a massive final state and
does not contribute to the proton wave function. The relevant
leading-order partonic subprocess is
\begin{equation}
g(p)+B^*(q)\,\raw\,\bar{Q}_m(k_1)+Q_M(k_2).
\label{eq:c7:m}
\end{equation}
The results presented in this Appendix are relevant both for the
calculation of electro-production cross sections,
where the virtual vector boson $B^*$ is assumed to be radiated by an
incoming lepton, and for hadroproduction, where the vector boson
is exhanged with a light quark. The process is depicted in
Fig.~\ref{fig:hHQprod}.
The amplitude in the 4F scheme is obtained from the diagrams displayed in figure~\ref{diag1}.
\noindent We define
\begin{align}
t &= (p-k_1)^2 = m^2-2p\cdot k_1,  
\nn\\
u &= (p-k_2)^2 = M^2-2p\cdot k_2, \nn\\
s &= (p+q)^2 = 2p\cdot q - Q^2.
\label{def:dis-mandelstam}
\end{align}
For a generic coupling between the vector boson $B$ and the heavy quarks,
\beq
g\Gamma^\mu \equiv g_R\gamma^\mu\frac{1+\gamma_5}{2}+g_L\gamma^\mu\frac{1-\gamma_5}{2}
\eeq 
the matrix element $\mathcal{M}^\mu$ reads
\beq
\mathcal{M}^\mu = -i\,g\,g_s\,t_c\bar{u}(k_2)
\Bigg[\Gamma^\mu\frac{\pslash-\kslash_1+m}{t-m^2}\gamma^\alpha
+\gamma^\alpha\frac{\kslash_2-\pslash+M}{u-M^2}\Gamma^\mu\Bigg]
v(k_1)\epsilon_\alpha(p).
\label{eq:dis-me}
\eeq
The transverse, longitudinal and axial components of the squared amplitude
can be isolated by suitable projection operators. We define
\begin{align}
\mathcal{M}_G&=-\frac{1}{2}g^{\mu\nu}\mathcal{M}_\mu\mathcal{M}^*_\nu
\\
\mathcal{M}_L&=\frac{4Q^2}{(s+Q^2)^2}p^\mu p^\nu\mathcal{M}_\mu\mathcal{M}^*_\nu
\\
\mathcal{M}_3&=-i\frac{2Q^2}{(s+Q^2)^2}
\varepsilon_{\mu\nu\sigma\rho}p^\sigma q^\rho\mathcal{M}_\mu\mathcal{M}^*_\nu.
\end{align}
We find
\begin{align}
\mathcal{M}_G=& \frac{2g_s^2C_F}{t_1^2u_1^2} \Bigg\lbrace 
(g_L^2+g_R^2) \Bigg[+2 m^2M^2 s_1^2 - 2s_1\left[m^4u_1+M^4t_1+(m^2+M^2)u_1t_1\right]
\nn\\
& +t_1u_1 \left(2 Q^4 -2s_1Q^2+t_1^2+u_1^2\right)+2Q^2\left[m^2u_1(2t_1+u_1)+M^2t_1(2u_1+t_1)\right]\Bigg]
\nn\\
& +16 g_L g_R m M\Bigg[s_1 (u_1m^2+t_1M^2)+u_1t_1(s_1-Q^2)\Bigg]\Bigg\rbrace
\label{mgt}
\\
\mathcal{M}_L=&\frac{16\,g_s^2C_F\,Q^2\left(g_L^2+g_R^2\right)}{s_1^2t_1 u_1} 
\Bigg[s_1 (u_1m^2+t_1M^2)+u_1t_1(s_1-Q^2)\Bigg]
\label{mgl}
\\
\mathcal{M}_3=&
\frac{4\,g_s^2C_F\,Q^2 \left(g_L^2-g_R^2\right)}{s_1^2 t_1^2 u_1^2}
\Bigg[ 2s_1^2(M^4t_1-m^4u_1)+2s_1m^2M^2(t_1^2-u_1^2) 
\nn\\
 & +2s_1Q^2(M^2t_1^2-m^2u_1^2)-2s_1u_1t_1(-s_1+2Q^2)(M^2-m^2)
\nn\\
& -t_1u_1(t_1-u_1)(2Q^4-2Q^2s_1+s_1^2)\Bigg]
\label{mg3}
\end{align}
where $t_1 = t-m^2$, $u_1=u-M^2$ and $s_1=s+Q^2$. The two--body
invariant phase space measure is given by
\beq
d\Phi_2 = \frac{d^3k_1}{(2\pi)^32E_1}\frac{d^3k_2}{(2\pi)^32E_2}
(2\pi)^4\delta^{(4)}(p+q-k_1-k_2)=\frac{1}{8\pi s_1} dt
\label{eq:phasespace}
\eeq
and therefore, by definition,~\cite{HarrisSmith}
\begin{align}
&\frac{d\hat\sigma_G^{\rm{4F}}}{dt}
=\frac{1}{64 s_1}\frac{1}{8\pi s_1}\mathcal{M}_T
\\
&\frac{d\hat\sigma_L^{\rm{4F}}}{dt}
=\frac{1}{32 s_1}\frac{1}{8\pi s_1}\mathcal{M}_L
\\
&\frac{d\hat\sigma_3^{\rm{4F}}}{dt}
=\frac{1}{64 s_1}\frac{1}{8\pi s_1}\mathcal{M}_3.
\end{align}

The partonic cross--section
\beq
\frac{d\hat\sigma_2^{\rm{4F}}}{dt}=\frac{d\hat\sigma_G^{\rm{4F}}}{dt}
+\frac{3}{2}\frac{d\hat\sigma_L^{\rm{4F}}}{dt}
\label{eq:s2}
\eeq
has singularities for $t\raw 0$ in the limit
$m\to 0$ and for $u\to 0$ in the limit $M\to 0$. The amplitudes
Eqs.~(\ref{mgt},\ref{mgl},\ref{mg3}) are manifestly symmetric
under the exchange $t_1\longleftrightarrow u_1$, $M\longleftrightarrow m$.\\
In case of DIS ($g_R=g_L=e_b\sqrt{4\pi\alpha_e}$, $M=m=m_b$) 
Eq.~\eqref{eq:s2} reads 
\begin{eqnarray}
\frac{d\hat\sigma_2^{4F}}{dt} 
&=& \frac{\as\,C_F\,\pi\alpha_e e_H^2}{8s_1^4t_1^2u_1^2} 
\Bigg[4m_b^2s_1^4+2m_b^2s_1^2(14t_1u_1Q^2-s)1^2Q^2-2s_1t_1u_1)\nn\\
&&-t_1u_1[24Q^2t_1u_1(s_1-Q^2)+s_1^2(2Q^4-2Q^2s_1+s_1^2-2t_1u_1)]\Bigg],
\label{app:sigmadis}
\end{eqnarray}
where $e_b=-1/3$ is the electric charge of the bottom quark.\\
In case of single top ($g_R=0$, $g_L=g_w/\sqrt{2}$, $M=M_t$ and $m = m_b$) 
Eq.~\eqref{eq:s2} reads
\begin{eqnarray}
\frac{d\hat\sigma_2^{4F}}{dt} &=& \frac{\as\,C_F\,g_w^2}{128s_1^4t_1^2u_1^2} 
\Bigg[2s_1^3(m_b^4u_1+M_t^4t_1)\,-2s_1^4m_b^2M_t^2 \nn\\
&& +4(M_t^2t_1+m_b^2u_1)\,(s_1^3Q^2-6Q^2t_1u_1s_1)+2s_1^2
t_1u_1(s_1-2Q^2)(m_b^2+M_t^2)\nn\\
&& + s_1^2u_1t_1(2Q^4-2Q^2s_1+s_1^2-2t_1u_1)-24Q^2t_1^2u_1^2(s_1-Q^2)\Bigg].
\label{app:sigmasingletop}
\end{eqnarray}

Choosing the partonic centre--of--mass frame of the $B^{*}g$ system,
the explicit expressions of the four-momenta are
\begin{eqnarray*}
p  &=&\left(\frac{s+Q^{2}}{2\sqrt{s}},0,0,-\frac{s+Q^{2}}{2\sqrt{s}}\right)\\
q    &=&\left(\frac{s-Q^{2}}{2\sqrt{s}},0,0,\frac{s+Q^{2}}{2\sqrt{s}}\right)\\
k_1  &=&\left(E_1,0,|\vec{k}|\sin\theta,|\vec{k}|\cos\theta\right)\\
k_2  &=&\left(E_2,0,-|\vec{k}|\sin\theta,-|\vec{k}|\cos\theta\right),
\end{eqnarray*}
where $E_1,E_2$ and $|\vec{k}|$ are fixed by energy-momentum
conservation and mass-shell relations:
\begin{eqnarray}
 |\vec{k}|^2 &=& \frac{\lambda(s,M^2,m^2)}{4s}
\nn\\
E_1 &=& \frac{s+m^2-M^2}{2\sqrt{s}}
\nn\\
E_2 &=& \frac{s-m^2+M^2}{2\sqrt{s}}
 \label{eq:energies}
\end{eqnarray}
where $\lambda(a,b,c)=a^2+b^2+c^2-2ab-2ac-2bc$.

The masses of the heavy quarks act as regulators of
collinear singularities, originated by poles in $t_1,u_1$
in the squared amplitudes. Since $t_-\leq t\leq t_+$, with
\beq
t_\pm=m^2-\frac{s+Q^{2}}{2s}
\left(s+m^2-M^2\mp\lambda^{1\over 2}(s,M^2,m^2)\right)
\eeq
we find
\beq
L_t=\int_{t_-}^{t_+}\frac{dt}{t-m^2} =
\log\frac{s+m^2-M^2+\lambda^{1\over 2}(s,M^2,m^2)}{s+m^2-M^2-\lambda^{1\over 2}(s,M^2,m^2)},
\label{eq:c7:logt}
\eeq 
and
\beq
L_u=\int_{t_-}^{t_+}\frac{dt}{u-M^2} =
\log\frac{s+M^2-m^2+\lambda^{1\over 2}(s,M^2,m^2)}{s+M^2-m^2-\lambda^{1\over 2}(s,M^2,m^2)}.
\label{eq:c7:logu}
\eeq
In the limit $m^2\ll M^2$, relevant for single top production, we have
\begin{equation}
L_t\raw
\log\left[\frac{s}{m^2}\left(1-\frac{M^2}{s}\right)^{2}\right]
\qquad L_{u}\raw\log\frac{s}{M^2},
\label{eq:c7:limit1}
\end{equation}
whereas for $m^2=M^2\raw 0$ as in the case of $b\bar{b}$ or $c\bar{c}$ production, we find
\begin{equation}
L_t=L_u\raw\log\frac{s}{m^2}.
\label{eq:c7:limit2}
\end{equation}

\subsection{Associated $Wb$ production}
\label{app:wb}
\begin{figure}
\includegraphics[width=0.9\textwidth]{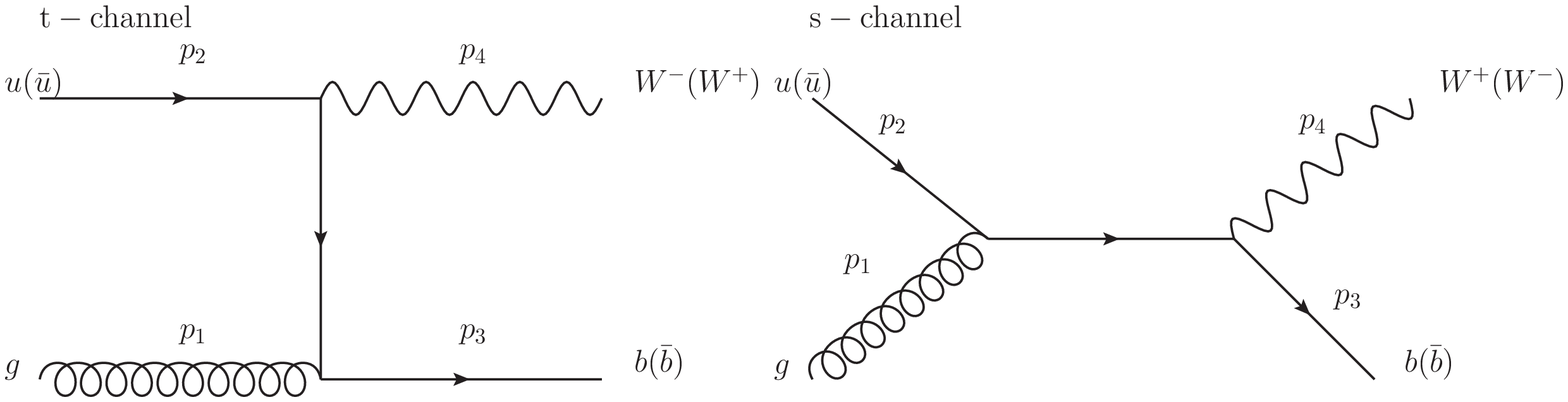}
\caption{\label{diag2}LO Feynman diagrams associated to the $t-$ and $s-$ channel $Wb$ production}
\end{figure}

We now consider associated $Wb$ production. In the simplified case 
in which only one light quark appears in the initial state, e.g. the $u$
quark, the relevant leading-order subprocesses are
\begin{eqnarray}
g(p_1) + \bar{u}(p_2)&\longrightarrow& \bar{b}(p_3)+W^{-}(p_4)\nn\\
g(p_1) + u(p_2)&\longrightarrow& b(p_3)+W^+(p_4)
\end{eqnarray}
and the LO Fewynman diagrams are shown in figure~\ref{diag2}.
\noindent We define
\begin{eqnarray*}
s &=& (p_1+p_2)^2 = 2p_1\cdot p_2\\
t &=& (p_1-p_3)^2 = m_b^2-2p_1\cdot p_3\\
u &=& (p_2-p_3)^2 = m_b^2-2p_2\cdot p_3.
\end{eqnarray*}
This process is related to the one discussed above by crossing, namely
\begin{align*}
& t \leftrightarrow s \qquad u \leftrightarrow t \qquad s\leftrightarrow u.
\end{align*}
Therefore the transversal, longitudinal and axial components of the squared matrix 
elements can be obtained from Eqs.~(\ref{mgt},\ref{mgl},\ref{mg3}) by setting
$g_R=0$, $g_L=g_w/\sqrt{2}$, $m=0$, $M=m_b$ and by crossing
the kinematic variables according to the above equation. 
The partonic differential cross section is given by
\begin{eqnarray}
\frac{d\hat\sigma^{4F}}{dt} &=& \frac{\alpha_s G_F}{12\sqrt{2} s^3 \left(t-m_b^2\right)^2}
\Bigg[m_b^8 -m_b^6 (2 s+t)-m_b^4 \left(2 M_W^4+2 M_W^2t-(s+t)^2\right)\nn\\
&&-m_b^2 \bigg(-4 M_W^6+2 M_W^4 t-2 M_W^2 \left(s^2-s t+2 t^2\right)+t (s+t)^2\bigg)\nn\\
&&-2 M_W^2t \left(2 M_W^4-2 M_W^2 (s+t)+s^2+t^2\right)\Bigg].\label{m0wb}
\end{eqnarray}
In the partonic centre of mass frame we can write the 4-momenta of the particles as
\begin{eqnarray*}
p_1&=&\left(\frac{\sqrt{s}}{2},0,0,\frac{\sqrt{s}}{2}\right)\\
p_2&=&\left(\frac{\sqrt{s}}{2},0,0,-\frac{\sqrt{s}}{2}\right)\\
p_3&=&(E_3,0,|\vec{p}|\sin{\theta},|\vec{p}|\cos{\theta})\\
p_4&=&(E_4,0,-|\vec{p}|\sin{\theta},-|\vec{p}|\cos{\theta})
\end{eqnarray*}
where, by the energy--momentum conservation, 
\begin{eqnarray}
 |\vec{p}| &=& \frac{\lambda^{1\over 2}(s,M_W^2,m_b^2)}{2\sqrt{s}}\nn\\
 E_3 &=& \frac{s+m_b^2-M_W^2}{2\sqrt{s}}\nn\\
 E_4 &=& \frac{s+M_W^2-m_b^2}{2\sqrt{s}}.
 \label{eq:c7:energies}
\end{eqnarray}
The 4F LO partonic cross section is readily obtained by integration in the range
$t_-\leq t\leq t_+$, with
\begin{align}
t_-&=m_b^2-\frac{1}{2}\left[s+m_b^2-M_W^2 +\lambda^{1\over 2}(s,m_b^2,M_W^2)\right]
\nn\\
t_+&=m_b^2-\frac{1}{2}\left[s+m_b^2-M_W^2 -\lambda^{1\over 2}(s,m_b^2,M_W^2)\right].
\label{eq:t-extr}
\end{align}
The dominant contribution in the collinear limit $t\to 0,m_b\to 0$
is therefore proportional to
\beq
L=\log\frac{t_+-m_b^2}{t_--m_b^2}
\to\log
\frac{(s-M_W)^2}{m_b^2 s}.
\label{logWb}
\eeq 

\section{Higher orders}
 \label{app:ho}
In Section~\ref{sec:general} we have considered the general class of processes
\beq
I(q)+g(p)\to b(k)+g(k_1)+\ldots +g(k_n)+X(P),
\label{gen}
\eeq
where $X$ is a generic heavy system, such that $P^2\geq M^2$.
By using a simple argument we have shown that transverse momentum of each of the $n$ gluons in the final state is
bounded from above by the same quantity and therefore expect a factor of
\beq
\log\frac{{\cal Q}^2(z)}{m_b^2}
\eeq
for each emitted gluon.

This simple argument has the disadvantage that it does not provide us
we the exact expression of the cross section in the collinear limit.
The argument can be made more precise by studying the explicit
expression of the squared amplitude and of the phase space measure for
multiparton emission in the collinear limit. In order to simplify
our discussion, we specialize to the case when $X$ is a real heavy particle,
such as a top quark or a $W$ boson. 

The dominant Feynman diagrams in the collinear limit are shown in 
Fig.~\ref{fig:ladder}.

It will therefore be convenient to express both the squared amplitude
and the phase space measure in terms of the variables
\begin{align}
t&=(p-k)^2
\nonumber\\
t_1&=(p-k-k_1)^2
\nonumber\\
&\ldots
\nonumber\\
t_n&=(p-k-k_1-\ldots-k_n)^2
\end{align}
because the leading contribution to the squared amplitude in the collinear 
limit will be proportional to
\beq
\frac{1}{t-m_b^2}
\frac{1}{t_1-m_b^2}\ldots\frac{1}{t_n-m_b^2}.
\label{collsing}
\eeq
The $(n+2)$-body phase space measure can be worked out
by the usual factorization technique~\cite{BK}. We obtain
\beq
d\Phi_{n+2}(p,q;k,k_1,\ldots,k_n,P)
=\int_{M^2}^{(\sqrt{s}-m_b)^2}
\frac{dM^2_n}{2\pi}\,d\Phi_2(p,q;k,P_n) d\Phi_{n+1}(P_n;k_1,\ldots,k_n,P),
\label{eq:phasespacet}
\eeq
where
\beq
P_n=k_1+\ldots+k_n+P;\qquad P_n^2=M_n^2.
\eeq
In the reference frame defined by 
Eqs.~\eqref{p} and~\eqref{q} we have
\beq
d\Phi_2(p,q;k,P_n)=\frac{1}{16\pi^2}\frac{|\vec k|}{\sqrt{s}}
d\phi d\cos\theta,
\label{dphi2}
\eeq
where $\theta,\phi$ are the polar and azimutal angles of $\vec{k}$,
and
\beq
|\vec{k}|^2=\frac{\lambda(s,m_b^2,M_n^2)}{4s}.
\label{K}
\eeq
Because
\beq
t=(p-k)^2=m_b^2-\frac{s+Q^2}{\sqrt{s}}\left(\sqrt{m_b^2+|\vec{k}|^2}
-|\vec{k}|\cos\theta\right),
\label{t}
\eeq
we may recast Eq.~\eqref{dphi2} in the Lorentz-invariant form
\beq
d\Phi_2(p,q;k,P_n)=\frac{1}{16\pi^2}\frac{dt}{s+Q^2}d\phi.
\eeq
The kinematic bounds on $t$ are easily read off Eqs.~(\ref{K},\ref{t}):
\begin{align}
&t^-\leq t\leq t^+
\nonumber\\
&\qquad t^\mp=m_b^2-\frac{s+Q^2}{2s}\left[s+m_b^2-M_n^2
\pm\lambda^{1\over 2}(s,m_b^2,M_n^2)\right].
\end{align}
The factorization procedure can be iterated until the whole
phase space is expressed as a product of two-body phase space measures.
Let us take one more step explicitly:
\beq
d\Phi_{n+1}(P_n;k_1,\ldots,k_n,P)
=\int_{M^2}^{M_n^2}\frac{dM_{n-1}^2}{2\pi}\,
d\Phi_2(P_n;k_1,P_{n-1})d\Phi_n(P_{n-1};k_2,\ldots,k_n,P),
\eeq
where
\beq
P_{n-1}=k_2+\ldots+k_n+P;\qquad P_{n-1}^2=M_{n-1}^2.
\eeq
Choosing
\begin{align}
&p-k=\left(\sqrt{t+\omega^2},0,0,\omega\right)
\label{pmk}
\\
&q=\left(\sqrt{q^2+\omega^2},0,0,-\omega\right)
\end{align}
the condition $(p-k+q)^2=M_n^2$ gives
\beq
\omega=\frac{\lambda^{1\over 2}(M_n^2,t,q^2)}{2M_n}.
\eeq
A straightforward calculation yields
\begin{align}
t_1=(p-k-k_1)^2
&=t-\frac{M_n^2-M_{n-1}^2}{2M_n^2}
\left[M_n^2+t-q^2-\lambda^{1\over 2}(M_n^2,t,q^2)\cos\theta_1\right]
\label{t1}
\end{align}
so that
\beq
d\Phi_2(P_n;k_1,P_{n-1})
=\frac{1}{16\pi^2}\frac{dt_1}{\lambda^{1\over 2}(M_n^2,t,q^2)}d\phi_1
\eeq
with
\begin{align}
&t_1^-\leq t_1\leq t_1^+
\nonumber\\
&\qquad t_1^\mp=t-\frac{M_n^2-M_{n-1}^2}{2M_n^2}
\left[M_n^2+t-q^2\pm\lambda^{1\over 2}(M_n^2,t,q^2)\right].
\end{align}
Thus
\begin{align}
d\Phi_{n+2}(p,q;k,k_1,\ldots,k_n,P)
=&\int_{M^2}^{(\sqrt{s}-m_b)^2}\frac{dM_n^2}{2\pi}\,
\frac{1}{(4\pi)^2}\frac{dt}{s+Q^2}d\phi
\nonumber\\
&\int_{M^2}^{M_n^2}\frac{dM_{n-1}^2}{2\pi}\,
\frac{1}{(4\pi)^2}\frac{dt_1}{\lambda^{1\over 2}(M_n^2,t,q^2)}d\phi_1
d\Phi_n(P_{n-1};k_2,\ldots,k_n,P).
\end{align}
The procedure can be iterated, to obtain
\begin{align}
&d\Phi_{n+2}(p,q;k,k_1,\ldots,k_n,P)
=\frac{1}{(2\pi)^n}\frac{1}{(4\pi)^{2n+2}}
d\phi d\phi_1\ldots d\phi_n
\nonumber\\
&\qquad \int_{M^2}^{(\sqrt{s}-m_b)^2}dM_n^2\int_{M^2}^{M_n^2}dM_{n-1}^2
\ldots \int_{M^2}^{M_2^2}dM_1^2
\nonumber\\
&\qquad
\frac{dt}{s+Q^2}
\int_{t^-}^{t^+}\frac{dt_n}{\lambda^{1\over 2}(M_1^2,t_{n-1},q^2)}
\int_{t_n^-}^{t_n^+}\frac{dt_{n-1}}{\lambda^{1\over 2}(M_2^2,t_{n-2},q^2)}
\ldots
\int_{t_1^-}^{t_1^+}\frac{dt_1}{\lambda^{1\over 2}(M_n^2,t,q^2)}
\label{ps}
\end{align}
with
\begin{align}
&t_j^-\leq t_j\leq t_j^+
\nonumber\\
&\qquad t_j^\mp=t_{j-1}-\frac{M_{n-j+1}^2-M_{n-j}^2}{2M_{n-j+1}^2}
\left[M_{n-j+1}^2+t_{j-1}-q^2\pm\lambda^{1\over 2}(M_{n-j+1}^2,t_{j-1},q^2)\right]
\end{align}
for $1\leq j\leq n$, where we understand that $t_0\equiv t$, $M_0\equiv M$.

Our next task is evaluating the most singular term
in the cross section, $\sigma^{\rm coll}$,
in the limit $m_b\to 0$. This is obtained by
integrating
the squared amplitude, Eq.~\eqref{collsing}, using the integration measure
Eq.~\eqref{ps}. Let us consider the case $n=1$ as an example.
In this case
\begin{align}
\sigma^{\rm coll}&=
\int_{M^2}^{(\sqrt{s}-m_b)^2}dM_1^2
\int_{t^-}^{t^+}dt\,\frac{1}{t-m_b^2}\frac{1}{\lambda^{1\over 2}(M_1^2,t,q^2)}
\int_{t_1^-}^{t_1^+}\frac{dt_1}{t_1-m_b^2}
\nonumber\\
&=\int_{M^2}^{(\sqrt{s}-m_b)^2}dM_1^2
\int_{t^-}^{t^+}dt\,\frac{1}{t-m_b^2}\frac{1}{\lambda^{1\over 2}(M_1^2,t,q^2)}
\log\frac{t_1^+-m_b^2}{t_1^--m_b^2},
\label{sigmacoll}
\end{align}
where
\begin{align}
&t_1^\mp=t-\frac{M_1^2-M^2}{2M_1^2}
\left[M_1^2+t-q^2\pm\lambda^{1\over 2}(M_1^2,t,q^2)\right]
\label{t1bounds}
\\
&t^\mp=m_b^2-\frac{s+Q^2}{2s}\left[s+m_b^2-M_1^2
\pm\lambda^{1\over 2}(s,m_b^2,M_1^2)\right],
\label{tbounds}
\end{align}
and we have omitted an overall proportionality constant which takes into
account, among other factors, the result of azimuthal integrations.
The collinear singularity occurs at the upper bound of integration, $t^+$,
which vanishes as $m_b\to 0$, as can be seen expanding
$t^\pm$ Eq.~\eqref{tbounds} in powers of $m_b^2$ to the first
non-trivial order:
\begin{align}
&t^-=-\frac{(s+Q^2)(s-M_1^2)}{s}+O(m_b^2)\label{tm}
\\
&t^+=-m_b^2\frac{M_1^2+Q^2}{s-M_1^2}
\label{tp}
+O(m_b^4).
\end{align}
We now note that both $\lambda^{-1/2}(M^2,t,q^2)$
and $t_1^-$ Eq.~\eqref{t1bounds}
are regular in $t=t^+\to 0$, and can therefore be
computed at $t=t^+$. Indeed
\begin{align}
&t_1^-=-\frac{(M_1^2-M^2)(M_1^2+Q^2)}{M_1^2}+O(t)
\label{t1m}
\\
&t_1^+=t\frac{M^2+Q^2}{M_1^2+Q^2}+O(t^2).
\label{t1p}
\end{align}
In order to set an upper bound for the most singular contribution to $\sigma^{\rm coll}$, Eq.~\eqref{sigmacoll},
we neglect  the $\mathcal{O}(t)$ and $\mathcal{O}(t^2)$ contributions in 
Eqs.~\eqref{t1m}-\eqref{t1p}, and note that the limits of integration in $t_1$ are bound by
\begin{align}
&t_1^- \ge -\frac{(s-M^2)(s+Q^2)}{s}
\label{t1m-disequality}
\\
&t_1^+ \le  t^+  \frac{M^2+Q^2}{M_1^2+Q^2} = 
- m_b^2\frac{M^2+Q^2}{s-M_1^2} \le - m_b^2\frac{M^2+Q^2}{s-M^2}  .
\label{t1p-disequality}
\end{align}
The first inequality is a consequence of the fact that
$t_1^-$ is an increasing function of $M_1^2$, and $M_1^2\le s$.
The second inequality follows from  $t\le t^+$ and $M_1^2\ge M^2$.
In so doing, the integrations in $t_1$ and $t$ become independent.
The integration in $t_1$ yields
\beq
\log \frac{t^-_1}{t^+_1-m_b^2} \le 
\log \frac{(s-M^2)^2}{s m_b^2}  = \log \frac{\mathcal{Q}^2(z)}{m_b^2}.
\eeq
The integration in $t$, up to less singular terms, gives rise to a logarithm 
which is also bounded by $\log \mathcal{Q}^2(z)/m_b^2$:
\beq
\log \frac{t^-}{t^+-m_b^2}  = 
\log\frac{(s-M_1^2)^2}{sm_b^2}
\le\log\frac{(s-M^2)^2}{sm_b^2} =  \log \frac{\mathcal{Q}^2(z)}{m_b^2}
\eeq
since $M_1^2\ge M^2$. Thus,
\beq
\sigma^{\rm coll}\leq
\int_{M^2}^{(\sqrt{s}-m_b)^2}\frac{dM_1^2}{M_1^2+Q^2}
\log^2 \frac{\mathcal{Q}^2(z)}{m_b^2}.
\label{sigmacoll2}
\eeq
This proves that the argument of the collinear logarithms 
is bounded from above by
$\frac{\mathcal{Q}^2(z)}{m_b^2}$ even in presence of one extra emission.
The proof can be
easily extended to the case of $n$ extra gluon emissions, just noting
that the integrations over $M^2_j$ in Eq.~\ref{ps} can
disentangled by using $M_n^2$ as the upper limit of integration for
all of them, neglecting $t_j$ in the
$\lambda^{\frac12}(M^2_j,t_k,q^2)$ denominators, and then recursively
building lower and upper bounds to the $t^{\mp}_j$ limits.

\begin{figure}[t]
\begin{center}
 \includegraphics[width=0.3\textwidth]{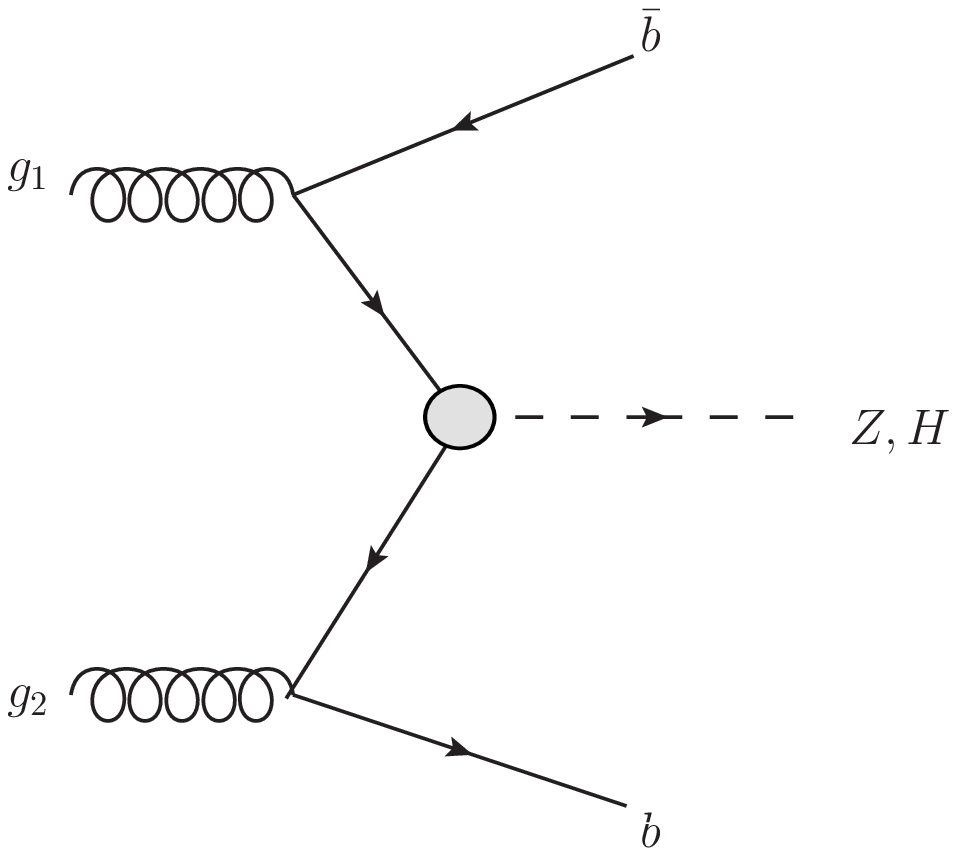}
\end{center}
\caption{$gg\to Z/H bb$ production.\label{fig:bb}
}
\end{figure}

The above proof can also be used with minor modifications to show that
$\frac{\mathcal{Q}^2(z)}{m_b^2}$ are universal kinematic terms
associated to the collinear logs in processes that can be described
via $b$-PDF for both initial legs, such as $Hbb$ and $Zbb$ production displayed in figure~\ref{fig:bb}.

In this case one can write:

\begin{eqnarray}
\sigma_{\rm coll} \propto
\int_{(m_b+M)^2}^{(\sqrt{s}-m_b)^2}  \frac{ds_1}{s_1} 
\int_{t^-_{2}}^{t^+_2}    \frac{1}{t_2 - m_b^2} dt_2 
\int_{t^-_{1}}^{t^+_1}  \frac{1}{t_1 - m_b^2} dt_1 
\label{eq:recursive_t}
\end{eqnarray}

with
\begin{eqnarray}
s_1 &=& (p_{\bar b}+ p_M)^2\\
t_1 &=& (p_1 -p_{\bar b})^2\\
t_2 &=& (p_2 -p_{b})^2\\
t_1^\pm  &=& m_b^2-\frac12 \frac{s_1-t_2}{s_1}  \left[ ( s_1 -M^2 + m_b^2) \pm  \lambda^{1\over 2}(s_1,M^2,m_b^2)\right] \\
t_2^\pm  &=& m_b^2  -\frac12   \left[  s-  s_1 + m_b^2  \pm \lambda^{1\over 2}(s,m_b^2,s_1)\right] 
\end{eqnarray}
Integration over $t_1$ directly leads to 
\beq
\log \frac{t^-_1-m_b^2}{t^+_1-m_b^2} \simeq \log \frac{(s_1-M^2)^2}{s_1 m_b^2} \,
\eeq
while integration over $t_2$ gives
\begin{equation}
\log \frac{t^-_2-m_b^2}{t^+_2-m_b^2} \simeq \log \frac {(s-s_1)^2}{s m_b^2} \,,
\end{equation}
leading to
\begin{eqnarray}
\sigma_{\rm coll} \propto
\int_{(m_b+M)^2}^{(\sqrt{s}-m_b)^2}  \frac{ds_1}{s_1} \log \frac{(s_1-M^2)^2}{s_1 m_b^2}  \log \frac {(s-s_1)^2}{s m_b^2}
\label{eq:hbb}
\end{eqnarray}
It is easy to see that the $s_1 \sim (\sqrt{s}-m_b)^2 $ and $s_1 \sim (m_b+M)^2$ integration regions each give rise to single logs of the form $\log \frac{\mathcal{Q}^2(z)}{m_b^2}\,$. The single logs correspond to the case where only one of the two $b$'s is collinear to the initial state while the other takes part to the genuine hard scattering (in the 5F language this configuration is described by $gb\to bh,bZ$). The double log (i.e.,  the dominant contribution that in the 5F approach leads to the leading order process $bb\to Z,h$)  is obtained from the intermediate region of integration for $s_1$, where   $s_1\simeq s_2$ and 
$s \gg s_1$ and $s_1\gg M^2$, a region that is always kinematically accessible due to the identity
\begin{equation}
s = s_{bb} - s_1 - s_2 + m_b^2 + m_b^2 + M^2\,.
\end{equation}
In such a region 
\begin{equation}
\log \frac {(s_1-M^2)^2}{s_1 m_b^2} \log \frac {(s-s_1)^2}{s m_b^2}  \simeq \log^2 \frac{\mathcal{Q}^2(z)}{m_b^2}\,.
\end{equation}

\section{Explicit NLO expression for $\tilde{b}$}

The coefficients $a_{j,b}^{(p)}$ (with $j=\Sigma,g$ and $p=1,2$), up to constant terms (that we set to zero in order to fulfil the boundary conditions $\tilde{b}(x,m_{b}^{2})=0$) are given by~\cite{Buza:matchPDF}
{\small
\begin{eqnarray}
a_{g,b}^{(1)}\left(z,\frac{\mu^{2}}{m_{b}^{2}}\right) &=& 2\,P_{qg}(z)\,\log\frac{\mu^{2}}{m_{b}^{2}} 
\end{eqnarray}
\begin{eqnarray}
a_{g,b}^{(2)}\left(z,\frac{\mu^{2}}{m_{b}^{2}}\right) &=& \bigg\lbrace C_{F}T_{R}\left[8\,P_{qg}(z)\log(1-z)-(2-4z+8z^{2})\log z -(1-4z)\right]\nn\\
&& - C_{A}T_{R}\left[8P_{qg}(z)\log(1-z) + (4+16z)\log z + \frac{8}{3z} + 2 + 16z - \frac{62}{3}z^{2}\right]\nn\\
&& + T_{R}^{2}\left[-\frac{8}{3}(z^{2}+(1-z)^{2})\right]\bigg\rbrace \log^{2}\frac{\mu^2}{m_b^2}\nn\\
&-& \bigg\lbrace C_{F}T_{R}\Bigg[ 8P_{qg}(z)[2\log z \log(1-z)-\log^{2}(1-z)+2\zeta(2)] \nn\\
&& \hspace*{1cm}- (2-4z+8z^{2})\log^{2}z-16z(1-z)\log(1-z) \nn\\
&& \hspace*{1cm}-(6-8z+16z^{2})\log z - 28 + 58z-40z^{2}\Bigg] \nn\\
&& + C_{A}T_{R}\Bigg[(8+16z+16z^2)[{\rm Li}_{2}(-z)+\log z\log(1+z)]+8P_{qg}(z)\log^{2}(1-z)\nn\\
&& \hspace*{1cm}+(4+8z)\log^{2}z+16z\zeta(2)+16z(1-z)\log(1-z) -\left(4+32z+\frac{176}{3}z^{2}\right)\log z\nn\\
&& \hspace*{1cm}-\frac{80}{9z}+8-100z+\frac{872}{9}z^2\Bigg]\bigg\rbrace \log\frac{\mu^2}{m_b^2}\,,
\end{eqnarray}
}for the gluon initiated splitting and by 
{\small
\begin{eqnarray}
a_{\Sigma,b}^{(2)}\left(z,\frac{\mu^{2}}{m_{b}^{2}}\right) &=&  \bigg\lbrace C_{F}T_{R}\left[-4(1+z)\log z-\frac{8}{3z}-2+2z+\frac{8}{3}z^{2}\right]\bigg\rbrace \log^{2}\frac{\mu^2}{m_b^2}\\
 &+&  \bigg\lbrace C_{F}T_{R}\left[-4(1+z)\log^{2}z+\left(4+20z+\frac{32}{3}z^2\right)\log z +\frac{80}{9z}-8+24z-\frac{224}{9}z^{2}\right]\bigg\rbrace \log\frac{\mu^2}{m_b^2}\,,\nn
\end{eqnarray}
}for the quark--initiated splitting. In the above expression we used the Altarelli-Parisi splitting function $P_{qg}$ defined as
\begin{equation}
P_{qg}(z) = T_{R}[z^{2}+(1-z)^{2}] \qquad T_R=\frac{1}{2}\,.
\label{eq:apqg}
\end{equation}

\bibliographystyle{JHEP}
\bibliography{main}

\end{document}